\documentclass{aa}

\usepackage{float} 

\usepackage{textcomp}
\usepackage[utf8]{inputenc}

\usepackage{amsmath, amssymb, amsfonts, amscd}
\usepackage{mathtools} 
\usepackage{array} 
\usepackage{cases} 
\usepackage{empheq} 
\usepackage{multirow} 
\usepackage{txfonts} 

\usepackage{graphicx} 
\usepackage{wrapfig} 

\usepackage[usenames,dvipsnames]{color} 
\usepackage{ulem} 
\usepackage[colorlinks=true,urlcolor=blue,citecolor=blue]{hyperref} 
\usepackage{natbib}

\usepackage[toc,page,titletoc]{appendix}
\usepackage{colortbl}

\begin{document}

\title{\Large Theoretical seismology in 3D : \\nonlinear simulations of internal gravity
  waves in solar-like stars}

\author{L. Alvan \inst{1}, A.S. Brun\inst{1}, S. Mathis\inst{1}}
\offprints{L. Alvan}

\institute{
Laboratoire AIM Paris-Saclay, CEA/DSM - CNRS - Universit\'e Paris Diderot, IRFU/SAp Centre de Saclay, F-91191 Gif-sur-Yvette, France
 \\{}\\{}
 \email{lucie.alvan@cea.fr, sacha.brun@cea.fr, stephane.mathis@cea.fr} 
}

\date{}

\abstract
{ Internal gravity waves (hereafter IGWs) are studied for their impact on
  the angular momentum transport in stellar radiation zones and the
  information they provide about the structure and dynamics of deep stellar
  interiors. We present the first 3D nonlinear numerical simulations of IGWs excitation and
  propagation in a solar-like star.
}
{ The aim is to study the behavior of waves in a realistic 3D
  nonlinear time-dependent model of the Sun and to characterize their properties.
}
{ We compare our results with theoretical and 1D predictions. It allows us
  to point out the complementarity between theory and simulation and to
  highlight the convenience, but also the limits, of the asymptotic and
  linear theories.
}
{ We show that a rich spectrum of IGWs is excited by the
  convection, representing about 0.4\% of the total solar luminosity. We
  study the spatial and temporal properties of this spectrum, the effect of
  thermal damping, and nonlinear interactions between waves. 
  We give quantitative results for the modes' frequencies, evolution with
  time and rotational splitting, and we discuss the
  amplitude of IGWs considering different regimes of parameters. 
}
{ This work points out the importance of high-performance simulation for its
  complementarity with observation and theory. It opens a large field of
  investigation concerning IGWs propagating nonlinearly in 3D spherical
  structures. The extension of this work to other types of stars, with
  different masses, structures, and rotation rates will lead to a
  deeper and more accurate comprehension of IGWs in stars.
}

\keywords{{Convection, Hydrodynamics, Stars: interiors, Sun: interior, Turbulence, Waves}}

\titlerunning{3D nonlinear simulations of internal gravity
  waves}
\authorrunning{L. Alvan, A.S. Brun, S. Mathis}

\maketitle

\section{Introduction}

IGWs are perturbations propagating in stably
stratified regions under the influence of gravity. Planetary atmospheres and
stellar radiation zones are therefore ideal places to find them. For example,
they can be observed in striated cloud structures in Earth's atmosphere where they are known
to produce large-scale motions such as the quasi-biennial oscillation
(QBO) in the lower stratosphere
\citep{1978JAtS...35.1827P,1997JGR...10226053D,2001RvGeo..39..179B,2002GeoRL..29.1245G}.
In stars, IGWs propagate in the radiative cores
of low-mass stars and the external envelopes of intermediate-mass and
massive stars \citep[e.g.,][]{aerts2010asteroseismology}. {High-frequency gravity modes have been
observed in solar-like stars \citep[e.g.,][]{1995ApJ...443L..29C} and more
massive stars}. IGWs are known for their
ability to mix {chemical} species and to transport angular momentum, affecting
the evolution of stars. 
They can be excited by several processes, depending on the
type of stars being considered. In single stars, three excitation processes have been
invoked. {First}, the
$\kappa$-mechanism is due to opacity bumps in ionization regions
 \citep[e.g.,][]{1989nos..book.....U,Gastine:2010ki}. {Next}, the $\epsilon$-mechanism
 occurring in massive evolved stars is a modulation of the nuclear reaction
 rate in the core \citep[e.g.,][]{2012ApJ...749...74M}. {Finally}, for solar-type
 stars, IGWs are mainly excited by stochastic motions such as the pummeling
 of convective plumes at the interface with adjacent radiative zones
 \citep[e.g.,][]{1986ApJ...311..563H,1990ApJ...363..694G,BBT2004,2005ApJ...620..432R,Belkacem:2009cc,Brun:2011bl,Shiode:2013kp,Lecoanet:ws}.\\\newline
{The propagation of IGWs in stellar radiative zones can affect
  their evolution on secular timescales.} They have been subject to intense theoretical studies,
invoking them to explain several physical mechanisms. With the large-scale
meridional circulation \citep{1992A&A...265..115Z,2004A&A...425..229M},
the different hydrodynamical shear and baroclinic instabilities \citep{1983apum.conf..253Z}, and
the fossil magnetic field
\citep{1998Natur.394..755G,BrunZahn2006,2008MNRAS.391.1239G,Duez:2010hg,2011AN....332..891S}, IGWs constitute
the fourth main process responsible for the angular momentum redistribution in
radiative interiors. Indeed, when they propagate, IGWs are
able to transport and deposit a net amount of angular momentum by radiative damping
\citep{Press1981,1993A&A...279..431S,ZahnTalonMatias1997}
and corotation resonances
\citep{Booker:1967wd,2013A&A...553A..86A}. Their action induces important changes in the internal rotation
profiles of stars during their evolution \citep{2008A&A...482..597T,Charbonnel:2013df,2013A&A...558A..11M}. In
the particular case of the Sun, IGWs are serious candidates to explain the solid body
rotation of its radiative interior down to 0.2$R_\odot$ \citep{1999ApJ...520..859K,CharbonnelTalon2005}. 
They may also provide the extra mixing required to answer the Li depletion question in F stars
\citep{1991ApJ...377..268G} and in the Sun
\citep{Montalban:1994wx}.\\\newline
{By interfering constructively, IGWs form standing modes also
known as gravity (g) modes.} Indeed, gravity waves' frequencies must be inferior to the Brunt-Väisälä (BV)
frequency deduced from the characteristics of the star (gravity, density,
and pressure profiles). For this reason, IGWs can propagate only in a
limited cavity and are susceptible to entering resonance, according to the
geometry of this cavity. {Such modes have became} the
object of study of astero- and helioseismology
\citep{aerts2010asteroseismology,Christensen-Dalsgaard97lecturenotes}, together
with acoustic (p) modes. Detecting and characterizing
g-modes is of great interest for obtaining informations about the inner
structure of different types of stars.\\\newline
For white dwarfs, \citet{1968ApJ...153..151L} was the first to observe a
rapid timescale oscillation in the single white dwarf now known as HL Tau
76. Four years later, \citet{1972NPhS..239....2W} and
\citet{1972NPhS..236...83C} were able to identify these oscillations with
nonradial gravity mode pulsations. Today, an abundance of reports of
high-frequency variability in white dwarf stars have been found and used to
understand the motions and internal composition of these stars
\citep{2005EAS....17..133V,Winget:2008dz}. In the case of subdwarf B (sdB) stars,
\citet{Green:2002va} observed a new class of sdB
pulsators with periods of about an hour corresponding to gravity
modes. And other reports have been made about detections of gravity modes
in the {upper} main-sequence (for example, in slowly pulsating B (SPB) and Be stars)
\citep{1991A&A...246..453W,DeCat:2011gq,2012A&A...546A..47N}. In the past few years, the importance of
g-modes have been underlined thanks to the CoRoT and Kepler
missions. In particular, the detection of mixed-modes{ that have the character of g-modes in the core region and of
p-modes in the envelope} has led to numerous
results in red giant seismology \citep[see][for a complete
review]{Mosser:2013vu}. For instance, \citet{Bedding:2011te} have used them as a way to distinguish
between hydrogen- and helium-burning red giants, and they also provide good
results for the deduction of the core rotation from the measurement of
their rotational splitting
\citep{2012Natur.481...55B,Mosser:2013uk,2012ApJ...756...19D}.\\\newline
However, g-modes remain hardly detectable in the Sun and solar-like stars
\citep{Anonymous:0GcaI-aC,TurckChieze:2004vw,2010A&ARv..18..197A}. Indeed,
these stars possess outer convective envelopes where IGWs are evanescent. They thus have a low
amplitude when they reach the photosphere level where one could have a chance to detect
the oscillations. In past years,
intense research have been invested in the quest for the
detection of g-modes in the Sun. {Both theoretical and numerical works
have been undertaken to estimate the solar g-modes' frequencies
\citep{1991SoPh..133..127B} and surface amplitudes} \citep{gough1986seismology,1990A&A...227..563B,1992A&A...257..763A,Anonymous:0GcaI-aC,1996A&A...312..610A,2009A&A...494..191B},
concluding that most powerfull modes should have amplitudes of about
$10^{-3}$ to $10^{-1}$ cm/s \citep{2010A&ARv..18..197A}. Detection of g-modes at the surface of the Sun was one
of the goals of the SOHO mission \citep{1995SSRv...72...81D}. Today,
asymptotic signatures of gravity modes have been found \citep{Garcia:2007iq} and used
to constrain the rotation of the core \citep{Mathur:2008hs}, but the
detection of individual g-modes at the surface of the Sun seems to elude
the community. \\\newline
In parallel with observational and theoretical works, numerical simulations
can help for understanding IGWs' properties and behavior in solar-like stars.
In the Sun, the main mechanism for
exciting IGWs is convective overshoot. Thus, a series of studies have been
performed to determine the extension of convective penetration zone and the
resulting excitation of IGWs in 2D
\citep{1984A&A...140....1M,1986ApJ...311..563H,1994ApJ...421..245H,2005ApJ...620..432R,Rogers:2006ks}
, and in 3D \citep{2000ApJ...529..402S,Brun:2011bl}. Some authors also
compared the spectrum of IGWs excited by convection and the energy flux 
carried by the waves with simpler parametric models of wave generation
\citep{1994SoPh..152..241A,1996A&A...312..610A,2003AcA....53..321K,Kiraga:2005wg,2005A&A...438..365D}. Finally,
the transport of angular momentum by waves has been studied with 1D stellar
evolution codes \citep{Talon:2005iu} but also in 2D
\citep{2006ApJ...653..756R}. 
{Here}, the use of a realistic
stratification in radiation zones is of great importance. Indeed, g-modes are very sensitive to the form of the cavity defined by the BV
frequency, particularly for the central region, under
0.2$R_\odot$ \citep{Brun1998,2012sf2a.conf..289A}. For instance, a slight modification of the
nuclear reaction rates in the model taken for calculating the BV frequency
can induce a frequency shift up to $2\mu$Hz in the range 50-300$\mu$Hz
where solar g-modes are expected to be found. Moreover, as shown by
\citet{RogersGlatzmaier2005} and \citet{Rogers:2008bl}, the effects of wave-wave and wave-mean-flow nonlinear
interactions have to be taken into account, which puts nonlinear codes in
the foreground. \\\newline
In the present work, we show results of 3D spherical
nonlinear simulations of a full sphere solar-like star. The computational
domain extends from 0 to 0.97$R_\odot$ by taking
the full radiative cavity into account. IGWs are naturally excited by penetrative
convection at the interface with the inner radiative zone and can propagate
and {give birth to standing modes} in the cavity. The paper is organized in four sections. After introducing the equations and
notations that define the numerical models, we show in
Sect.~\ref{sec:excit-penetr-conv} that
a rich spectrum of IGWs is excited by convective penetration. In Sect.~\ref{sec:waves-properties}, we
examine the properties of this spectrum precisely, highlighting its richness where both modes and
propagating waves are present. We give quantitative results about the group
velocity of such waves, we measure their period spacing, their lifetime, and
the splitting induced by the rotation. Lastly, Sect.~\ref{sec:amplitudes}
presents our results concerning the waves' amplitude and the effect of the
radiative damping affecting their propagation. In particular, we discuss
the effect of the diffusivities on the amplitude of waves and the nonlinear
wave-wave interactions.

\section{Numerical model}
\label{sec:numerical-model}
\subsection{Equations}
\label{sec:equations}
\begin{figure*}
  \begin{center}
  \includegraphics[width=1.\textwidth]{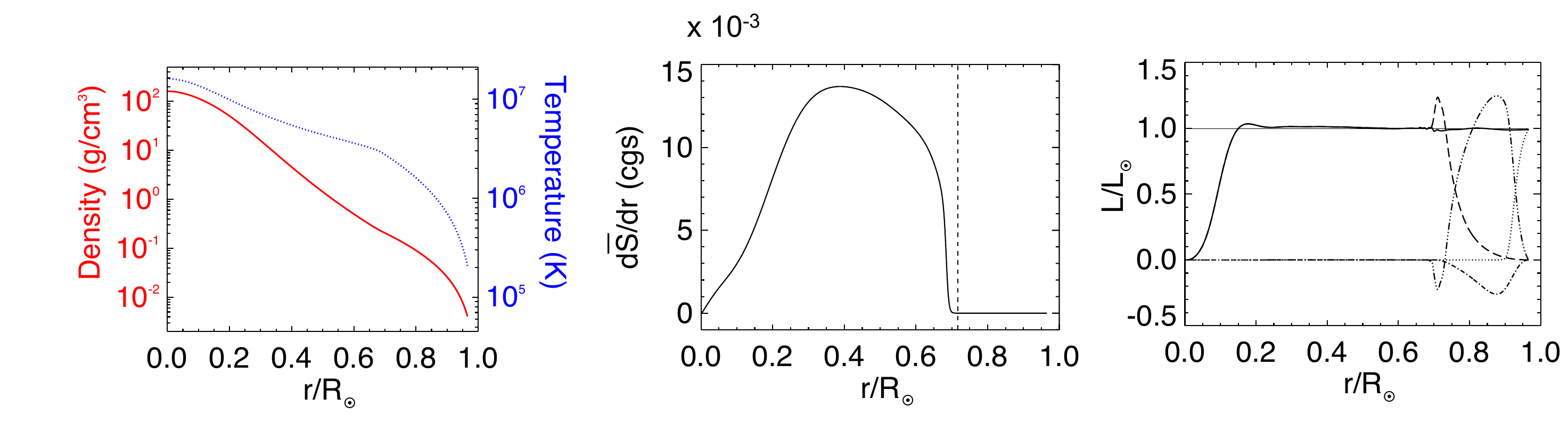}
  \caption{{\textbf{Left}:} Radial profiles of the reference
    density $\bar\rho$ (red / solid line) and temperature $\bar T$ (blue /
    dotted line) as a function of the
    normalized radius. \textbf{Middle:} Radial profiles of the entropy
    gradient. The vertical dotted line marks the radius where $d\bar{S}/dr$
    changes sign. It corresponds to the initial limit between convective and
    radiative zones. \textbf{Right: } Radial energy flux balance
   converted into luminosity and normalized to the solar luminosity $L_\odot$. The
   values have been averaged over latitude, longitude and time {($\approx$ 30 days)}. We show the
   contribution to the energy flux from radiative diffusion (long dashes),
   enthalpy (three-dot-dashed), kinetic energy (dash-dot), modeled SGS
   processes (dot), and viscous diffusion (dashes). The solid line is the
   sum of all these components.}
  \label{fig:profiles}
\end{center} 
\end{figure*}

Following \citet{Brun:2011bl}, we use the hydrodynamic ASH code
\citep{CluneAl1999, BMT2004} to solve the full set of 3D anelastic
equations in a rotating star, treating radiative and
convective regions and their interface simultaneously. 
These equations are fully nonlinear in velocity, and thermodynamic variables are linearized with
respect to a spherically symmetric and evolving mean state.
We note $\bar\rho$, $\bar{{P}}$, $\bar {{T}}$, and $\bar {{S}}$
the reference density, pressure, temperature and specific
entropy. \\
Fluctuations about this reference state are denoted by  $\rho$,
${P}$, ${T}$, and ${S}$. We assume a linearized equation of state
\begin{equation}
  \label{eq:10}
  \frac{\rho}{\bar\rho} = \frac{{P}}{\bar{{P}}} -
  \frac{{T}}{\bar{{T}}} = \frac{{P}}{\gamma\bar{{P}}} -\frac{{S}}{c_p}\hbox{,}
\end{equation}
{and the zeroth-order ideal gas law}
\begin{equation}
  \label{eq:11}
  \bar{{P}} = \mathcal{R} \bar\rho \bar{{T}} \hbox{,}
\end{equation}
where $\gamma$ is the adiabatic exponent, $c_p$ the specific heat per
unit mass at constant pressure, and $\mathcal{R}$ the gas constant. The continuity equation in the anelastic approximation is
\begin{equation}
  \label{eq:27}
  \vec\nabla.\left(\bar\rho\vec {{\mathrm{v}}}\right) = 0 \hbox{,}
\end{equation}
where ${\mathrm{v}} = \left({\mathrm{v}}_r,{\mathrm{v}}_\theta,
  {\mathrm{v}}_\varphi \right)$ is the local velocity expressed in
spherical coordinates $\left( r,\theta,\varphi \right)$ in the frame rotating at constant angular velocity
$\vec\Omega_0=\vec\Omega_0\vec{e_z}$. The usual momentum equation is
\begin{equation}
  \label{eq:288}
  \bar\rho\left(\frac{\partial \vec {{\mathrm{v}}}}{\partial t} + \left(\vec
    {{\mathrm{v}}}.\vec\nabla\right)\vec {{\mathrm{v}}}\right) =-\vec\nabla {P} + \rho\vec{{g}}-2\bar\rho \vec\Omega_0 \times \vec {{\mathrm{v}}} -
\vec\nabla . \vec{\mathcal{D}} -\left[\vec\nabla\bar{{P}} - \bar\rho
  \vec{{g}}\right] \hbox{,}
\end{equation}
where $\vec{{g}}$ is the gravitational acceleration, and $\mathcal{D}$ the
viscous stress tensor defined by
\begin{equation}
  \label{eq:9}
  \mathcal{D}_{ij} = -2\bar\rho\nu\left(e_{ij} - 1/3\left(\vec\nabla  . \vec{{\mathrm{v}}}\right)\delta_{ij}\right)\hbox{,}
\end{equation}
with $e_{ij}=1/2\left( \partial_j {\mathrm{v}}_i+\partial_i {\mathrm{v}}_j  \right)$ the strain rate tensor and $\delta_{ij}$ the Kronecker
symbol. The bracketed term on
the righthand side of Eq.~\eqref{eq:288} is initially zero because the system
begins in hydrostatic balance. Then, as the simulation evolves, the
reference state is driven away from this balance by turbulent
pressure.\\\newline
{\citet{Brown:2012bd} have shown that depending on the
  used anelastic formulation, the quality of the conservation of energy in
  stably-stratified atmospheres varies. In particular, when modeling highly stratified
  radiative interiors, the energy in waves may be overestimated. As a
  consequence, instead of the classical formulation, \citet{Brown:2012bd} advocate implementing the
Lantz-Braginsky-Roberts (LBR) \citep[e.g.,][]{LantzPHD,1995GApFD..79....1B} equations that treat the reduced
pressure 
$\widetilde{\omega}=P/\bar\rho$ instead of the fluctuating pressure $P$. In ASH, the new momentum equation is thus}
\begin{equation}
  \label{eq:289}
\bar\rho\left(\frac{\partial \vec {{\mathrm{v}}}}{\partial t} + \left(\vec
    {{\mathrm{v}}}.\vec\nabla\right)\vec {{\mathrm{v}}}\right) = -\bar{\rho}\vec\nabla
  \widetilde\omega -\bar{\rho}\frac{S}{c_p}\vec{g} - 2\bar\rho \vec\Omega_0 \times \vec
  {{\mathrm{v}}} - \vec\nabla . \vec{\mathcal{D}} \hbox{,}
\end{equation}
{where only the contribution of entropy fluctuations remains
in the buoyancy term, the contribution due to pressure perturbations being included in the reduced pressure gradient. \\
It is important to note that, in
this formulation, we neglect the extra buoyancy term relative to the nonadiabatic reference state in the radiative region. This assumption is based on
energy conservation arguments developed in \citet{Brown:2012bd}.} Finally,
the equation of conservation of energy remains unchanged 
\begin{eqnarray}
  \label{eq:8}
\bar\rho \bar{{T}}\frac{\partial {S}}{\partial t} +
\bar\rho\bar{{T}}\vec{{\mathrm{v}}}.\vec\nabla\left({S}+\bar{{S}}\right)
&=& \bar\rho \epsilon + \vec\nabla . \left[ \kappa_r\bar\rho c_p
  \vec\nabla\left({T}+\bar{{T}}\right) \right.\\\nonumber
\left. +\kappa\bar\rho\bar{{T}}\vec\nabla{S} +
  \kappa_0\bar\rho\bar{{T}}\vec\nabla\bar{{S}} \right] &+&
2\bar\rho\nu\left[e_{ij}e_{ij}-1/3\left(\vec\nabla
    . \vec{{\mathrm{v}}}\right)^2\right] \hbox{,}
\end{eqnarray}
where $\kappa_r$ is the radiative diffusivity based on a 1D solar structure
model. As perturbations and
motions can occur on smaller scales than our grid resolution, the effective
eddy diffusivities $\nu$ and $\kappa$ represent momentum and heat transport
by subgrid-scale (SGS) motions that are unresolved by the
simulation. Their profiles are functions of radius chosen for each model
depending on its objectives. The functions chosen in this work are detailed
in Sect.~\ref{sec:model} and represented in Fig.~\ref{fig:Knu}. \\
The
diffusivity $\kappa_0$ is part of the SGS treatment in the convective zone.
It is set such as to have the unresolved eddy flux (entropy flux) carrying the solar flux outward the top of the
domain (see left panel of Fig.~\ref{fig:profiles}). It drops off
exponentially with depth to ensure that it does not play any role in the
radiative zone \citep{2000ApJ...532..593M}. In Eq.~\eqref{eq:8}, a volume-heating term
$\bar\rho \epsilon$ is also included, representing
energy generation by nuclear burning. We have assumed a simple
representation of the nuclear reaction rate by setting $\epsilon =
\epsilon_0 {T}^k$, with $\epsilon_0$ a constant determined such that
the radially integrated heating term equals the solar luminosity at the
base of the convection zone and $k = 9$. The exponent $k$ is chosen to
obtain a heating source term in agreement with that of a 1D standard model
\citep{BrunAl2002}, considering both contributions of the p-p chains and CNO cycles.

\subsection{Boundary conditions {and time-step control}}

In this paper, we have compared various models of the Sun. For all of
them, the computational domain extends from $r_{\mathrm{bot}}$ = 0 to $r_{\mathrm{top}}$ = 0.97$R_\odot$ where
$R_\odot =6.96\times10^{10}$cm is the solar radius. For the problem
to be well posed, we thus need to define boundary conditions. At 
the top of the domain, we have opted for torque-free 
velocity conditions {and constant heat flux} \citep{Brun:2011bl}: 
\begin{enumerate}
\item rigid: $\left. \mathrm{v}_r\right\lvert_{r_{\mathrm{top}}}= 0$,\newline
\item stress-free: $\left.\displaystyle\frac{\partial }{\partial
    r}\left (\frac{\mathrm{v}_\theta}{r} \right)\right\lvert_{r_{\mathrm{top}}} = \left.\displaystyle\frac{\partial}{\partial
    r}\left (\frac{\mathrm{v}_\varphi}{r} \right)\right\lvert_{r_{\mathrm{top}}} = 0 $,\newline
\item constant mean entropy gradient: \\ 
$\displaystyle\frac{\partial \bar{S}}{\partial
    r}\biggr\lvert_{r_{\mathrm{top}}}= -1.5 \times 10^{-7}$cm.K$^{-1}$.s$^{-2}$.
\end{enumerate}

The inner boundary conditions are special because another new feature of
the ASH code is that we are now able to extend our 
computational domain to $r=0$. Indeed, the central
singularity requires special attention. Following
\citet{2007PhRvE..75b6303B}, we have implemented regularity conditions
imposing that only $\ell=1$ mode can go through the center {and adapted
  the thermal conditions}. In the code, we
use the poloidal-toroidal decomposition 
\begin{equation}
  \label{eq:29}
  \bar\rho\vec{\mathrm{v}} =
  \vec\nabla\times\vec\nabla\times\left(W\vec{e_r}\right) + \vec\nabla\times\left(Z \vec{e_r}\right)\hbox{,}
\end{equation}
which ensures that the mass flux remains divergenceless (see Eq.~\eqref{eq:27}). The conditions in
$r=0$ thus translate to
\begin{itemize}
\item $Z=0$ for all $\ell$,\newline
\item $W=0$ and $\displaystyle\frac{\partial W}{\partial r}=0 $ for $\ell=1$,\newline
\item $W=0$ and $\displaystyle\frac{\partial^2 W}{\partial r^2}=0 $ 
  for $\ell \ne 1$,\newline
\item $\bar{S}=0$ for $\ell \ne 0$,\newline
\item $\displaystyle\frac{\partial \bar{S}}{\partial r}=0 $ for $\ell=0$. \newline
\end{itemize}

{The detail of the calculation is developed in Appendix A. Since the number of constraints is higher than the number of conditions, we explain
  our choice and show that another set of boundary conditions gives the same result at 0.1\%.}

{Owing to the convergence of the mesh size as we get closer to $r=0$,
  the horizontal Courant-Friedrichs-Lewy (CFL) condition
  \begin{equation}
    \label{eq:444}
    \tau^{\displaystyle\mathrm{h}}_{\mathrm{CFL}}=\sqrt{\frac{r^2}{\ell_{\mathrm{max}}(\ell_{\mathrm{max}}+1)\mathrm{v}_\mathrm{h}^2}}\hbox{,}
  \end{equation}
with
$\mathrm{v}_\mathrm{h}=\sqrt{\mathrm{v}_\theta^2+\mathrm{v}_\varphi^2}$, becomes too extreme
  if we retain all the $\ell$ values. We thus apply a filter as we get closer
  to the center that asymptotes to $\ell=1$ since only this
component of the flow is allowed to go through. {The thermodynamic variables are not affected by this modification.} To assess the
radial dependence of the filter and rather than imposing a functional
shape, we evaluate the horizontal velocity spectrum at each depth and time step
and retain scales with $p_{\mathrm{tol}}$ of the peak value, using the usual
choice $p_{\mathrm{tol}} =10^{-3}$ . The horizontal CFL condition
thus becomes
\begin{equation}
    \label{eq:445}
    \tau^{\displaystyle\mathrm{h}}_{\mathrm{CFL}}=\sqrt{\frac{r^2}{\ell_{\mathrm{eff}}(\ell_{\mathrm{eff}}+1)\mathrm{v}^2_{\mathrm{h}}}}\hbox{,}
  \end{equation}
with $\ell_{\mathrm{eff}} \underset{r\to 0}{\longrightarrow} 1$.
We tested
several values of $p_{\mathrm{tol}}$ and did not noticed significant changes in the wave
spectrum. We finally impose a maximum time step
\begin{equation}
  \label{eq:446}
  \tau_{\mathrm{CFL}} = \mathrm{min} \left(\tau^{\displaystyle\mathrm{r}}_{\mathrm{CFL}},\tau^{\displaystyle\mathrm{h}}_{\mathrm{CFL}}\right)\hbox{,}
\end{equation}
with 
 \begin{equation}
   \label{eq:447}
  \tau^{\displaystyle\mathrm{r}}_{\mathrm{CFL}} = \frac {\mathrm{min}\left({\Delta r}\right)} {\mathrm{max}(|\mathrm{v}_r|)}\hbox{.}
 \end{equation}
}

\begin{table*}
  \centering
  \begin{tabular}[center]{cccccc}\hline \hline
    Parameter & \textit{ref} & \textit{therm1} & \textit{therm2} & \textit{turb1} & \textit{turb2}\\\hline
    ($N_r$,$N_\theta$,$N_\varphi$) & (1581,256,512) &  (1581,256,512) &  (1581,256,512) &  (1581,256,512) &  (1581,512,1024) \\
    $\nu_{\mathrm{top}} (\mathrm{cm}^2/$s) & $4 \times 10^{12}$ &$4 \times 10^{12}$ & $4 \times 10^{12}$ & $2 \times 10^{12}$ & $1 \times 10^{12}$ \\  
    $\nu_{\mathrm{exp}}$ & $10^{-3}$ &$10^{-3}$ & $10^{-3}$ & $10^{-3}$ & $10^{-3}$ \\     
    $\kappa_{\mathrm{top}} (\mathrm{cm}^2$/s) & $3.2 \times 10^{13}$ &$3.2 \times 10^{13}$ & $3.2 \times 10^{13}$ & $1.6 \times 10^{13}$ & $0.8 \times 10^{13}$ \\  
    $\kappa_{\mathrm{exp}}$ & $10^{-4}$ &$10^{-3}$ & $10^{-2}$ & $10^{-4}$ & $10^{-4}$ \\
    $P\mathrm{r}$ in CZ & 0.125 & 0.125 & 0.125 & 0.125 & 0.125 \\    
    $P\mathrm{r}$ in RZ &1.25 & 0.125 & 0.0125 &1.25 &1.25 \\
    $Re$ & 169 & 169 & 169 & 338 & 675\\
\end{tabular}
  \caption{$N_r$, $N_\theta$, and
    $N_\varphi$ are the radial, latitudinal, and longitudinal mesh
    points. \textit{turb2} is the more turbulent model so we doubled its horizontal
    resolution to ensure that the fluid motions are well resolved. When it is convenient, we distinguish between convection zone
    (CZ) and radiative zone (RZ). $P\mathrm{r}=\nu/\kappa$ is the Prantdl number,
    $R\mathrm{e} = VL/\nu$ the Reynolds number where $L = 0.97R_\odot$ and $V$ is
    the rms convective velocity.}
  \label{tab:param_models}
\end{table*}

\subsection{Numerical resolution}

To initialize the 3D simulation, we specify a reference state derived from
a 1D solar structure model \citep{BrunAl2002}. We impose the entropy gradient $d\mathrm{\bar{S}}/dr$ and
the gravitational acceleration $g$ based on the 1D model and then deduce the
reference density $\bar\rho$ from the equation of hydrostratic equilibrium
and the ideal gas law (Eq.~\eqref{eq:11}) using a Newton-Raphson method. The left and middle panels of Fig. \ref{fig:profiles} show the
reference density, temperature, and
entropy gradient. In the middle panel, if
$d\bar{S}/dr >0$ then the BV frequency (see following sections) is real and positive and IGWs can
propagate in this region. In the convective region, $d\bar{S}/dr <0$,
{and that translates into} IGWs being evanescent. During the simulation,
these reference values are updated using the spherically averaged perturbation
fields. After having evolved the model over several convective overturning
times, we obtain the flux balance represented in the
righthand panel of Fig.~\ref{fig:profiles}. The different contributions to the energy
flux are represented. In the convective envelope, the inward kinetic energy flux due to the asymmetry
between up- and downflows is balanced by the outward enthalpy flux that exceeds the solar luminosity and carries the main
part of the energy in this zone. We note the
penetration of convective motions below the convection zone. The
system is expected to adjust to a new equilibrium by modifying the
background thermal stratification \citep[e.g.,][]{1991A&A...252..179Z} but the relaxation timescale is too long
(about $10^5$ yr) to be achieved in the simulation. For this reason, we
speed up the relaxation process by increasing the radiative diffusivity
$\kappa_r$ {and the associated radiation flux} at the base of
the convective zone to balance the inward enthalpy flux
\citep{2000ApJ...532..593M,Brun:2011bl}. In the radiative zone, the main
contribution to the total flux is brought by the radiative flux. Finally, the entropy flux
represents the flux carried by the unresolved motions {and is confined to the upper layers}.\\\newline
For the numerical resolution, the velocity and thermodynamic variables are
expanded in spherical harmonics $Y_{\ell,m}(\theta,\varphi)$ for their
horizontal structure. For the radial structure we use a
finite-difference approach on a non uniform grid, unlike
\citet{Brun:2011bl} where the variables were expanded in two Chebyshev polynomials $T_n(r)$ in the radial
direction. This new feature of the code has been tested by comparing
results in simpler setting with the previous version (with Chebyshev
decomposition) and with the anelastic benchmark problems of
\citet{Jones:2011in}. The agreements are as good, if not better than with
the Chebyshev expansion {(Featherstone et al. 2013, private communication)}. Then, following \citet{1984JCoPh..55..461G}, we use an explicit
Adams-Bashforth time integration scheme for the
advection and Coriolis terms, and a semi-implicit Crank-Nicholson treatment for
the diffusive and buoyancy terms \citep{CluneAl1999}.

\subsection{Models}
\label{sec:model}
All of the models described in this paper are based on the same reference
state. We distinguish five models of the Sun where we have chosen
different diffusion coefficients. The radial profiles of $\nu$ and $\kappa$
are
\begin{eqnarray}
  \label{eq:35}
  \nu(r) &=& \nu_{\mathrm{top}}\left[ \nu_{\mathrm{exp}} + f(r)(1-\nu_{\mathrm{exp}}) \right] \hbox{,} \\
  \kappa(r) &=& \kappa_{\mathrm{top}}\left[ \kappa_{\mathrm{exp}} + f(r)(1-\kappa_{\mathrm{exp}}) \right] \hbox{,}
\end{eqnarray}
where
\begin{equation}
  \label{eq:36}
  f(r) =
  \frac{1}{2}\left[\tanh\left(\frac{r-r_t}{\sigma_t}\right)+1\right] \hbox{.}
\end{equation}
{The radius $r_t$ = $4.86 \times 10^{10}$cm and stiffness $\sigma_t = 0.04 \times 10^{10}$cm
are identical for all models.} {The difference concerns} the choice of $\nu_{\mathrm{top}}$,
$\nu_{\mathrm{exp}}$, $\kappa_{\mathrm{top}}$, and $\kappa_{\mathrm{exp}}$ referenced in
Tab~\ref{tab:param_models}. 

\begin{figure}[h]
  \begin{center}
  \includegraphics[width=0.45\textwidth]{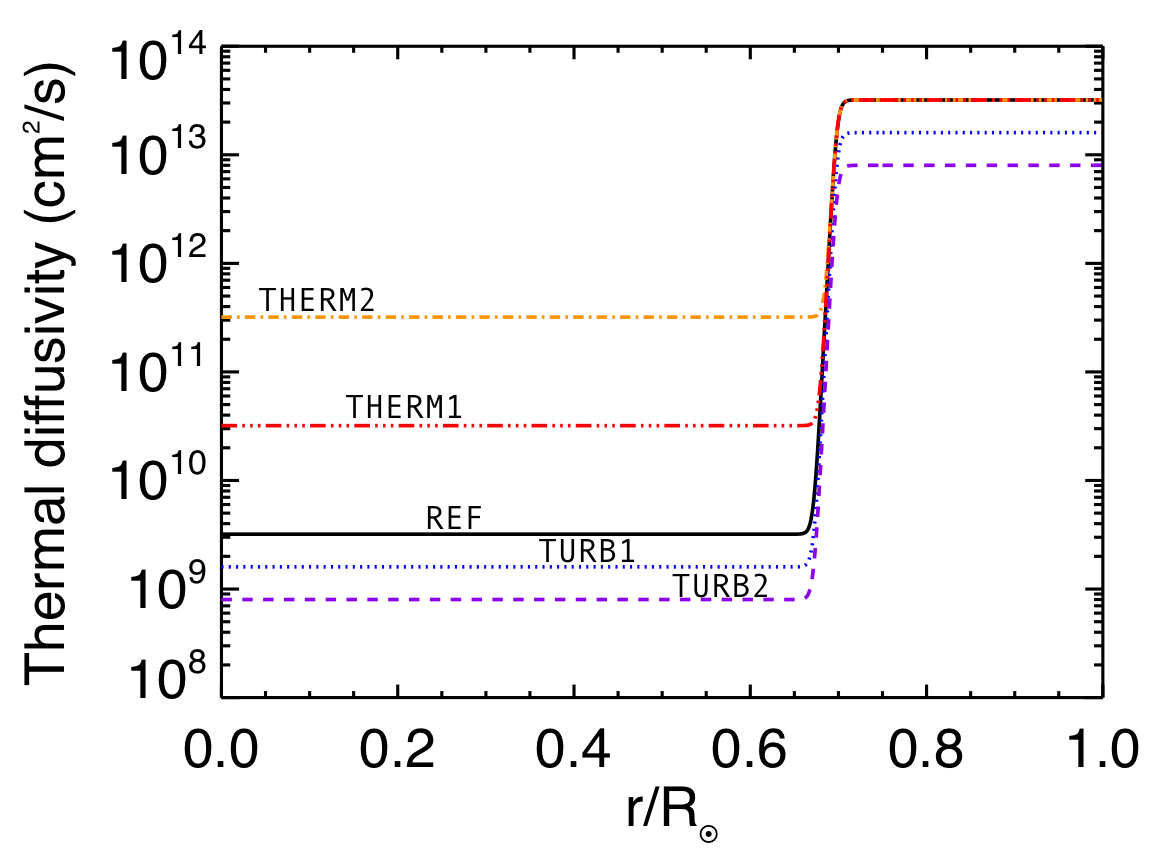}
 \includegraphics[width=0.45\textwidth]{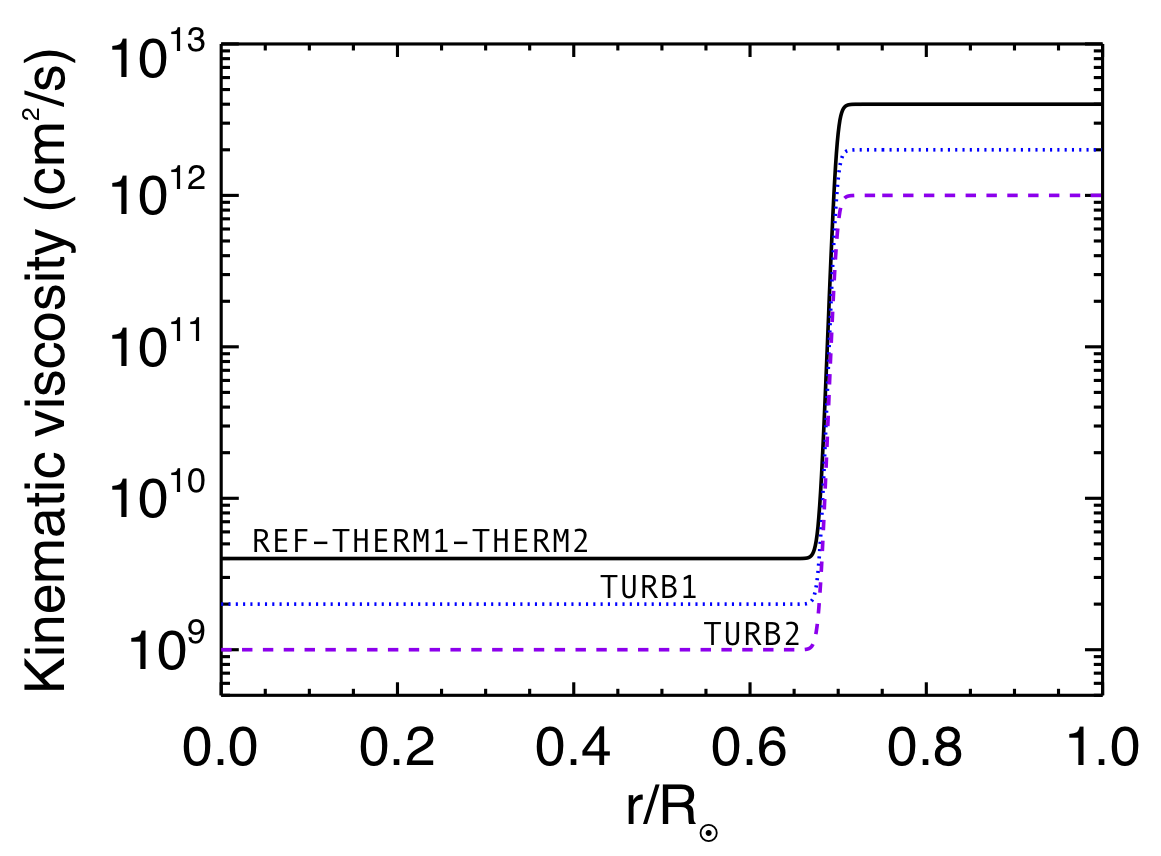}
  \caption{Diffusion coefficients $\nu$ and $\kappa$ (cm$^2$/s) for the
    five models presented in the article. The corresponding parameters are
    detailed in Tab~\ref{tab:param_models}.} 
  \label{fig:Knu}
\end{center} 
\end{figure}

\begin{figure}[h]
  \centering
  \includegraphics[width=0.45\textwidth]{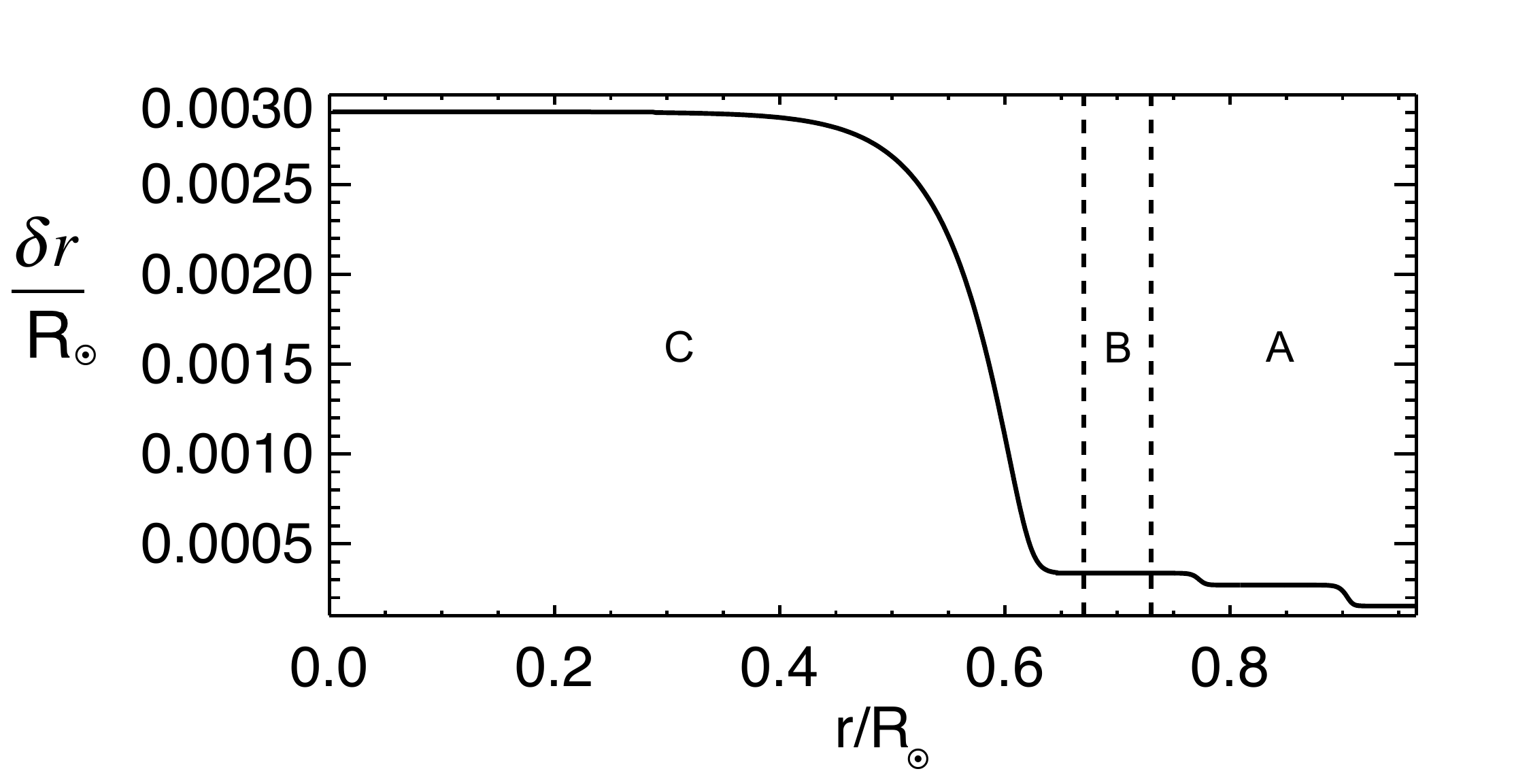}
  \caption{{Spacing (in solar radius) between two consecutive radial levels as a function of the normalized radius. Zone A (1012 points, $0.24R_\odot$) represents the main part of
    the convective zone. Zone B (179 points, $0.06R_\odot$) corresponds to the interface region
    where the kinematic viscosity and the thermal diffusivity drop abruptely. And zone C (390 points, $0.67R_\odot$) is the radiative zone.}}
  \label{fig:radgrid}
\end{figure}
\begin{figure}
  \centering
  \includegraphics[width=0.5\textwidth]{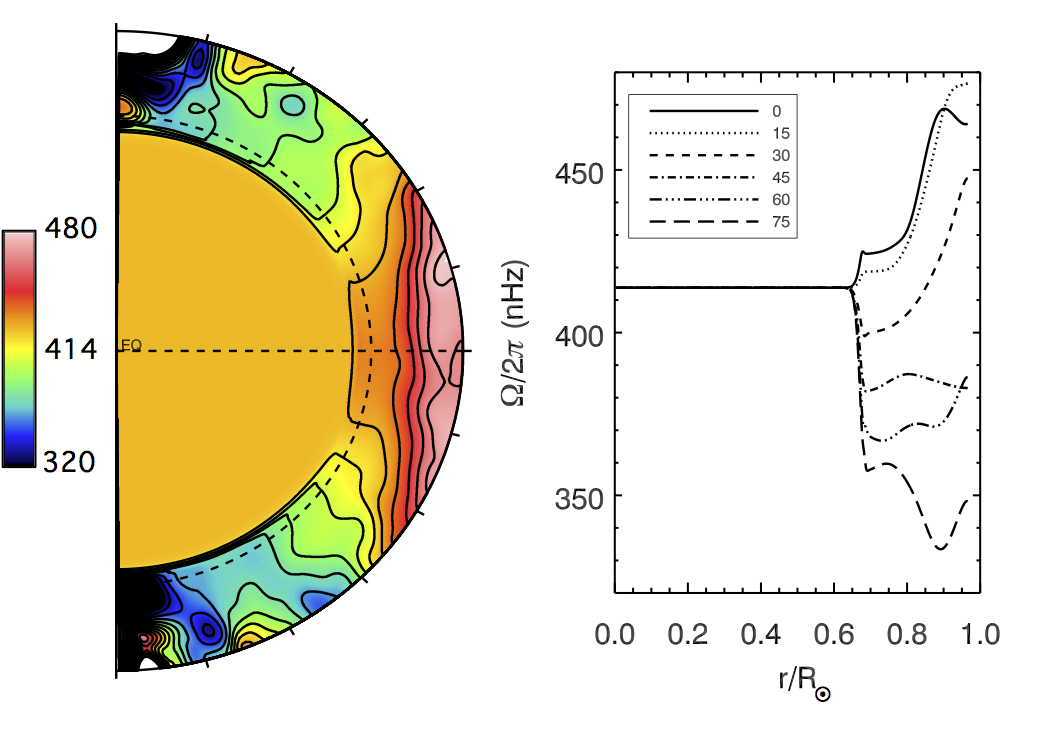}
  \caption{Differential rotation profile of the model
    \textit{ref}. \textbf{Left}: Angular velocity $\Omega(r,\theta)=\displaystyle\frac{V_\varphi}{r\sin\theta}+\Omega_0$ averaged over 
    longitude and time. Dashed lines represent the equator and the interface between
    convective and radiative zones. \textbf{Right}: Radial cuts of $\Omega$ at selected
    latitudes. We clearly see the differential rotation affecting the
    convective zone, the shear layer at the base of the convection
    zone (tachocline), and the flat rotation profile in the radiative zone.}
  \label{fig:omega}
\end{figure}

These profiles are chosen to study the thermal
and viscous effects of the fluid on the waves. The main model that will be used in the
following sections is called \textit{ref}. Its diffusion coefficients were
selected to obtain the clearer possible pattern and spectrum of gravity
waves. Indeed, it is known that gravity waves are damped during their propagation by a factor
depending on the fluid's radiative diffusivity
\citep{ZahnTalonMatias1997}. Models \textit{therm1} and \textit{therm2}
were computed to study the effect of this damping. They differ from
\textit{ref} only in the radiative zone where their $\kappa$ coefficient is
10 (\textit{therm1}) and 100 (\textit{therm2}) times higher (see Fig.~\ref{fig:Knu}). 
On the other hand, we expect a stronger wave's excitation with a
more turbulent convection. For this reason, we discuss two other
models called \textit{turb1} and \textit{turb2} where the Reynolds number $R\mathrm{e}=VL/\nu$
($V$ and $L$ are characteristic velocity and length scale) is increased by factors 2
and 4 in comparison with \textit{ref}. An overview of these diffusivities is
given in Fig. \ref{fig:Knu}. \\
In the convective zone, all models have the
same Prandtl number $P\mathrm{r}=\nu/\kappa=0.125$. In the radiative zone,
however, these values differ (see Tab~\ref{tab:param_models}) and have
different impacts on the amplitude of the waves observed in the radiative
zone. We discuss this point further in Sect.~\ref{sec:sens-phys-param}. 
{The horizontal and radial resolutions of the models are also indicated in Tab~\ref{tab:param_models}. In particular, the choice of the radial
grid requires attention in order to deal with the strong entropy and diffusivity gradients at the interface between convective and radiative zones. The total
number of radial points in the five models is $1581$. We show their distribution in Fig. \ref{fig:radgrid}. The number of points in zone C (radiative
zone) allows a good compromise between the resolution needed to deal with gravity waves, the stability of the models near the center, and
the cost of the total simulation.}\\
Finally, we note that all models rotate at the solar rotation rate,
$\Omega_\odot=2.6\times10^{-6}$ rad/s. About 130 turnover times after the
beginning of the simulation, we observe a differential rotation in the
convective zone as shown in Fig.~\ref{fig:omega} for model \textit{ref}.
The equator rotates faster than the
poles, and we retrieve a conical shape at mid-latitude, as deduced by
helioseismology \citep[e.g.,][]{2003ARA&A..41..599T}. {Since model
$\textit{ref}$ is more turbulent than the one published in
\citet{Brun:2011bl}, the overall $\Delta \Omega$ contrast is about 130
nHz. The sharp transition to solid body rotation in
the radiative zone (i.e., the tachocline).} {This rotation profile is due to our uniform initial conditions and is maintained during the simulation
because the total computed time is shorter than the radiative spreading time \citep{Spiegel:1992tr}. For more details concerning the
confinement of the solar tachocline, see \citet{BrunZahn2006} and \citet{2011A&A...532A..34S}. }This bulk rotation have a
visible effect on IGWs that is discussed in
Sect. \ref{sec:rotational-splitting}.

\section{Excitation of gravity waves}
\label{sec:excit-penetr-conv}

\begin{figure}
  \centering
  \includegraphics[width=0.45\textwidth]{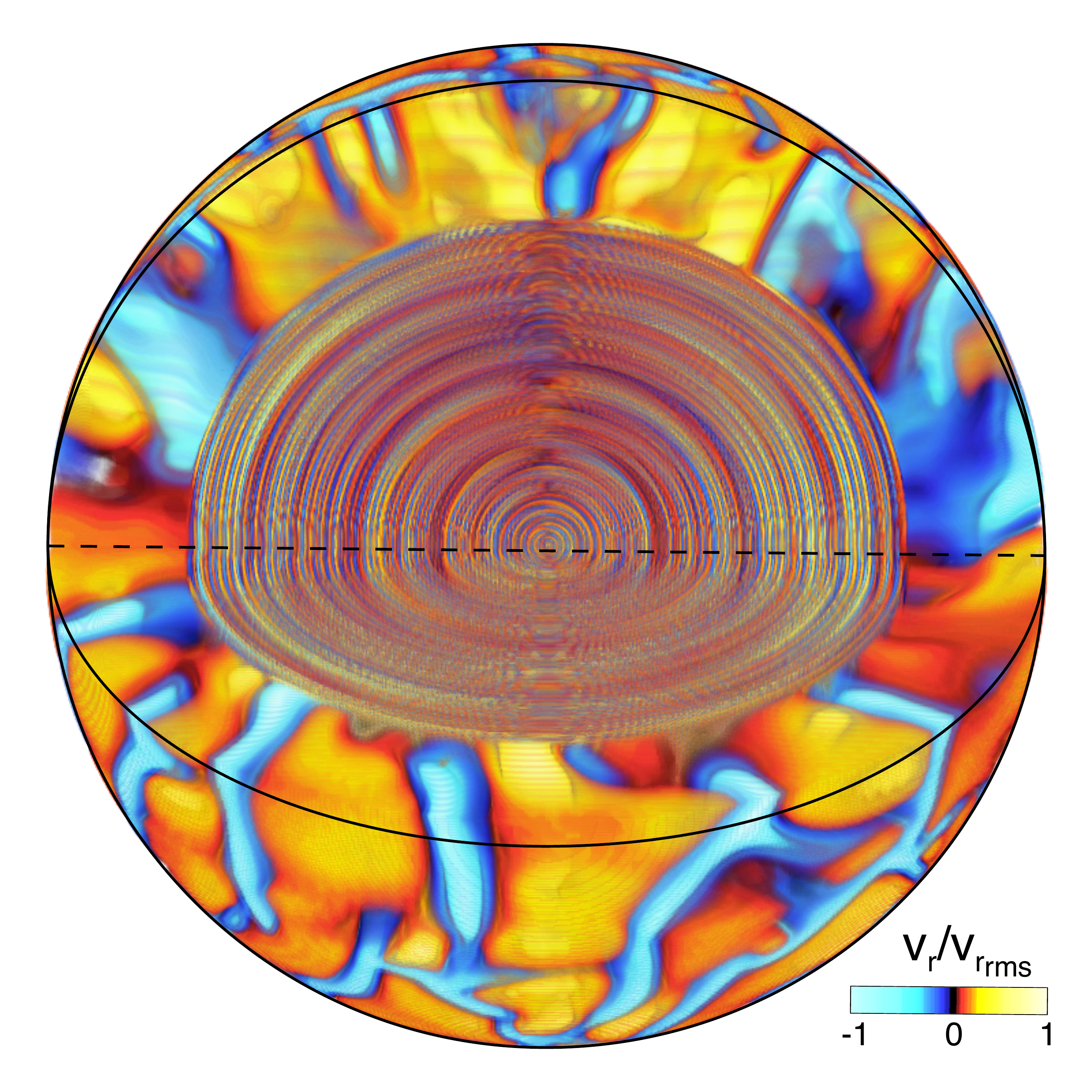}
  \caption{3D representation of the normalized radial velocity in the full simulated star (model \textit{ref}). A quadrant of the
    sphere has been removed to show the wave pattern in the
  radiative zone.}
 \label{fig:3d_eqsl}
\end{figure}

Due to the coupling between convective and radiative zones, waves are excited and propagate in the
inner radiative zone. Figure \ref{fig:3d_eqsl} shows a 3D view of model
\textit{ref} where we clearly see these waves by removing a quadrant of the
sphere. Colors correspond to radial
velocity. The convective pattern in the outer zone is visible with blue
downward flows and red upward flows. In the radiative zone, spherical patterns
correspond to the wavefronts of gravity waves. For the sake of the
visualization, the amplitude of $\mathrm{v}_r$ has been normalized by its root mean
square at each radius, making the waves appear as if their amplitude was about the same as the velocity in the convective zone. In reality, there is a
drop of amplitude of six to ten orders of magnitude between both zones,
depending on the model, as we will discuss in
Fig.~\ref{fig:comp_models}. In this section, we show how the penetration
of convective plumes in the radiative zone excites the rich spectrum of
gravity waves that is observed.

\subsection{Penetrative convection}
\label{sec:penetr-conv}

\begin{figure}
  \centering
  \includegraphics[width=0.5\textwidth]{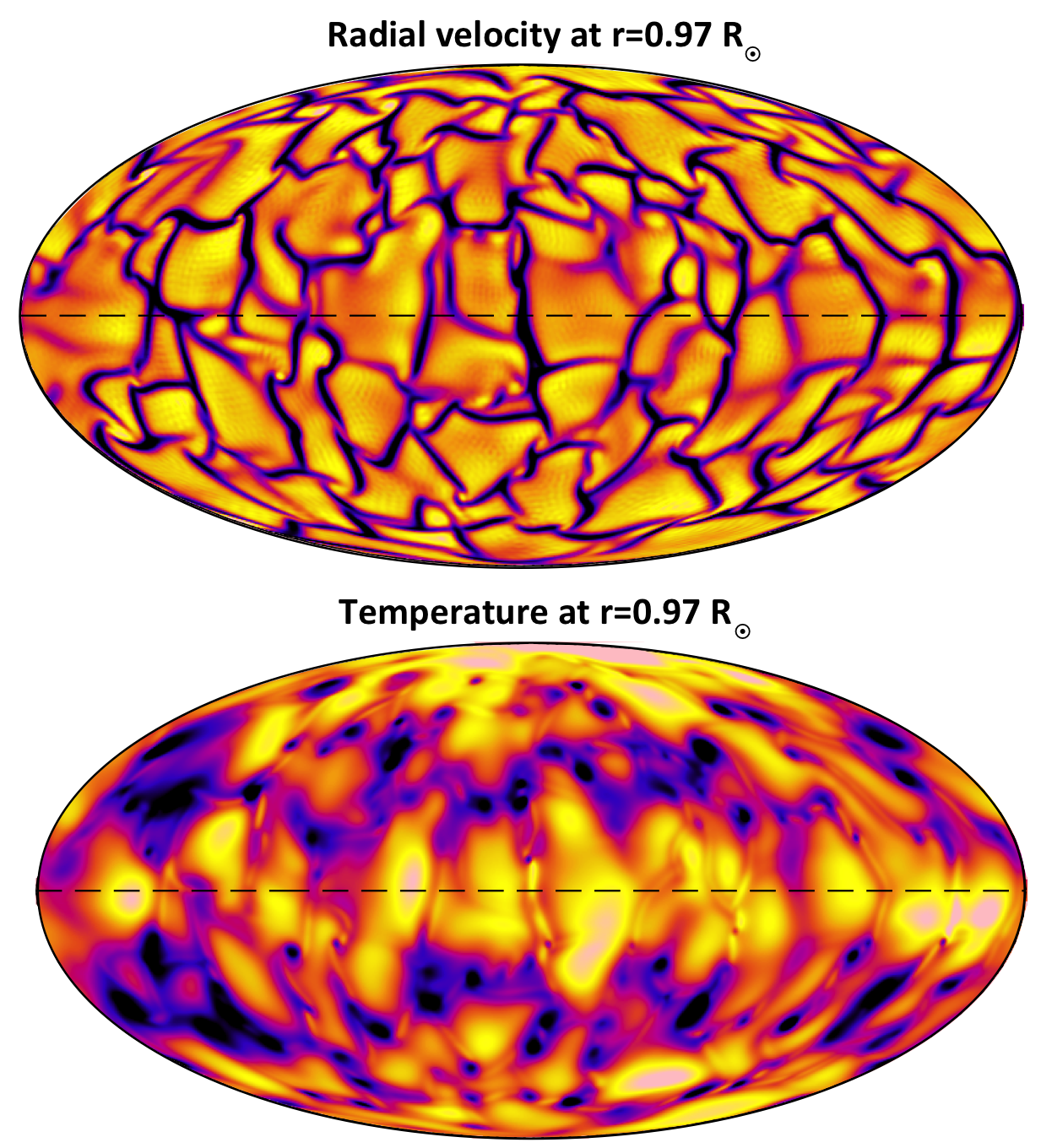} \\
  \caption{Mollweide projection of the radial velocity and temperature fluctuations
    at r=0.97$R_\odot$ for the model \textit{ref}. The horizontal dashed line corresponds to the equator. Dark
    tones denotes negative (inward) velocities and temperature fluctuations.}
  \label{fig:shsl}
\end{figure}

Several processes have been invoked to explain the
excitation of gravity waves in stars. Their relative efficiency differ
with the type of star. In the case of the
Sun and solar-like stars, the main excitation process is the
pummeling of convective plumes at the interface between
convective and radiative zones. Indeed, the convection does not stop abruptly at
the interface with the radiative zone. When convective plumes reach this boundary,
their inertia makes them penetrate the radiative region. Then they are forced to
slow down by buoyancy, and the loss of kinetic energy is converted into
gravity waves, as discussed in detail in \citet{Brun:2011bl}.\\\newline
In Fig. \ref{fig:shsl} we represent the radial velocity and the temperature fluctuations
realized in the model \textit{ref} at the top of the computational domain
(r=0.97$R_\odot$). Convective motions are apparent as a network of narrow
cool downflow lanes (dark/blue) surrounding broader warmer upflows (red). This
pattern varies with time, convective cells continuously emerging
and merging with one another or splitting into several distinct
structures. If we move deeper into the convection zone \citep{Brun:2011bl},
isolated plumes appear, corresponding to 
the strongest downflows that managed to go through the entire convective
envelope. \\\newline

\begin{figure}[h]
  \centering
  \includegraphics[width=0.5\textwidth]{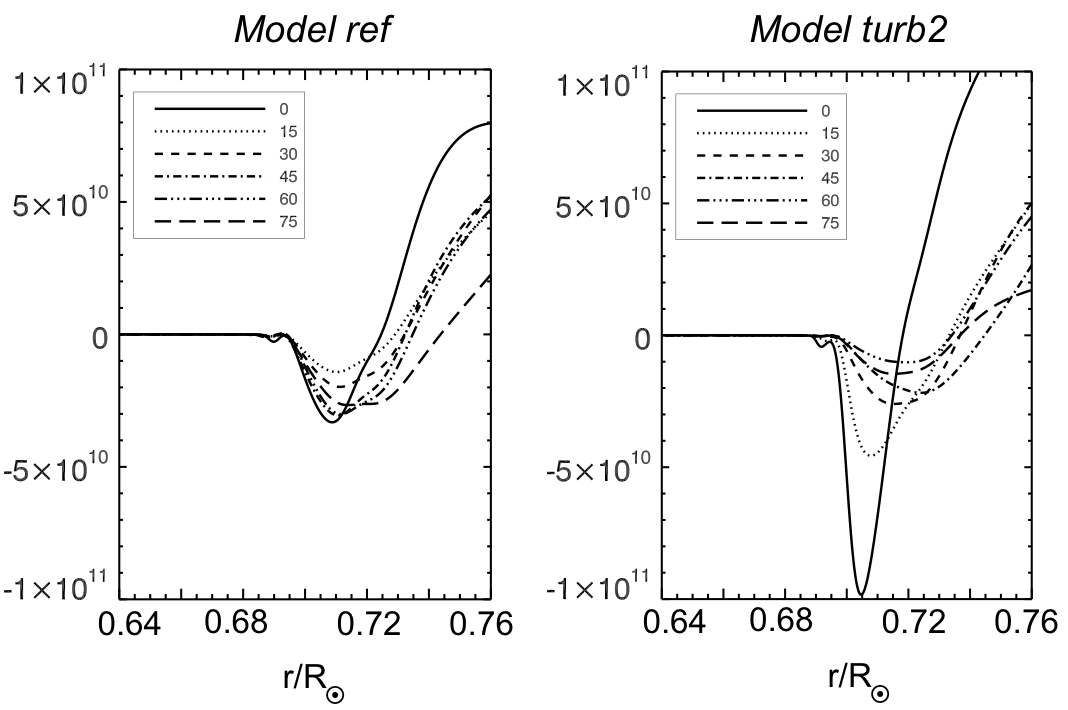}
  \caption{Radial cuts of azimuthal and temporal averages of the radial enthalpy flux in
    models \textit{ref} and \textit{turb2} at specified latitudes zooming
    into the overshoot region.}
  \label{fig:enthalpy}
\end{figure}

As explained in Sect.~\ref{sec:model}, we have computed several models with
different diffusivity coefficients. In particular, the convective
turbulence increases from model \textit{ref} to \textit{turb2} (their
Reynolds numbers are given in Tab~\ref{tab:param_models}). We see
later that IGWs are excited in \textit{turb2} with a higher amplitude than
in \textit{ref}. Indeed, the radial enthalpy profile at the interface with
radiative zone is different in these two models. We represent a radial cut
of an azimuthal and temporal average of the radial enthalpy flux for both
models in Fig. \ref{fig:enthalpy}. The negative peaks of enthalpy flux at the base of the convection
zone characterize the buoyant braking of convective plumes. We clearly see
that this peak is thinner and more pronounced for \textit{turb2}. Moreover,
the measure of the rms
velocity of the fluid just above the radiative zone (the limit is defined
by the change of sign of the entropy gradient $d\bar{S}/dr$) shows that
the plumes in \textit{turb2} are quicker than in \textit{ref}.

\subsection{Excitation of gravity waves: St Andrew's cross}
\label{sec:excit-grav-waves}

\begin{figure}[h]
  \centering
\includegraphics[width=0.42\textwidth]{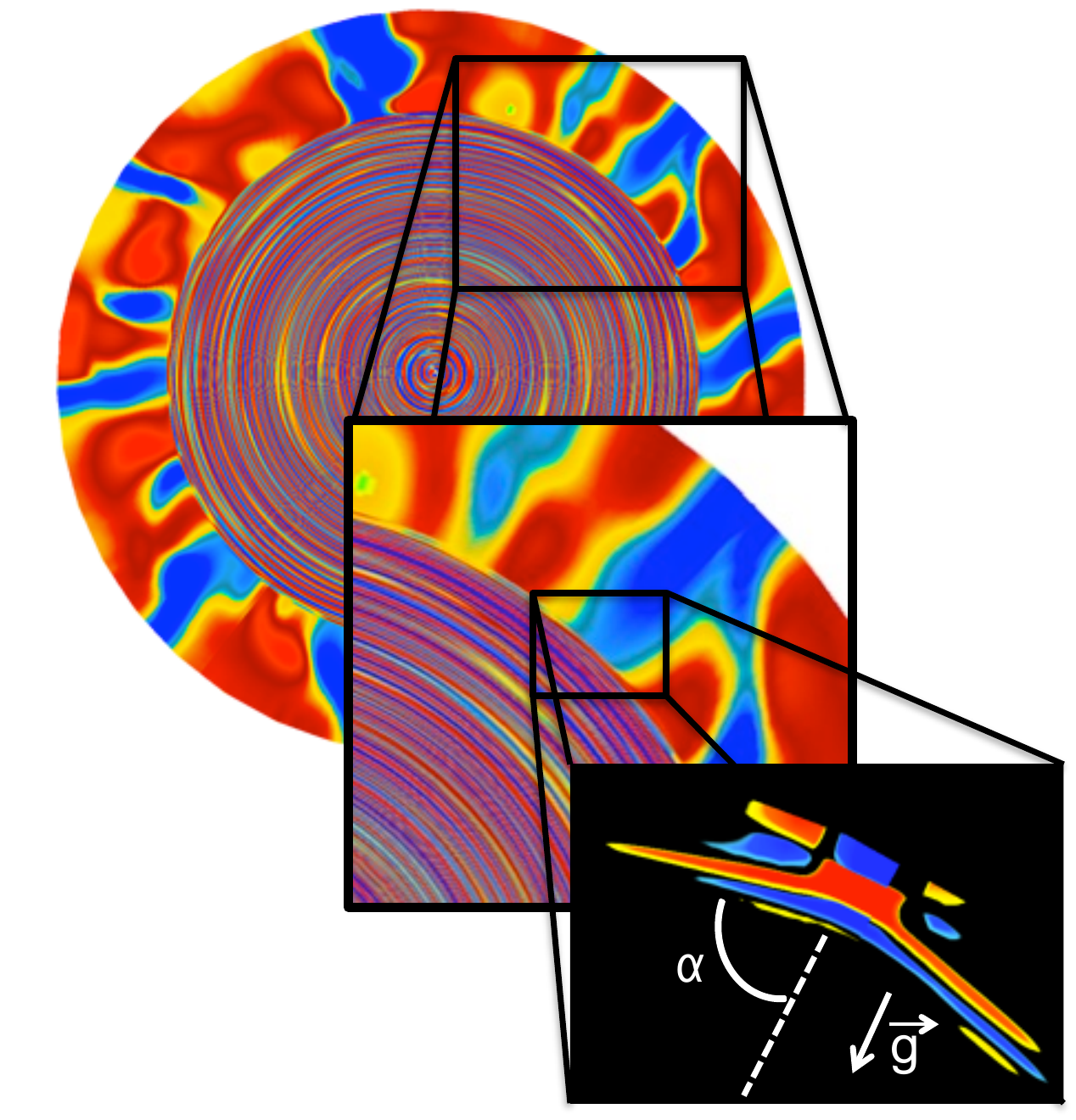}
\caption{Zoom in the region of excitation {of a wave in model
    \textit{ref}. The plane shown is the one perpendicular to the plume of
    interest (blue zone in the center of the middle panel), where a St
    Andrew's cross is produced (third panel)}. In the last panel, the
  radius is extended, {and the background color changed to black} to make the cross more apparent.}
  \label{fig:cross}
\end{figure}

Thanks to a zoom in the region of penetration shown in Fig.~\ref{fig:cross}, we retrieve a classical
result of fluid mechanics concerning the excitation of IGWs by a localized
disturbance in a stably stratified fluid \citep{1978cup..book.....L,1991JFM...231..439V}. When we neglect the rotation, the
linearized dispersion relation for gravity waves is 
\begin{equation}
  \label{eq:1}
  \omega = \frac{N k_h}{\sqrt{k_r^2+k_h^2}}\hbox{,}
\end{equation}
where $\omega$ is the frequency of the wave {(in Hz)},
and $\vec k = k_r\vec e_r + \vec k_h$ the wavevector decomposed into
its radial and horizontal parts. The Brunt-Väisälä frequency ({given in Hz})
\begin{equation}
  \label{eq:16}
N = \frac{1}{2\pi}\sqrt{-\bar{g}\left(\frac{1}{\bar\rho} \frac{\partial \bar\rho}{\partial
  r}-\frac{1}{\Gamma_1\bar{P}}\frac{\partial \bar{P}}{\partial r}\right)}
\end{equation}
describes the stratification in density in the radiative zone.

\begin{figure}
  \centering
  \includegraphics[width=0.3\textwidth]{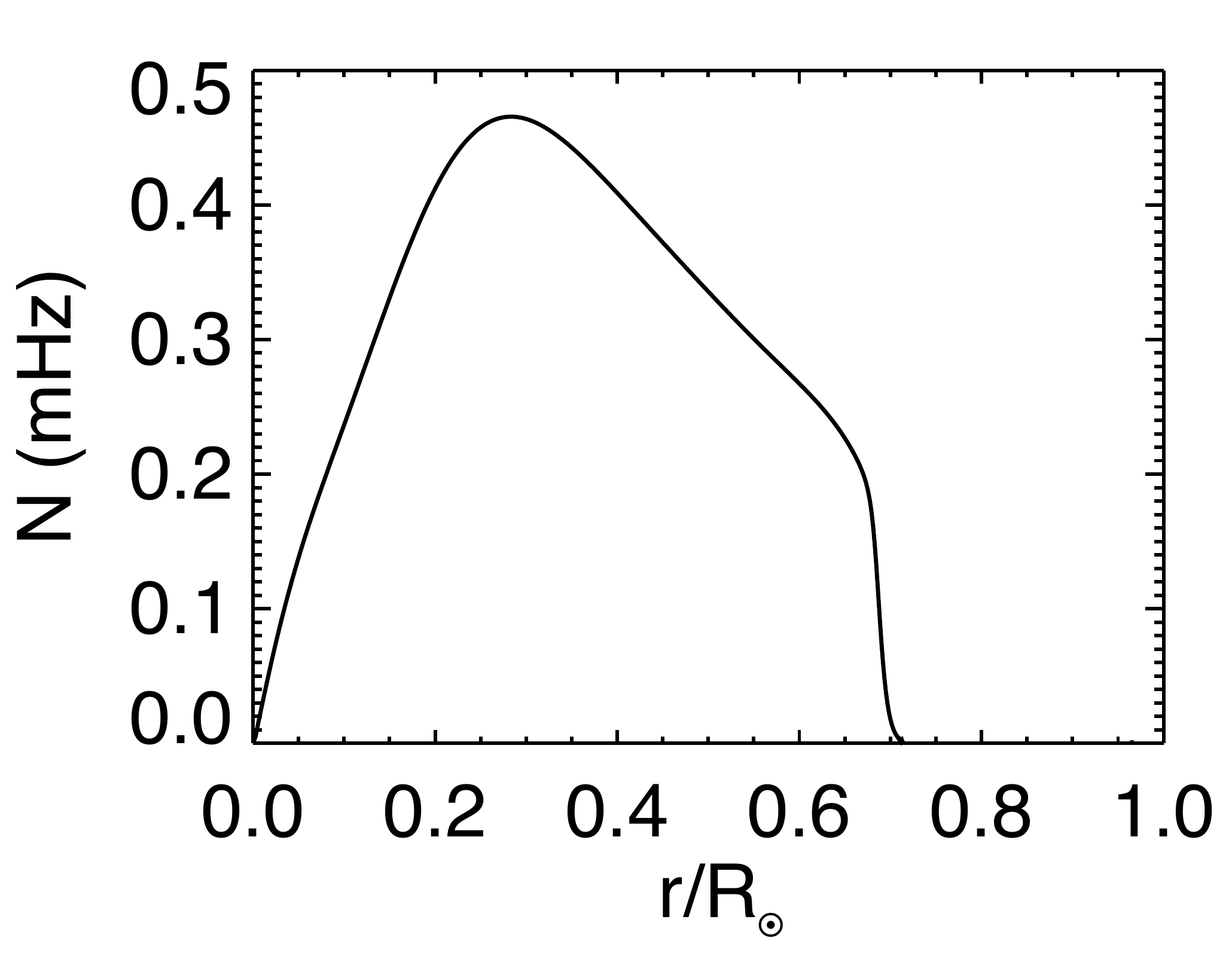}
  \caption{Radial profile of the Brunt-Väisälä frequency $N$ {(common to all
   models)}.}
  \label{fig:BVfreq}
\end{figure}

The profile of $N$ is shown in Fig.~\ref{fig:BVfreq}. It is real and positive in the
radiative zone (that corresponds to the positive entropy gradient shown in
Fig. \ref{fig:profiles}) and becomes purely imaginary for $r>0.717R_\odot$, where the
entropy gradient is negative. It defines a cavity
where gravity waves can propagate and resonate. The maximum value of
$N$ is about 0.45 mHz, and according to the dispersion relation given in
Eq. \eqref{eq:1}, this is the maximum frequency allowed for IGWs.\\\newline

\begin{figure}[h]
  \centering
  \includegraphics[width=0.42\textwidth]{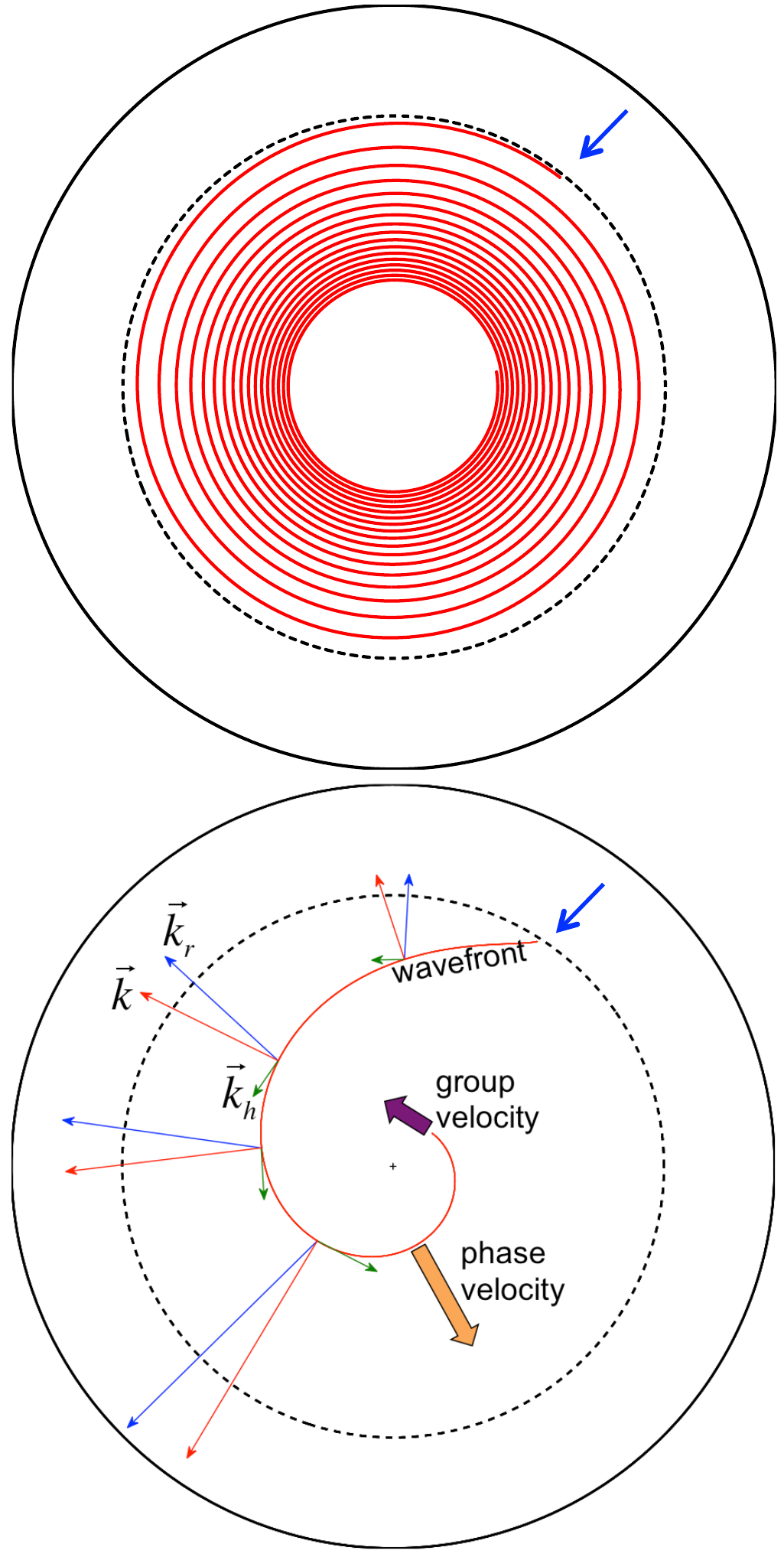}
  \caption{Propagation of two gravity waves calculated by raytracing. The
    top panel shows a wave at the low-frequency $\omega_1= 3 \times 10^{-3}$ mHz. The ray
    spirals toward the center with an almost radial wavevector, i.e., $k_h
    \ll k_r$. In the bottom panel, a higher frequency wave with $\omega_2 = 0.2$ mHz is
    represented with arrows indicating the directions of $\vec{k}$, $k_r$,
    and $k_h$. The scale is respected, so $k_h \approx \frac{1}{5}
    k_r$. {Blue arrows in the top right of each panel point out the
      place where both waves are excited.}}
  \label{fig:schema}
\end{figure}

In the Boussinesq approximation, {where the variation in density
is only considered in the buoyant term,} waves produced by a localized
time-monochromatic perturbation are known to propagate inside beams \citep{LighthillBook}, which
develop around a St. Andrew's cross in two dimensions. The energy is radiated around
an angle $\alpha$ to the vertical such that 
\begin{equation}
  \label{eq:2}
  \alpha = \arccos{\frac{\omega}{N}}. 
\end{equation}
Figure \ref{fig:cross} shows the St Andrew's cross produced by the
penetration of a plume in the radiative zone of our model \textit{ref}. In the third panel, we have
extended the radius in order to highlight the cross. In fact, the angle
$\alpha$ is close to 90°, which corresponds to very low frequency
 waves.\\
To clarify the relation between this measurement of the angle $\alpha$
and the wave pattern visible in
Figs.~\ref{fig:3d_eqsl} and \ref{fig:cross}, we show the ray paths of two gravity waves obtained using the
raytracing method in
Fig. \ref{fig:schema}. This linear theory \citep{goughHouches} defines the Hamiltonian
$W(\vec{x},\vec{k},t)=\omega$ and uses the dispersion relation
(Eq.~\eqref{eq:1}) to obtain the equations governing the ray path of one
gravity wave of frequency $\omega$, {along which the energy is conveyed}. In our spherically symmetrical case,
these equations are reduced to
\begin{equation}
  \label{eq:38}
  \left\{
      \begin{array}{ll}
 \displaystyle\frac{\mathrm{d}r}{\mathrm{dt}}=\displaystyle\frac{\partial
   W}{\partial k_r}\hbox{,}\\
\\
  \displaystyle\frac{\mathrm{d}\theta}{\mathrm{dt}}=\displaystyle\frac{1}{r}\displaystyle\frac{\partial W}{\partial k_h}
\hbox{,}\\
\\
k_h^2 = \displaystyle\frac{\ell(\ell+1)}{r^2}\hbox{,}\\
    \end{array}
    \right.
\end{equation}
and completed by the dispersion relation. We here neglect the rotation, which is
justified by the fact that $\omega \gg 2\Omega_\odot$. Figure \ref{fig:schema} shows the
curves obtained for $\omega_1 = 3 \times 10^{-3}$ mHz (top panel) and $\omega_2 = 0.2$
mHz (bottom panel), starting from the same initial conditions ($r_0,
\theta_0$). {Since gravity waves are transverse (unlike
  acoustic waves)}, the ray propagates perpendicularly to the wavevector $\vec{k}$ as
shown by the arrows in the bottom panel, where the ratio between $k_r$ and
$k_h$ is respected. It is clear that the top panel with the low-frequency
wave is closer to the wave pattern observed in ASH and beginning with the
St Andrew's cross shown in Fig. \ref{fig:cross}.

\subsection{Wavefronts in 3D}

\begin{figure*}
  \centering
  \includegraphics[width=0.8\textwidth]{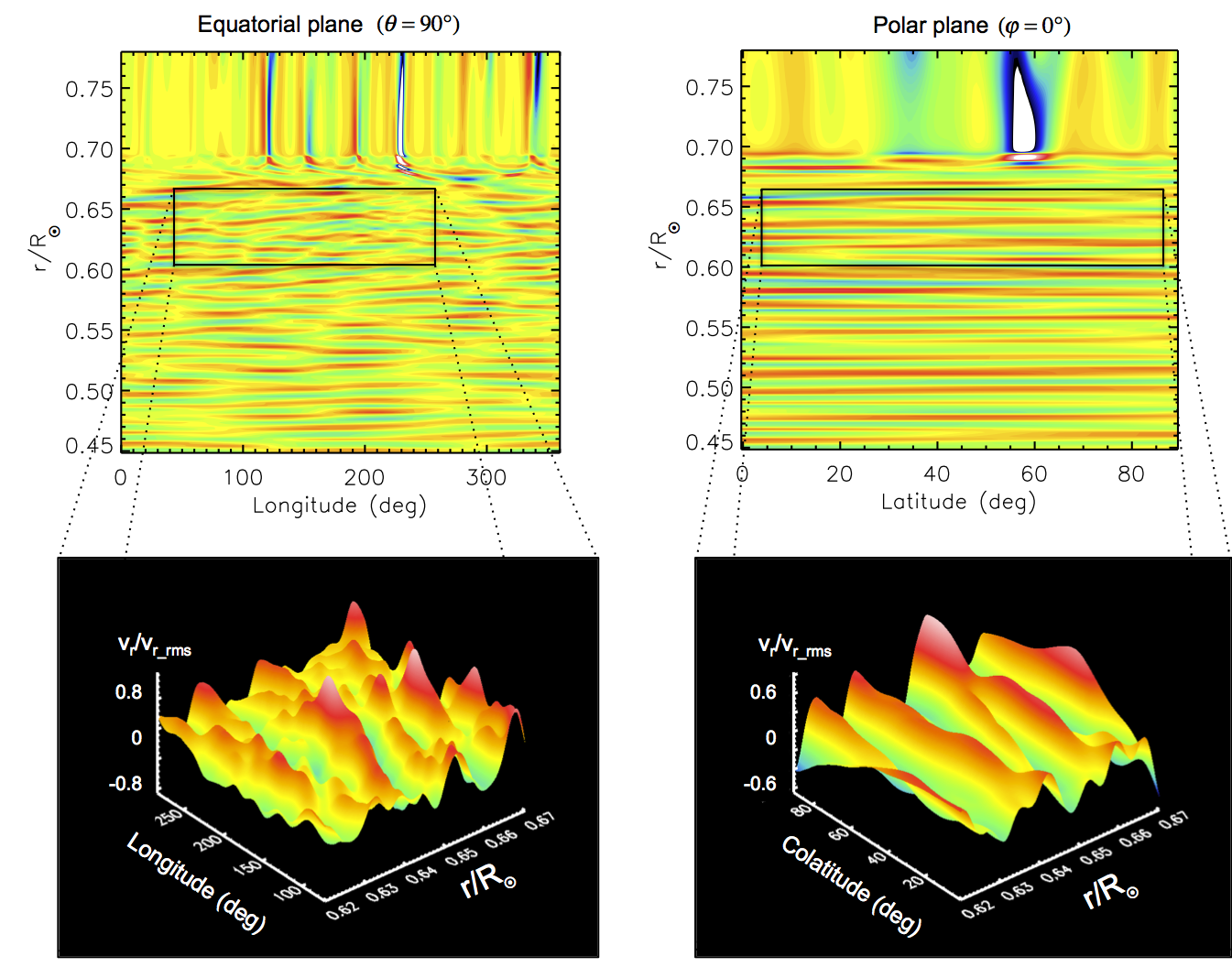}
  \caption{Form of the wavefronts in the equatorial and
      polar planes. \textbf{Top}: Contour plots of the normalized radial velocity in
      the equatorial (left) and polar (right) planes zoomed in the upper
      region of the radiative zone. \textbf{Bottom}: Surface plots
      corresponding to the region delimited by the black rectangles in the
      top panel that show the amplitude of the wavefronts varying with
      longitude (left) and colatitude (right). {The velocity has been divided by its rms value at each radius, in order to visualize the form of
      the wavefront despite the attenuation of the amplitude.}}
  \label{fig:polar}
\end{figure*}

{We now understand how IGWs are excited by the penetration
  of convective plumes in the radiative zone. An interesting question could
be the orientation of the plumes and the way waves propagate in the 3D
sphere. Indeed, looking at Fig. \ref{fig:3d_eqsl}, it seems that wavefronts
of IGWs fill the whole radiative region without distinction between
longitudinal and latitudinal directions. In Fig. \ref{fig:polar}, however,
we show that both planes are not equivalent for propagating waves. The lefthand
panel shows the radial velocity as a function of normalized radius and
longitude $\varphi$ for colatitude $\theta=\pi/2$ (equatorial plane). We
see convective plumes between $r=0.69$ and $0.78R_\odot$ that form St
Andrew's crosses as discussed in the previous section. The wavefronts are thus
inclined with respect to the horizontal. In the righthand panel, we represent
$\mathrm{v}_r$ as a function of the colatitude $\theta$ for
$\varphi=0$. This time, the wavefronts are almost parallel to the horizontal.
By following the transition from equatorial to polar plane, we understand
that the waves are mainly excited in the region close to the equator, but
then propagate throughout the whole sphere. We see later that the region of
propagation of IGWs depends on their azimuthal number $m$.}

\subsection{Spectrum}
\label{sec:spectrum}
We are therefore able to see low-frequency IGWs excited
by convective penetration and propagating in the radiative zone. However, observing the waves in the physical
space is not sufficient for characterizing 
them because only the largest perturbations are visible. Indeed, other
waves with lower amplitudes could be excited but not directly observable. \\\newline
Starting from a
temporal sequence of the radial velocity field $V_r(r_0,\theta,\varphi,t)$
at a given depth $r_0$, we successively 
apply a spherical harmonic transform at each time step, which gives
$\hat{V_r}(r_0,\ell,m,t)$, followed by
a temporal Fourier transform on the whole sequence of ($\ell$,$m$)
spectra. {This transformation into spherical harmonics allows us to
quantitatively compare our results to seismic observations and
  oscillation calculations, which could not be possible in 2D.} We thus obtain
a new field $\tilde{V_r}(r_0,\ell,m,\omega)$, which can be represented as a
function of $\omega$, $\ell$, and $m$. {The maximum degree
  $\ell_{\mathrm{max}}$ is related to the horizontal resolution $N_\theta$ of the model \citep{CluneAl1999}} 
\begin{equation}
  \label{eq:4}
  \ell_\mathrm{max} \le \frac{2N_\theta-1}{3} \hbox{.}
\end{equation}
{For models
  \textit{ref}, \textit{therm1}, \textit{therm2}, and \textit{turb1}, we have
$\ell_\mathrm{max}=170$ and $\ell_\mathrm{max}=340$ for models \textit{turb2} and
\textit{sem-lin} (see Sect. \ref{sec:non-line-inter-1})} . We discuss the effect of rotation
later, and for the moment, we add all contributions in $m$ quadratically in
order to create a power spectrum in $\omega$ and $\ell$. This results in the
following quantity:
\begin{equation}
  \label{eq:5}
  E(r_0,\ell,\omega) = \sum_{m} |\tilde{V_r}(r_0,\ell,m,\omega)|^2 \hbox{.}
\end{equation}

\begin{figure*}
  \centering
   \includegraphics[width=1.\textwidth]{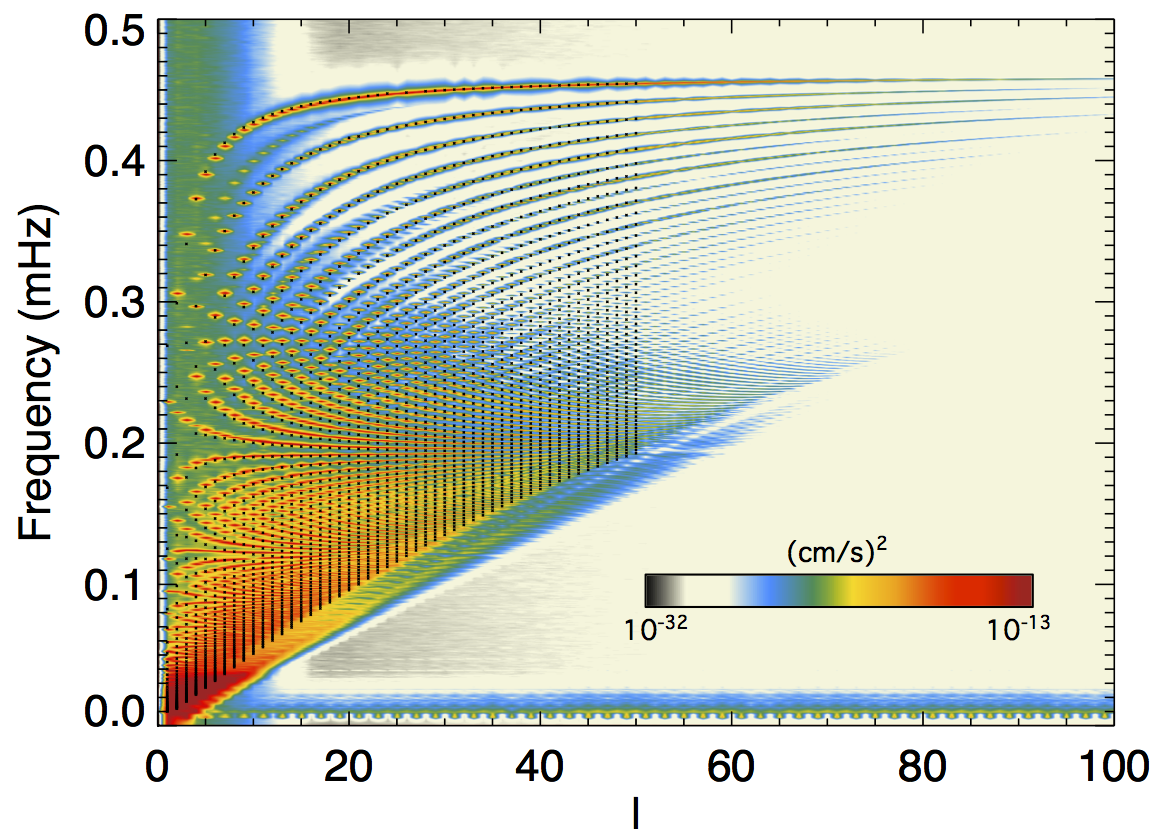}
  \caption{Energy $E$ in (cm/s)$^{2}$ plotted as function of the degree $\ell$ and
    the frequency $\omega$ at the radius $r_0=0.26$ $R_\odot$ (model \textit{ref}). Ridges are visible at high frequency, tending to
 the maximum Brunt-Väisälä frequency (about 0.45 mHz). Black crosses
 represent the frequencies predicted by the oscillation code ADIPLS for
 order $\ell \in [1,50]$ and radial number $n \in [0,58]$. {We have cut
 the horizontal axis to $\ell=100$ since no ridges are visible farther out.}}
  \label{fig:omega_l}
\end{figure*}

In Fig.~\ref{fig:omega_l}, we plot $E$ as a function
of $\ell$ and $\omega$ for $r_0=0.26 R_\odot$. The figure obtained looks very similar
to the one predicted by linear theory
\citep[e.g.,][]{1986A&A...165..218P,Christensen-Dalsgaard97lecturenotes}. Black crosses superimposed on the colored background mark the position of the frequencies predicted by
the oscillation code
ADIPLS\footnote{http://users-phys.au.dk/jcd/adipack.n}. We computed 
these frequencies using the ASH background state for $\ell=1$ to 50.
{The agreement with ASH results varies from 99\% for low frequencies to 92\% for high frequencies
(that corresponds to lower radial order $n$). This could be because
 the volume of the cavity where modes are formed is submitted to slight
  variations during time. Indeed, the interface between convective and
  radiative zones is time-dependent. We estimate that this could affect the
  modes' frequencies by about 1\% to 3\%. The BV frequency is very close to zero
  at this depth that limits the impact. We also consider that we measure the frequencies
  in ASH using finite (about 100 days) temporal sequences, and finally,
  nonlinear interactions and radiative effects are not taken into account in
  ADIPLS code and are possibly responsible for small changes in the modes' frequencies.} Modes with the same radial
order $n$ - essentially
given by the number of zeros in the radial direction in the eigenfunctions
- form ridges, particularly visible at high frequency. As imposed by the
dispersion relation (Eq. \eqref{eq:1}) and invoked in Sect. \ref{sec:excit-grav-waves}, the maximum frequency
corresponds to the maximum value of the BV frequency represented
in Fig~\ref{fig:BVfreq}, i.e., $\sim$ 0.45 mHz.
The modes' frequencies are known to decrease with increasing radial order $n$. The theoretical
spectrum extends to zero frequency at all degrees $\ell$, but the radial resolution
of our simulation imposes an upper limit to the order $n$ (here $n_{max}
\sim 58$). The richness of this spectrum proves that a large set of
waves is actually excited, and not only the low-frequency IGWs
visible in the real space. We then discuss the detailed properties
of this spectrum in the following section.

\section{Waves' properties}
\label{sec:waves-properties}

The properties of internal gravity waves have been studied in detail using
linear-and asymptotic theories. In this section, we show that the waves
observed in our simulations verify these properties but also provide
further information that is not accessible to linear theory. Here, we describe only the
model \textit{ref}. For the study of the waves' frequencies, all models are
equivalent. The differences lie in the amplitude of waves that are
discussed in Sect. \ref{sec:amplitudes}.

\subsection{Phase and group velocities}
\label{sec:phase-group-veloc}

For the moment, let us come back to the physical domain. To describe the propagation of a wave, we define the group velocity
$\vec{\mathrm{v}}_g=\vec\nabla_k \omega$, the speed at which the envelope of the
wave (and thus the energy) propagates through space,
and the phase velocity $\vec{\mathrm{v}}_p=\displaystyle\frac{\omega}{k}\vec{\hat k}$, which
characterizes the propagation of wavefronts (constant phases). We denote $\vec{\hat k}$ as the unit vector in the direction $\vec{k}$, and $\vec\nabla_k \omega$ the
gradient of the frequency $\omega$ as a function of the wavevector
$\vec{k}$. From Eq.\eqref{eq:1}, we can deduce the vertical and horizontal components
of the group and phase velocities of a gravity wave:
\begin{equation}
  \label{eq:6}
  \left\{
      \begin{array}{l}
V_{\mathrm{pr}} = \displaystyle\frac{\omega}{k^2}k_r = N \left(\displaystyle\frac{k_r
  k_h}{k^3}\right)\hbox{,}\\
 V_{\mathrm{ph}} = \displaystyle\frac{\omega}{k^2}k_h = N
 \left(\displaystyle\frac{k_h^2}{k^3}\right)\hbox{,}\\
  V_{\mathrm{gr}} = \displaystyle\frac{\partial\omega}{\partial k_r} = -N
  \left(\displaystyle\frac{k_r k_h}{k^3}\right)\hbox{,}\\
  V_{\mathrm{gh}} = \displaystyle\frac{\partial\omega}{\partial k_h} =\displaystyle \frac{N}{k}\left(1-\frac{k_h^2}{k^2}\right)  \hbox{.}
      \end{array}
  \right.
\end{equation}

\begin{figure}
  \centering
  \includegraphics[width=0.45\textwidth]{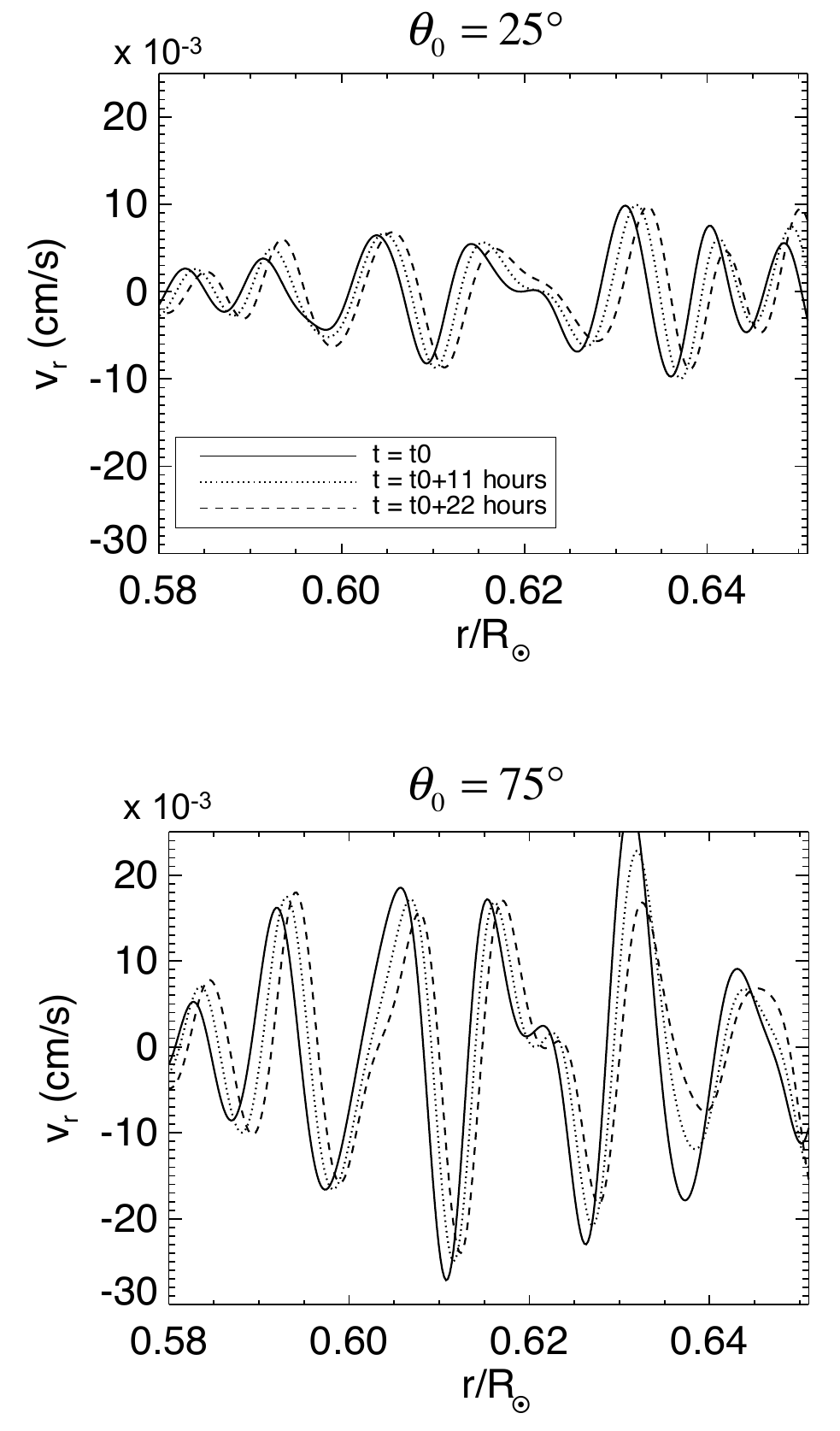}
  \caption{{Radial velocity fluctuations $\mathrm{v}_r(r,\theta_0,\varphi_0)$ as function of normalized radius for
    $\theta_0=35^o$ (top) and $\theta_0=75^o$ (bottom) represented for three consecutive
    instants (model \textit{ref}). The longitude $\varphi_0=150^o$ is the
    same for both panels. The wavefronts move from the left to the right, allowing a radial phase velocity $V_{\mathrm{pr}} \approx
    2\times10^{3}$cm/s to
    be measured, independent of $\theta_0$ and the
    same order of magnitude than the one calculated for a ray at $\omega_1=3\times10^{-3}$ mHz
    (see Fig. \ref{fig:velo_ray}).} }
  \label{fig:vpr}
\end{figure}

\begin{figure*}
  \centering
  \includegraphics[width=1\textwidth]{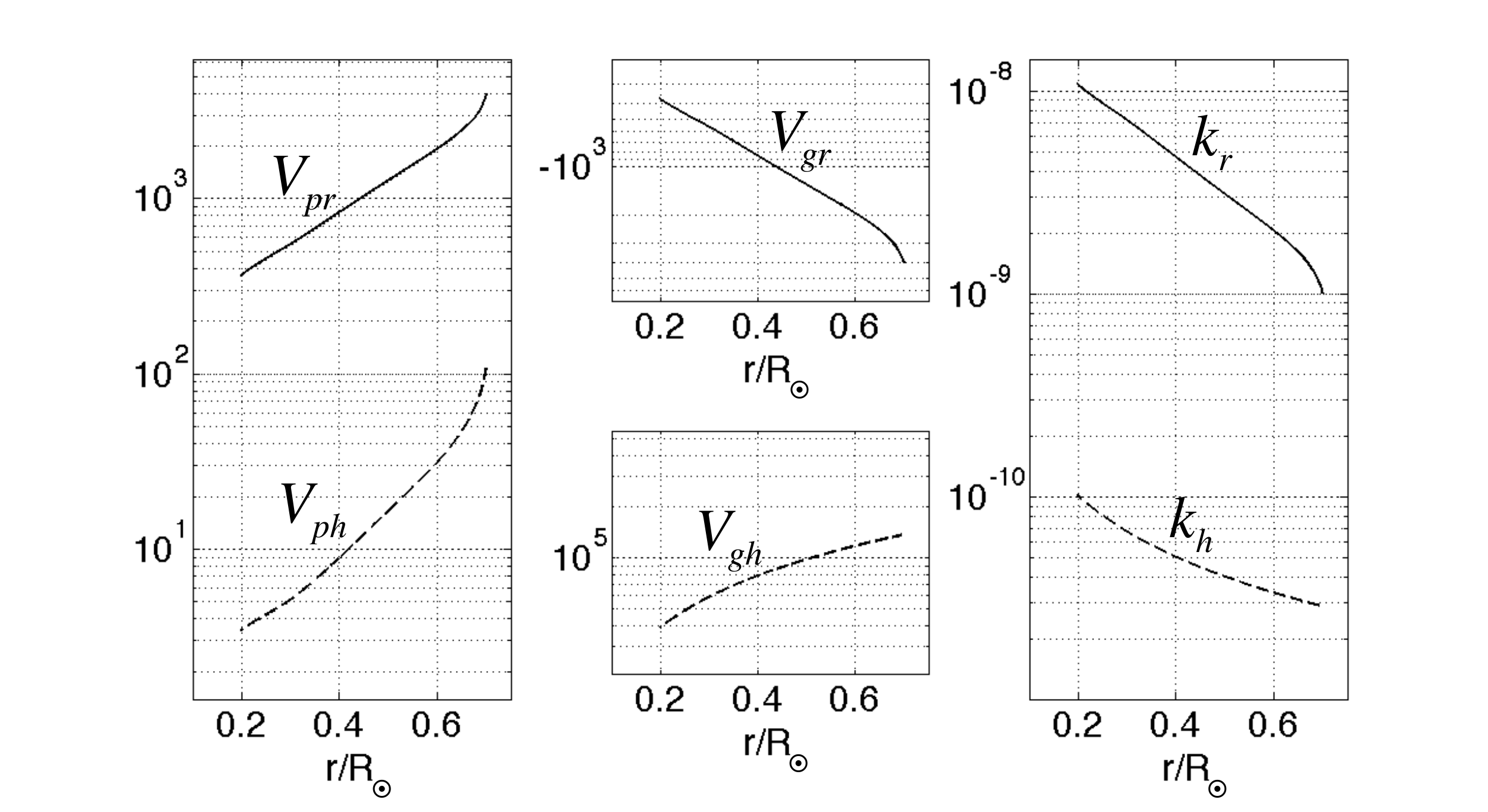}
  \caption{Components of phase velocity, group velocity, and wavevector
    computed by raytracing for the frequency $\omega_1= 3\times10^{-3}$ mHz. The
  corresponding raypath is shown in the top panel of Fig.~\ref{fig:schema}. The
  velocities are expressed in cm/s and the wavevector components in
  cm$^{-1}$. We observe that the ratio between $k_r$ and $k_h$ is about
  $10^{2}$ which explains the almost circular spiral observed in Fig. \ref{fig:schema}.}
  \label{fig:velo_ray}
\end{figure*}

A simple scalar product shows the orthogonality of $\vec{\mathrm{V}}_{\mathrm{p}}$ and
$\vec{\mathrm{V}}_{\mathrm{g}}$, and we also notice that $V_{\mathrm{pr}}=-V_{\mathrm{gr}}$. As already presented
in Sect. \ref{sec:excit-grav-waves}, Fig.~\ref{fig:schema} obtained
with our raytracing code provides an
illustration of the directions of $\vec{k}$, $\vec{\mathrm{V}}_{\mathrm{g}}$, and
$\vec{\mathrm{V}}_{\mathrm{p}}$. For the low-frequency waves visible in Fig.~\ref{fig:3d_eqsl},
we can measure their phase velocity by plotting the
variations in radial velocity (for instance) as a function of the normalized radius for three consecutive
instants (Fig.~\ref{fig:vpr}). The signal translates with time from inward to outward. Wavefronts are easy to locate in this figure because their propagation is mainly
radial (as explained in Fig. \ref{fig:schema}). The Brunt-Väisälä frequency $N$ is a function of the radius so the
phase velocity is not constant. Nevertheless, we can give an estimation of
its value by measuring the mean distance travelled by the wavefronts during a
given time. We find $V_{\mathrm{pr}} \approx 2\times 10^{3}$ cm/s. \\
Measuring the
horizontal phase velocity is more difficult. Thus, we here use our
raytracing code to calculate the theoretical values corresponding to the frequency
$\omega_1=3 \times 10^{-3}$ mHz. Figure
\ref{fig:velo_ray} shows the evolution of $\vec{\mathrm{V}}_{\mathrm{p}}$, $\vec{\mathrm{V}}_{\mathrm{g}}$, and
$\vec{k}$ along the ray. Radial components of $\vec{\mathrm{V}}_{\mathrm{p}}$ and $\vec{k}$ are about two orders of
magnitude higher than their horizontal parts. This is fully coherent with
the almost circular spiral observed and with the asumption $k_h \ll k_r$ in the
literature concerning low-frequency gravity waves ($\omega \ll
N$).

\subsection{Spectrum}
\label{sec:spectrum-1}
The low-frequency waves that we see in real space are not the only
ingredients of the excited spectrum. We have given an overview of the
richness of this spectrum in Sec.~\ref{sec:spectrum}. We now propose a more
detailed analysis.

\subsubsection{Temporal and spatial dependencies}
\label{sec:fits}

\begin{figure*}
  \centering
  \includegraphics[width=0.9\textwidth]{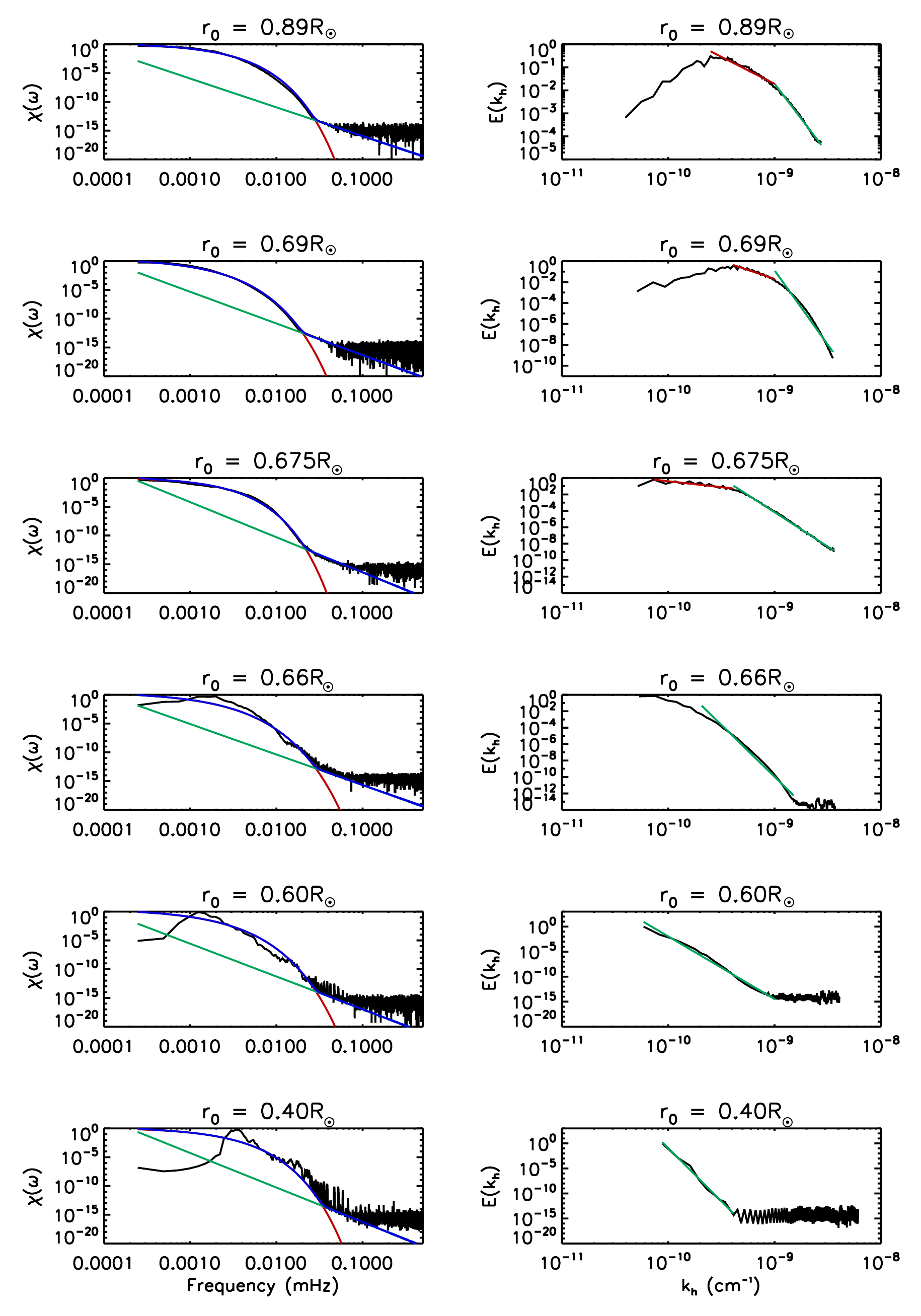}
  \caption{Decomposition of the spectrum $E(r_0,k_h,\omega)$ of model \textit{ref} into its temporal and spatial
    dependencies. The best fit for $\chi(\omega)$ is a combination of a
    Gaussian-like and a Lorentzian-like functions since $E(k_h)$ decreases
    as $k_h^{-\eta}$ with $\eta$=5-7.}
  \label{fig:fits}
\end{figure*}

We recall that the quantity studied here is given by Eq.~\eqref{eq:5}.
Since $E(r_0,\ell,\omega)$ depends on both $\ell$ and $\omega$, we would like to be
able to distinguish between these two dependencies. The frequency $\omega$
corresponds to the temporal variations. We thus introduce the horizontal
wavevector, $k_h = \displaystyle\frac{\sqrt{\ell(\ell+1)}}{r}$, to characterize the spatial
variations. A first method should be to choose a value of $\omega$ to study
the variations of $E$ with $k_h$ and vice versa. That is the method employed by
\citet{2009A&A...494..191B} to obtain the function $\chi(\omega)$ with a
wavenumber $k_{h0}$ corresponding to the maximum of energy. The
disadvantage of this method is that it does not consider the whole
spectrum. For this reason, we chose to follow the idea of \citet{Rogers:2013ui} by
computing a singular value decomposition (SVD) of $E(r_0,k_h,\omega)$. The
concept of the SVD is to decompose $E$ into its separable part and a leftover part such that
\begin{equation}
  \label{eq:13}
  E(r_0,k_h,\omega) = c_1 E(k_h) \chi(\omega) + \sum_{i>1}c_i
  E_i(k_h,\omega) \hbox{.}
\end{equation}
\newline
Of course, this is meaningful only if the initial function $E$ is
separable. We compute this decomposition for several depths $r_0$. In the
convective zone ($r_0 \ge 0.69$ $R_\odot$ if we take the overshoot region into account), the ratio $c_1/\sum_{i} c_i$
characterizing the separability of $E(r_0,k_h,\omega)$ varies
between 60\% and 66\% and is superior to 84\% in the radiation zone. Figure
\ref{fig:fits} shows the result of these calculations where we 
have superimposed the best fit for each curve. In the convection zone,
gravity waves are evanescent so the spectrum is mainly a turbulence
spectrum associated with thermal convection. The chosen fit is a combination of a Gaussian-like function
\begin{equation}
  \label{eq:14}
  \mathcal{G}(\omega) =
  \frac{1}{\sqrt{\pi}\omega_{G}}\exp\left({-\left[\frac{\omega}{\omega_G}\right]^{\alpha_G}}\right) \hbox{,}
\end{equation}
and a Lorentzian-like function
\begin{equation}
  \label{eq:15}
  \mathcal{L}(\omega) = \frac{2}{\pi
    \omega_L}\left[\frac{1}{1+\left(\frac{\omega}{\omega_G}\right) ^{\alpha_L}}\right]
  \hbox{,}
\end{equation}

\begin{figure*}
  \centering
\includegraphics[width=1\textwidth]{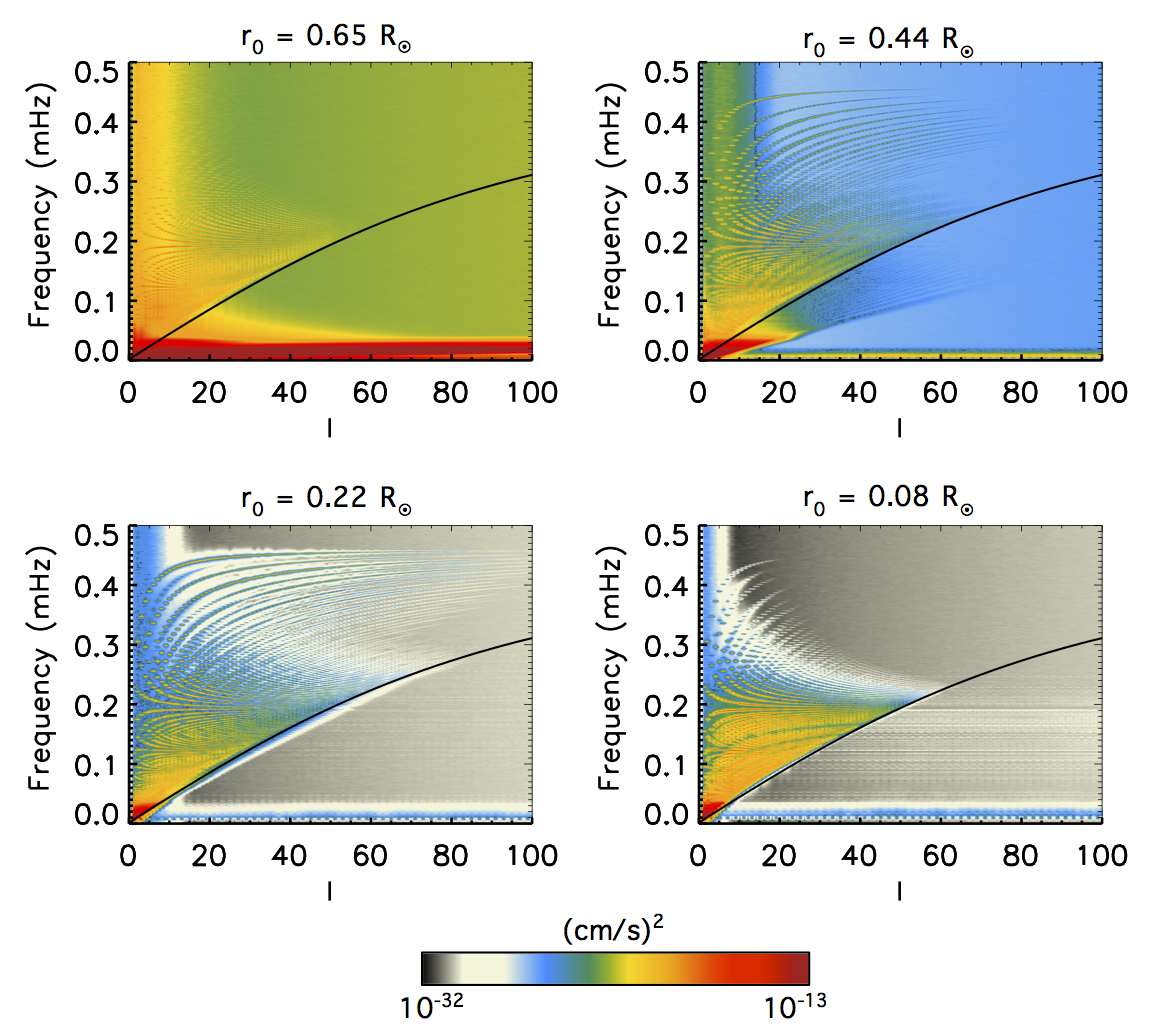}
  \caption{{Variation in the spectrum shape as function of depth $r_0$
    (model \textit{ref}). The black line marks the last resolved ridge
    $n=58$ common to all depths. We have volontarily cut the horizontal
    axis to $\ell=100$ to show the part where ridges are the most visible and
    used the same color table for the four depths, although the minimum and
    maximum amplitudes vary greatly. In particular, the change in the
    background color shows the increase in the background noise when
    reaching the excitation zone ($r_0=0.65R_\odot$) and then entering the
    convective region.}}
  \label{fig:omegal_depth}
\end{figure*}

where $\omega_G = 3\beta$ and $\omega_L = \beta/3$. The parameters vary
with $r_0$ in the range $\alpha_G \in$ [0.67,0.8], 
$\alpha_L \in$ [5,6], and $\beta \in$ [0.06,0.13]. For the eddy-time
function $\chi(\omega)$, we retrieve the results found by
\citet{2009A&A...494..191B} 
in the convective zone, showing that the best fit is not a pure Gaussian function as
used in \citet{1994ApJ...424..466G}, but rather a combination with a
Lorentzian-like function. However, we notice that we had to increase
$\alpha_L$ from 2 to 5-6 to fit the strong slope formed by $\chi(\omega)$ at high
frequency. \\
These results are similar if we apply the SVD to the more
turbulent models \textit{turb1} and \textit{turb2}. Because
  \citet{2009A&A...494..191B} studied spectra coming from a purely convective shell, the difference may be due to the coupling with
the radiative interior. This fit remains approximatively correct in the
radiative zone, although the energetical peaks relative to gravity waves make
the spectrum more noisy at high frequency. For $E(k_h)$, we also observe a
modification of the form when passing from convective to radiative
interior. For $r_0>0.66R_\odot$, we fit the curve with two straight
lines. The red line (left) corresponds to the slope $k_h^{-5/3}$ of a
Kolmogorov turbulence spectrum. This result agrees with
\citet{Samadi:2003ir}. For higher horizontal wavenumbers, we fit the
curve with a second straight line (green) that is much more inclined ($k_h^{-5}$ to $k_h^{-7}$). This result seems to
be closer to the one presented by \citet{Rogers:2013ui}. In the radiative zone, the spectrum is no longer turbulent, and only
the strongest slope remains. To conclude, the SVD reveals that the
temporal part of the spectrum behaves more like a Lorentzian, whereas its
spatial part have a power that decreases as $k_h^{-5}$ to $k_h^{-7}$.

\subsubsection{Variation with depth}
\label{sec:evolution-with-depth}

We now have a look at the differences between spectra calculated at
different depths $r_0$. We recall that the spectrum shown in Fig. \ref{fig:omega_l} was
taken at $r_0=0.26R_\odot$. In Fig. \ref{fig:omegal_depth}, we display
four spectra calculated with the same temporal sequence of model
\textit{ref}. The first one is taken near the region of excitation
($r_0=0.65R_\odot$), the last one near the center ($r_0=0.08R_\odot$), and
the two others in the middle of the radiative zone ($r_0=0.44R_\odot$ and
$r_0=0.22R_\odot$). We do not show the convective zone since no modes are visible there.
It is clear that the spectrum's aspect is different depending on
the depth. 
First of all, we notice that a common feature
between all these spectra is the inferior limit underlined by a black line. It is situated exactly at the same place in all spectra and
corresponds to the ridge number $n=58$. Since the order $n$ represents the number of
nodes of the radial eigenfunctions (see Sect. \ref{sec:period-spacing}), we
can understand that this boundary is due to the radial 
resolution of our model.
At 0.65$R_\odot$, only the low-frequency
part of the spectrum is visible. If we move deeper in radius, higher
frequencies appear. For $r_0=0.44R_\odot$, there is a region that looks
different, under the black line. The energy in this region does not form peaks that are regularly spaced
in period, such as g-modes should. Consequently, we interpret this
region as propagating gravity waves that do not form g-modes. This zone reduces and disappears when we move down in depth reinforces
this hypothesis since it corresponds to the action of the radiative
damping on these waves (discussed in
Sect. \ref{sec:radiative-damping}). 

\subsection{g-modes}
\label{sec:g-modes}
Until now, we have looked at the overall shape of the spectrum, considering
both propagative and standing IGWs. In this
section, we concentrate on the high-frequency part of the spectrum
corresponding only to g-modes. They are identifiable by their radial order
$n$ defined by the number of nodes of their eigenfunction.

\subsubsection{Period spacing and eigenfunctions}
\label{sec:period-spacing}

\begin{figure*}
  \centering
  \includegraphics[width=1.\textwidth]{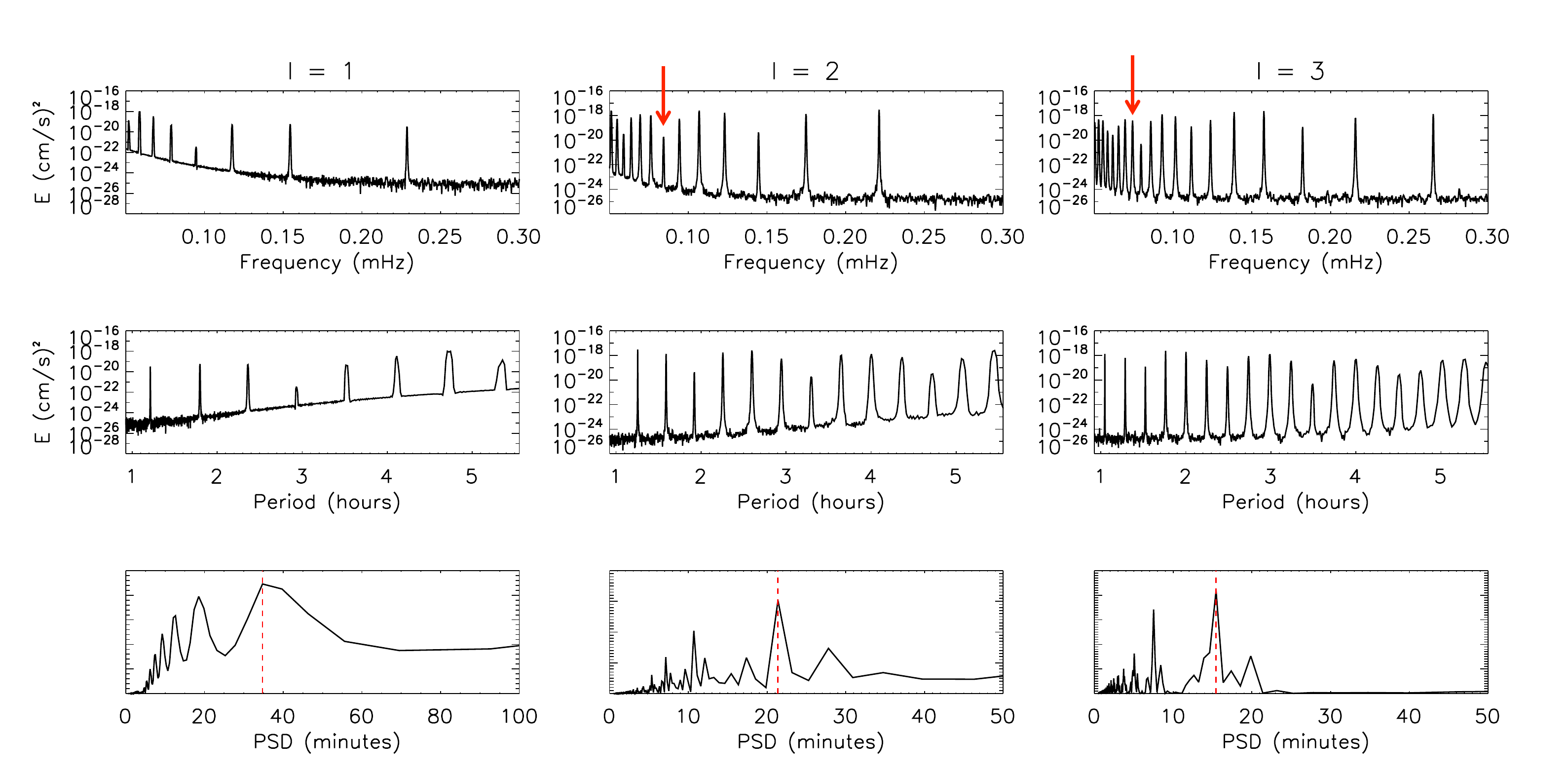}
  \caption{\textbf{Top and middle}: Spectrum of gravity waves for $\ell=1,2,3$ as function
  of frequency and period. \textbf{Bottom}: Fourier transform of the middle
  spectrum that shows the constant period spacing between peaks. We find $\Delta P_1 = 37.1$ min,
  $\Delta P_2 = 21.2$ min, and $\Delta P_3 = 14.8$ min as pointed out
    by vertical red dotted lines.}
  \label{fig:power_freq}
\end{figure*}

We first focus on the behavior of $E$ for a given order $\ell$. One of the
main asymptotic properties of g-modes - used to detect their signatures in the
Sun \citep{Garcia:2007iq} and stars \citep{Bedding:2011te} - is that they are supposelly equally spaced 
in period \citep[e.g.,][]{aerts2010asteroseismology}. To check this property, we represent the variations in $E(r_0,\ell,\omega)$ in
Fig.~\ref{fig:power_freq} as a function
of $\omega$ and $P=1/\omega$ for $\ell$=1, 2, and 3. We
volontarily limit the frequency to the range [0.05,0.30] mHz for $\ell$=1,
and [0.05,0.45] mHz for $\ell$=2 and 3 to
focus only on well defined peaks corresponding to radial orders $n \le 20$. The bottom panel represents the Fourier
transform of the period spectrum, and we observe that a main peak appears, indicating the value of
$\Delta P_\ell$, the period spacing between modes. We find $\Delta P_1 = 37.1$ min,
  $\Delta P_2 = 21.2$ min, and $\Delta P_3 = 14.8$ min.

The asymptotic theory predicts that $\Delta P_\ell$ must be given by
\begin{equation}
  \label{eq:12}
  \Delta P_\ell = \frac{\pi}{\sqrt{\ell(\ell+1)}\displaystyle\int_{r_1}^{r_2}\displaystyle\frac{N}{r}dr} \hbox{,}
\end{equation}
where $r_1$ and $r_2$ are the turning points defined by $N(r_1) = N(r_2) =
\omega$ \citep[e.g.,][]{Christensen-Dalsgaard97lecturenotes}.
We compare our measures to the values given by Eq.~\eqref{eq:12}
for several $\ell$. By taking for $N$ the
profile defined in ASH and represented in Fig.~\ref{fig:BVfreq}, we obtain
on average agreement of about 5\% between the theoretical and the measured
values. This small difference is due, on one hand, to the error made by measuring $\Delta P_\ell$
with a finite time sequence and on the other, from the fact that Eq.~\eqref{eq:12}
has been obtained assuming $n \gg \ell$. As recall above, \citet{Garcia:2007iq} have used the equally spacing between modes
to detect their signature in the GOLF data from the Sun. For $\ell$=1, they
found a peak corresponding to $\Delta P_1$ located between 22 and 26
minutes. For $\Delta P_2$, they were expecting 9-15 minutes and 5-11
minutes for $\Delta P_3$. In Eq. \eqref{eq:12}, we see that $\Delta P_\ell$ is inversely proportional to
$\displaystyle\int_{r_1}^{r_2}\displaystyle\frac{N}{r}dr$. Then, we can show that 60\%
of the total value of this integral is built on the value of $N$ in the
inner 0.2 $R_\odot$ \citep{Brun1998,2012sf2a.conf..289A}. Consequently, a slight difference in the BV profile
between our model and the one used by \citet{Garcia:2007iq} can easily
explain the observed bias. \\\newline

\begin{figure*}
  \centering
 \includegraphics[width=0.99\textwidth]{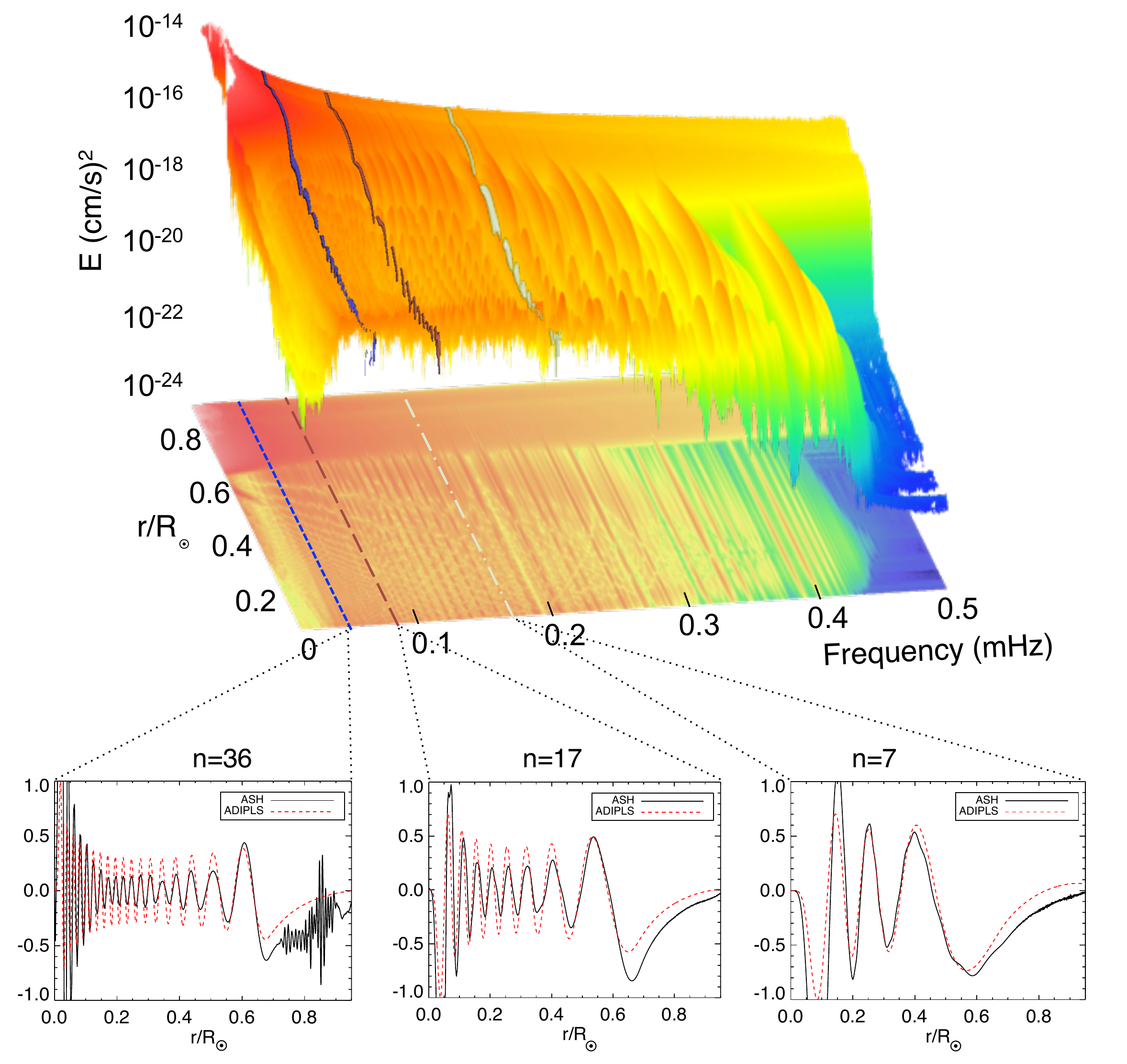}
  \caption{{\textbf{Top:} Energy $E$ as a function of the normalized radius and frequency for
    the order $\ell$=5 (model \textit{turb2}). \textbf{Bottom:} Extraction of three normalized eigenfunctions corresponding to
    $n=36$, $n=17$, and $n=7$. Their positions in the top panel are indicated by blue dashed, red long dashed and green dot-dashed lines. The
    eigenfunctions computed by the oscillation code ADIPLS are represented by red dotted lines.}}
  \label{fig:wave_fun2}
\end{figure*}

As shown in Sect.~\ref{sec:evolution-with-depth}, we are able to calculate
the spectrum at different depths. To visualize the radial evolution of some
g-mode amplitudes, we represent a map of the energy $E(r,\ell =5,\omega)$ Fig.~\ref{fig:wave_fun2} as a function
of the normalized radius $r/R_\odot$ and the frequency $\omega$. 
This figure was obtained by calculating the spectrum for each radial point.  
    We can count the number $n$ of nodes in the radial
    eigenfunctions and see that it increases with decreasing frequencies. The signal is projected on the bottom of the figure, which allows to precisely see the
    distribution of the nodes as a function of the radius and the frequency.
The bottom panels provide a comparison between the eigenfunctions in ASH and in ADIPLS for three different values
  of $n$.

\subsubsection{Rotational splitting}
\label{sec:rotational-splitting}

\begin{figure}[h]
  \centering
  \includegraphics[width=0.45\textwidth]{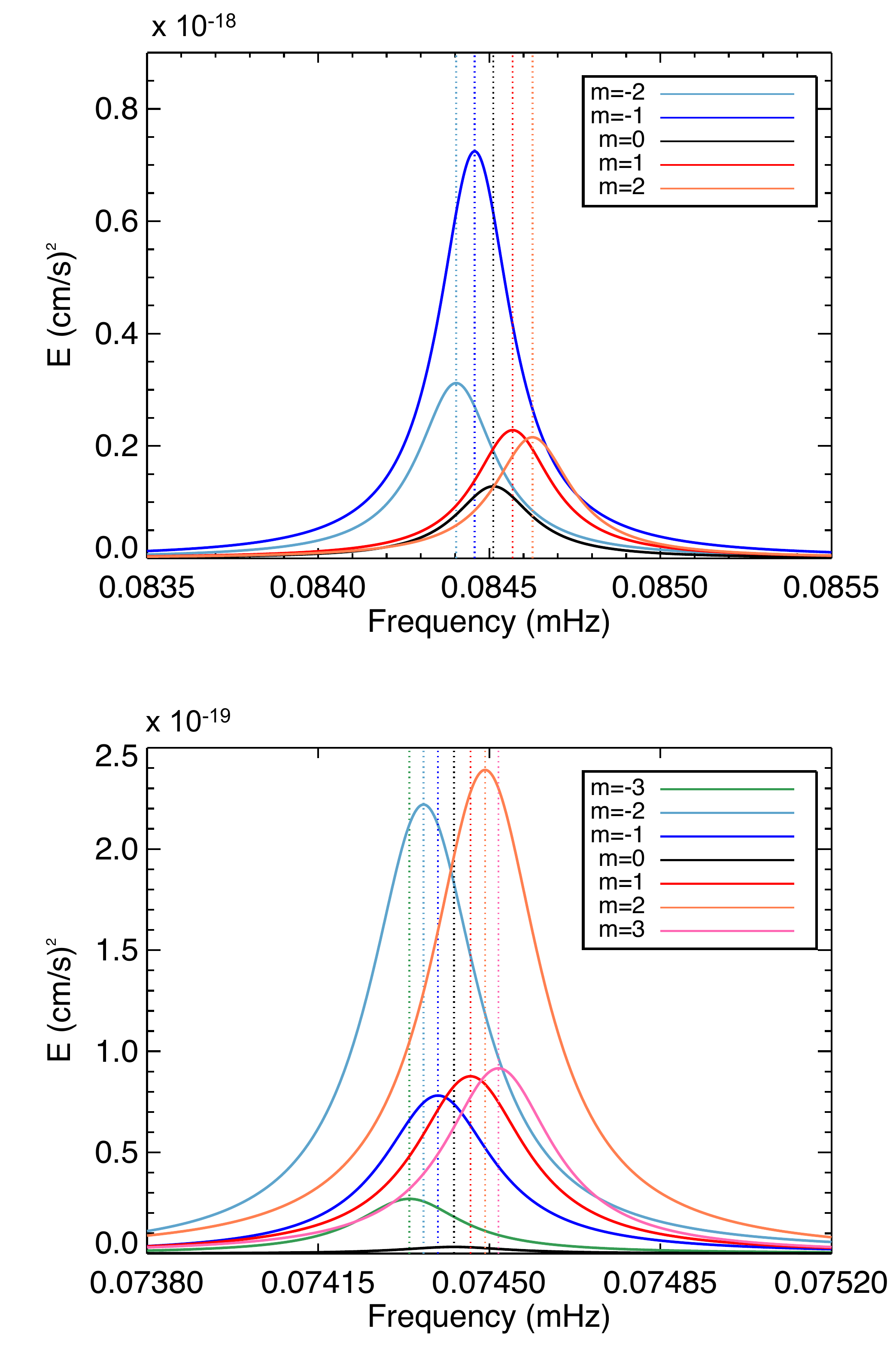}
  \caption{Rotational splitting for $\ell=2$ and $-2<m<2$ (top) and  $\ell=3$
    and $-3<m<3$ (bottom). {The chosen frequencies are identified by
      red vertical arrows in Fig. \ref{fig:power_freq}.} Here, the peaks have
    been fitted with a Lorentzian function to determine the position of
    their maximum precisely.}
 \label{fig:splitting}
\end{figure}

Until now, we have put aside the effects of rotation by
summing over all the $m$ components of a given $\ell$ mode. Without rotation,
although $m$ modes are not located at the same place ($m$=0 lies in the
meridional plane and the more high $|m|$ the more inclined the plane), the
frequencies are degenerated. That is to say that modes identified by the
same pair ($\ell$,$n$), but different $m$ are merged in the same peak in the spectrum. But we do not
forget that all models presented in this paper rotate at the solar rotation rate
$\Omega_\odot$ (see Fig.~\ref{fig:omega}). To look at rotational effects we thus need to distinguish one $m$
component from one another. One must first establish the difference between
prograde (propagating in the direction of rotation) and retrograde
waves. Thus, rotation increases the
phase speed of prograde waves and decreases the one of retrograde
waves. This results in a separation of their frequencies. Figure \ref{fig:splitting} is the superposition of peaks with same values
of $\ell$, but values of $m$ vary between $-\ell$ and $+\ell$. We recall that the
energy $E$ is obtained from the radial velocity $V_r(r_0,\theta,\varphi,t)$ thanks to a spherical
harmonic transform followed by a temporal Fourier transform. To obtain negative
values of $m$, we took the temporal Fourier transform of $\hat V_r^*
(r_0,\ell,m,t)$, which is the complex conjugate of $\hat V_r(r_0,\ell,m,t)$ (see
Sect.~\ref{sec:spectrum}). We observe that the
peaks move from left to right as $m$ increases. We thus retrieve the
phenomenon called rotational splitting
\citep[e.g.,][]{aerts2010asteroseismology} that
allows asteroseismologists to reconstruct the internal rotation profile of
stars \citep{2012ApJ...756...19D}. \\\newline
The theory of stellar oscillations \citep[e.g.,][]{Christensen-Dalsgaard97lecturenotes}
predicts that the frequency splitting must be given by
\begin{equation}
  \label{eq:18}
  \delta_{n\ell m} = m\beta_{n\ell}\displaystyle\int_0^{R_\odot}K_{n\ell}(r)\Omega(r)\mathrm{d}r\hbox{,}
\end{equation}
where 
\begin{equation}
  \label{eq:19}
  K_{n\ell}(r) = \frac{\left(\xi_r^2+L^2\xi_h^2-2\xi_r\xi_h-\xi_h^2\right)r^2\bar\rho}{\displaystyle\int_0^{R_\odot}\left(\xi_r^2+L^2\xi_h^2-2\xi_r\xi_h-\xi_h^2\right)r^2\bar\rho\mathrm{d}r}\hbox{,}
\end{equation}
and
\begin{equation}
  \label{eq:20}
  \beta_{n\ell}=\frac{\displaystyle\int_0^{R_\odot} \left(\xi_r^2+L^2\xi_h^2-2\xi_r\xi_h-\xi_h^2\right)r^2\bar\rho\mathrm{d}r}{\displaystyle\int_0^{R_\odot}\left(\xi_r^2+L^2\xi_h^2\right)r^2\bar\rho\mathrm{d}r}\hbox{,}
\end{equation}
are functions of the radial and horizontal
displacements ($\xi_r$ and $\xi_h$) and of the reference density $\bar\rho$, and $L^2=\ell(\ell+1)$. Moreover, the rotational kernel $K_{n\ell}$ is unimodular, i.e.,
\begin{equation}
  \label{eq:21}
  \int_0^{R_\odot} K_{n\ell}(r) \mathrm{d}r = 1 \hbox{.}
\end{equation}
For high-order g modes, we can neglect the terms containing $\xi_r$, so
that
\begin{equation}
  \label{eq:22}
  \beta_{n\ell} \approx 1-\frac{1}{\ell(\ell+1)}\hbox{,}
\end{equation}
and for a uniform rotation $\Omega_S$ (in the radiative zone), we obtain
\begin{equation}
  \label{eq:23}
  \delta_{n\ell m} = m\beta_{n\ell}\Omega_S\hbox{.}
\end{equation}
Finally, in the frame rotating with the star, it becomes
\begin{equation}
  \label{eq:24}
  \delta_{n\ell m} = -m(1-\beta_{n\ell})\Omega_S\hbox{.}
\end{equation}

\begin{figure}
  \centering
 \includegraphics[width=0.45\textwidth]{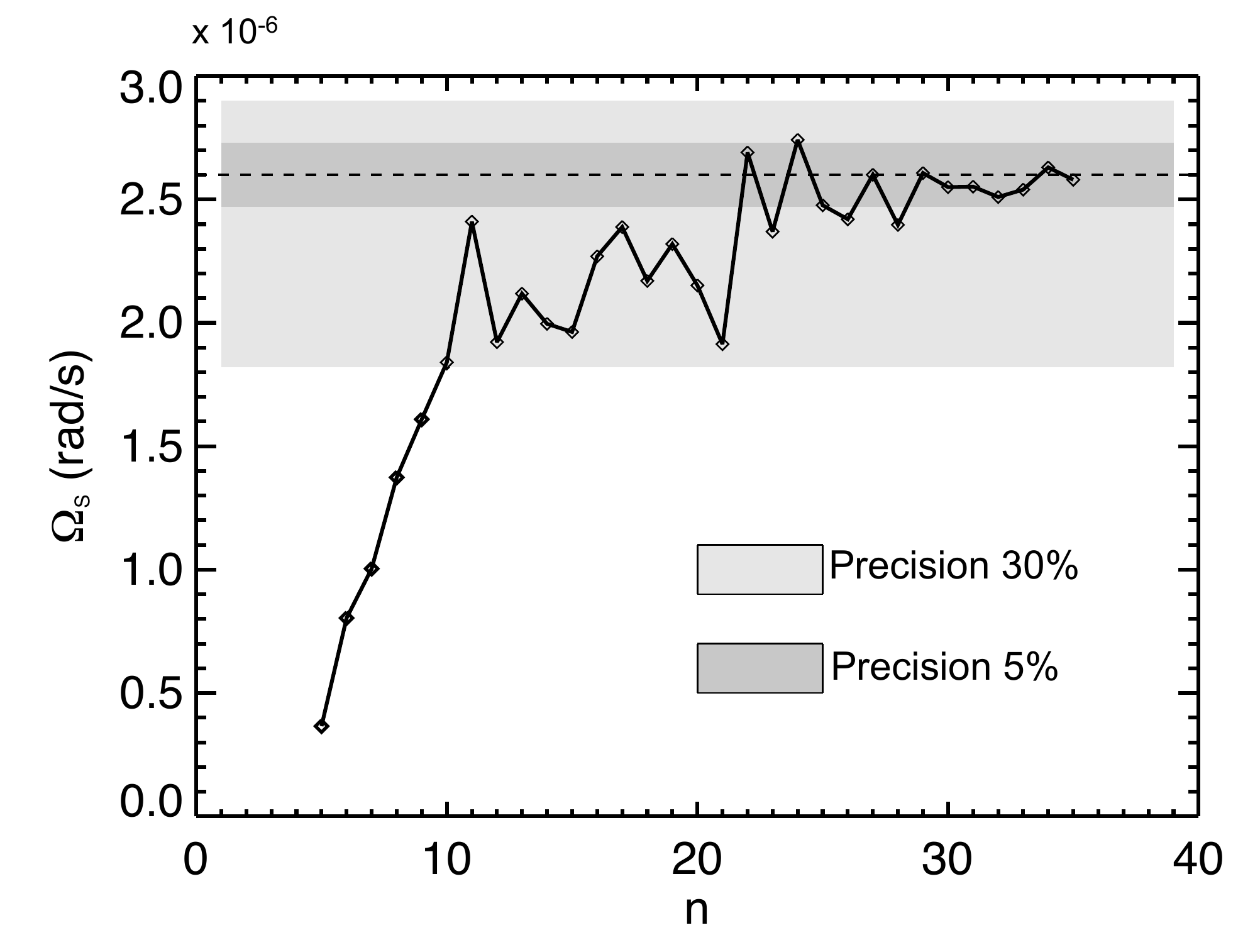}
  \caption{Estimation of the rotation rate from measuring the
    rotational splitting for $\ell$=3. The same tendency is obtained for other
    values of $\ell$. Two gray zones indicate the precision of the measure
    with respect to the right rotation rate that is imposed. We obtain less than 5\% error for $n > $ 30.}
   \label{fig:split_tendance}
\end{figure}

The usual way to use this relation is to deduce the rotation rate
$\Omega_S$ from the measure of $\delta_{n\ell m}$. Here, since we already know
the value of $\Omega_S=\Omega_\odot$, we can evaluate the precision of this
method. In Fig. \ref{fig:split_tendance}, we represent the values of $\Omega_S$ obtained
by inserting different measures of $\delta_{n\ell m}$ in Eq.~\eqref{eq:24}. As
expressed by Eq. \eqref{eq:21}, these
values do not depend on the depth at which we extract the spectrum. We
observe that we approach the real solar rotation rate imposed in the simulation
($2.6\times10^{-6}$ rad/s) when $n$ increases. Indeed, although the convective zone is submitted
to a differential rotation, IGWs are not sensitive to it because they do
not propagate into this zone. This result confirms that
Eq.~\eqref{eq:24} is mostly valid for asymptotic modes. Unfortunately, the
more we increase $n$, the more the frequency decreases, so peaks get
very close to each other. For this reason, the identification of peaks
corresponding to the same couple ($\ell$,$n$) becomes imprecise, if not
impossible, for very high $n$ and we have to stop around $n=35$. In spite of
that, the convergence is quick enough to estimate the rotation with an
accuracy of 30\% from $n=10$ (corresponding to frequencies in the range
[0.05,0.1] mHz for $\ell \sim 2-4$, which can be observed in stars) knowing that the rotation rate
$\Omega_S$ is mostly underestimated by this method. To increase the
precision to 5\%, we have to look at modes with $n>25$ that corresponds to frequencies around 0.01 mHz. \\\newline

{Another interesting piece of information supplied by}
Fig. \ref{fig:splitting} is the asymmetry of amplitude between prograde and
retrograde modes. It is clear here that the usual assumption of energy equal distribution is not verified \citep{Belkacem:2009wl} and that one should take
this bias into accound in asteroseismic and stellar evolution studies. We also notice in both panels that
the $m=0$ peak is much lower than the other ones. For $\ell=3$, $m=0$ is hardly
visible because it is very close to the horizontal axis. In contrast, higher peaks
correspond to higher $m$. As explained above, in a spherically symmetrical
star, IGWs propagate in planes inclined with respect to the meridional
plane as a function of $m$. We thus understand that the most energetic
modes lie in planes close to the equatorial plane, and we might be more able
to detect them in this area.

\begin{figure}
  \centering
   \includegraphics[width=0.5\textwidth]{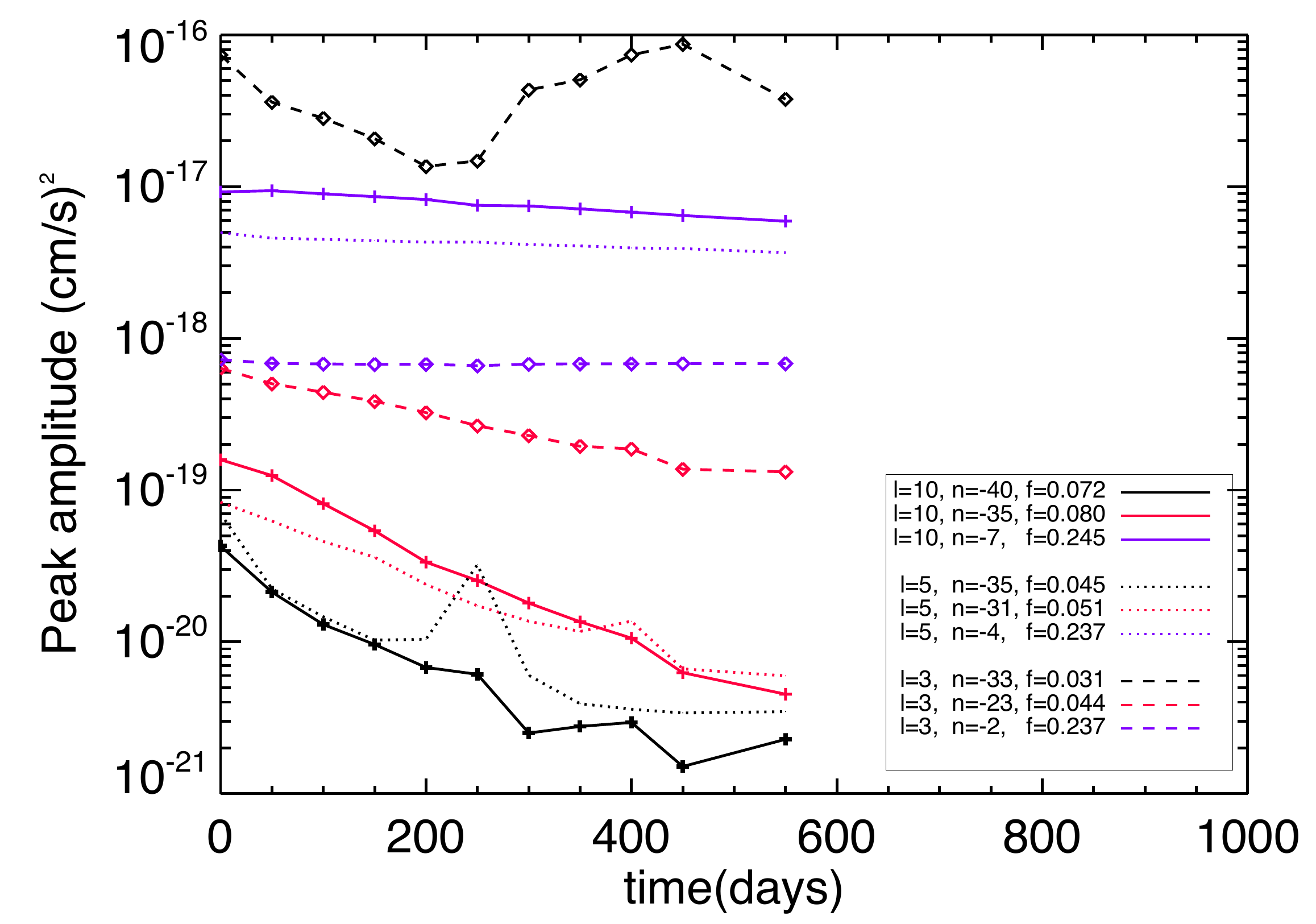}
  \caption{Evolution of the peak's amplitude as a function of time. High-frequency waves (violet) have a lifetime much greater than 550 days since
    their amplitude does not vary over this time interval. Intermediate-frequency waves (red)
    are damped along their propagation and for low-frequency waves (black)
    their amplitude sometimes increases suddenly, due to a re-excitation. The
    legend indicates the values of ($\ell$,$n$) and the corresponding
    frequency $f$ in mHz for each curve.}
  \label{fig:lifetime}
\end{figure}

\subsubsection{Lifetime}
\label{sec:lifetime}

The knowledge of g-mode lifetimes is very important for
detecting them in the Sun. \citet{1990ApJ...363..694G} find mode
lifetimes of about 106 years, while \citet{2010A&ARv..18..197A} give about 1 million years. Thus a large incertainty remains in the
literature about this value. The standard method for obtaining the lifetime
of modes is the measure of the half width at half maximum (HWHM) of the
peaks. This implicitly supposes that the time series used to calculate the
spectrum are much longer than the lifetime. In this work, a timescale of several hundred
years is out of reach because we have to deal with a time step of about
100s. Our maximum currently availabe time series is about 550 days. To
skirt the problem, we have cut this main sequence into several consecutive subsequences (11 times
50 days) and measured the amplitude of some peaks in each
subsequence. Figure \ref{fig:lifetime} represents these amplitudes as a
function of time. Three regimes are identifiable. High-frequency waves are almost constant in amplitude or slightly
decreasing. This shows that the
lifetime of high-frequency modes is indeed much greater than 550 days. Intermediate-frequency waves are damped (thermal effects) along their
propagation. \\
A comparison shows a disagreement between this temporal damping and linear theoretical predictions. Indeed, when using the linearized equations for the
gravity waves propagation \citep[e.g.,][]{ZahnTalonMatias1997}, the wave's amplitude
is expected to decrease as predicted by \citet{1998A&A...333..343F}
\begin{equation}
  \label{eq:17}
  e^{-\displaystyle\kappa_r \displaystyle\frac{k_h^2}{2}\frac{N^2}{\omega^2}t} \hbox{,}
\end{equation}
which gives a much steeper slope than the one observed in Fig.~\ref{fig:lifetime}.
We discuss the role played by nonlinearities in mitigating the damping effects in
Sect.~\ref{sec:radiative-damping}. Finally, low-frequency waves seem to be re-excited during this temporal window since their amplitude increases abruptly at a some
instants, for example at $t=250$ days. This
excitation may be due to the arrival of a new plume exciting a wave at the
same frequency. This would be coherent with our observations of
Sect. \ref{sec:excit-grav-waves} showing that convection excites
high-amplitude low-frequency waves. The other possibility for explaining this
re-excitation process could be related to the nonlinear interactions
between two other waves (cf. Sect. \ref{sec:non-line-inter-1}). 

\begin{figure*}
  \centering
  \includegraphics[width=0.95\textwidth]{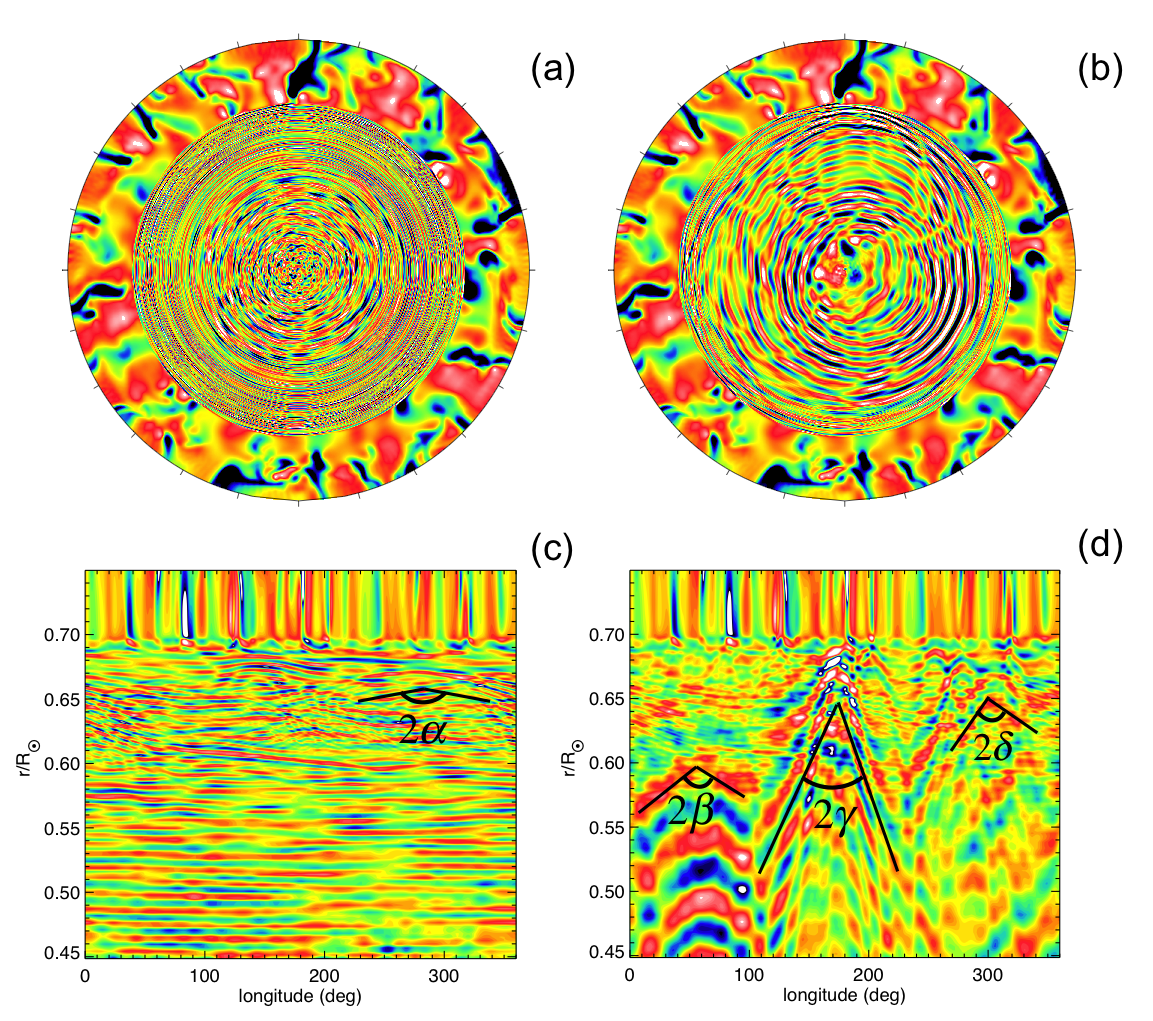}
  \caption{Comparison between fully nonlinear model \textit{turb2} in
    panels (a) and (c) and semi-linear model \textit{sem-lin} in panels (b)
  and (d). {The quantity represented is the normalized radial
    velocity. Both top panels are equatorial slices where we see the
    outer convective region (red/blue tones denote positive/negative
    values). The cross shape visible in  the radiative zone in
    top left panel is a Moiré pattern. In the bottom panels, we have zoomed in the upper part of the radiative zone to highlight the departure of wavefronts from the base
    of plumes.} We can measure the angle formed by the wavefronts in both
  situations and deduce that nonlinear interactions favor
  low-frequency waves.}
  \label{fig:eqsl_zoom}
\end{figure*}

\subsection{Nonlinear wave interactions}
\label{sec:non-line-inter-1}
\begin{figure}[h]
  \centering
  \includegraphics[width=0.4\textwidth]{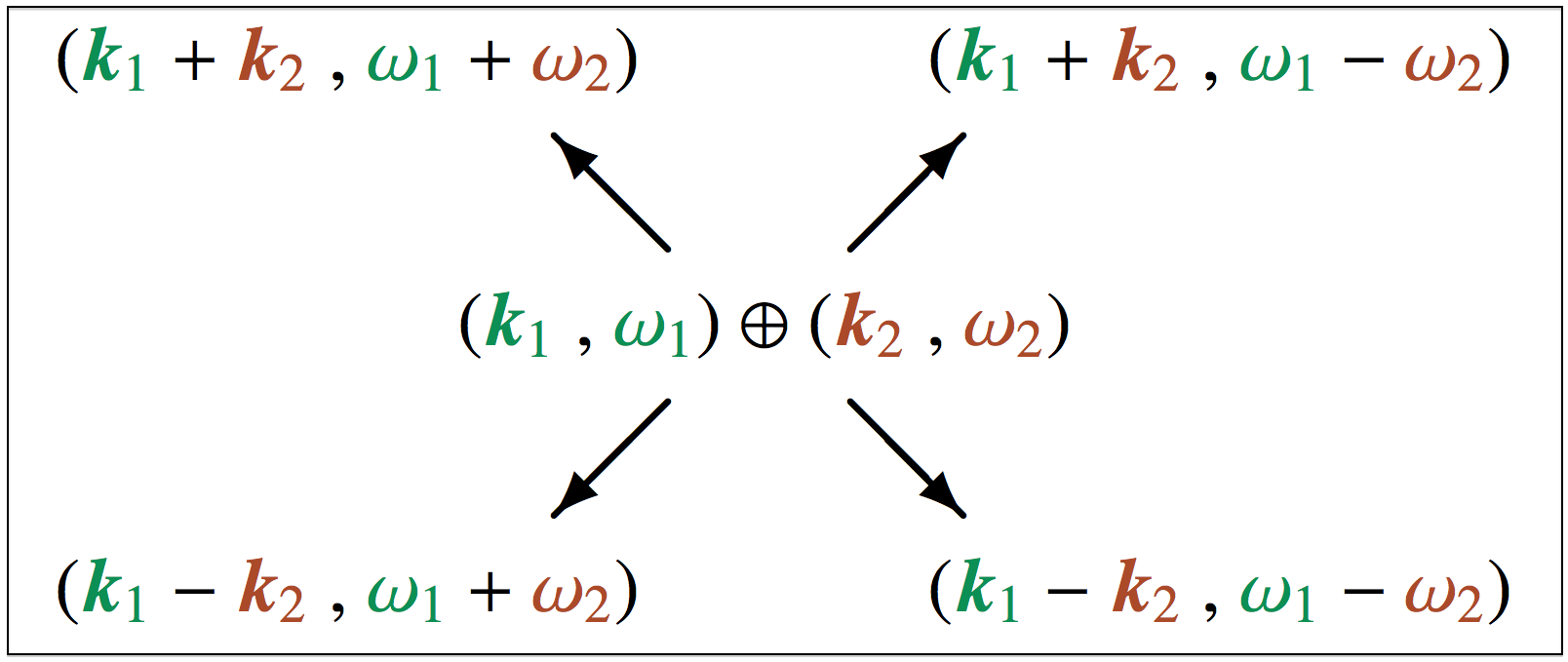}
  \caption{Diagram showing the possibilities for two waves $(\vec{k}_1,\omega_1)$ and $(\vec{k}_2,\omega_2)$ to interact and give birth to a third wave.}
  \label{fig:triad}
\end{figure}

In this section, we compare the previous results - obtained thanks to a nonlinear
resolution of the hydrodynamical equations Eq. \eqref{eq:288} and
\eqref{eq:8} - and a
model called \textit{sem-lin} where we have set
the nonlinear terms in the radiative region to zero, as inspired by a similar
approach in \citet{RogersGlatzmaier2005}. We multiplied
the nonlinear terms by a function equal to one if $r>0.73$R$_\odot$, to zero if
$r<0.71$R$_\odot$, and decreasing linearly from 1 to 0 between these
points. Except for this filter, the model  \textit{sem-lin} is identical to \textit{turb2}. We
activated the filter at time $t_0$ when the model is relaxed and
stable. Figure \ref{fig:eqsl_zoom} compares the dynamics in the
radiative interiors of both models. We
have represented the normalized radial velocity $v_r/v_{rms}$ at colatitude
$\theta=\pi/2$ (equatorial plane) as a
function of longitude $\varphi$ and normalized radius. Lefthand panels (a) and
(c) correspond to the fully nonlinear model 
\textit{turb2} and righthand panels (b) and (d) to the model
\textit{sem-lin}. We observe in (b) that the wavefronts are much more
inclined, looking like the high-frequency ray represented in the bottom
panel of Fig.~\ref{fig:schema}. We thus retrieve the same result as
\citet{RogersGlatzmaier2005} in their 2D simulations. In panels (c) and 
(d), we have zoomed in the interface between the convective and radiative
zones in order to measure the angles formed by the wavefronts. Again, it is clear
that in the nonlinear case, plumes excite very low-frequency waves
with wavefronts almost horizontal, as discussed in
Sect. \ref{sec:excit-grav-waves}. For example, we measure $\alpha \approx 80^o$ at $r \approx
0.65R_\odot$, which gives $\omega=0.036$ mHz according to Eq.~\eqref{eq:2}. In
panel (d), however - corresponding to the semi-linear model \textit{sem-lin}
- the St Andrew's crosses are much more pronounced. In the figure shown
here, we can measure three angles $\beta \approx 55^o$, $\gamma \approx
22^o$, and $\delta \approx 45^o$, which correspond to
frequencies 0.15 mHz, 0.19 mHz, and 0.14 mHz respectively. We could explain these
observations by considering the rules governing the wave-wave
interactions. As shown in the interaction diagram (Fig. \ref{fig:triad}) , there are four possible
combinations for two waves to excite a third one \citep[e.g.,][]{ROG:ROG1096}. Wavevectors are
vectorially added, so the two possible horizontal wavevectors are $\vec
k_{h3} = \vec k_{h1}+\vec k_{h2}$ and $\vec k_{h3} = \vec k_{h1}  - \vec
k_{h2}$. The waves with the biggest wavenumber is rapidly damped and only the
one with the smallest wavenumber remains. Since the frequency is linked to the
wavenumber by Eq. \eqref{eq:1}, we understand that nonlinear interactions favor low frequencies.





\section{Waves' amplitude and energy}
\label{sec:amplitudes}
In this paper, we wish to discuss two important questions concerning solar gravity waves: their
precise frequencies and
their amplitudes. In this section, we analyze the energetical
aspects of the spectrum.

\subsection{Energy transfer from convective zone to waves}
\label{sec:energy-transfer-from-1}
\begin{figure}
  \centering
  \includegraphics[width=0.5\textwidth]{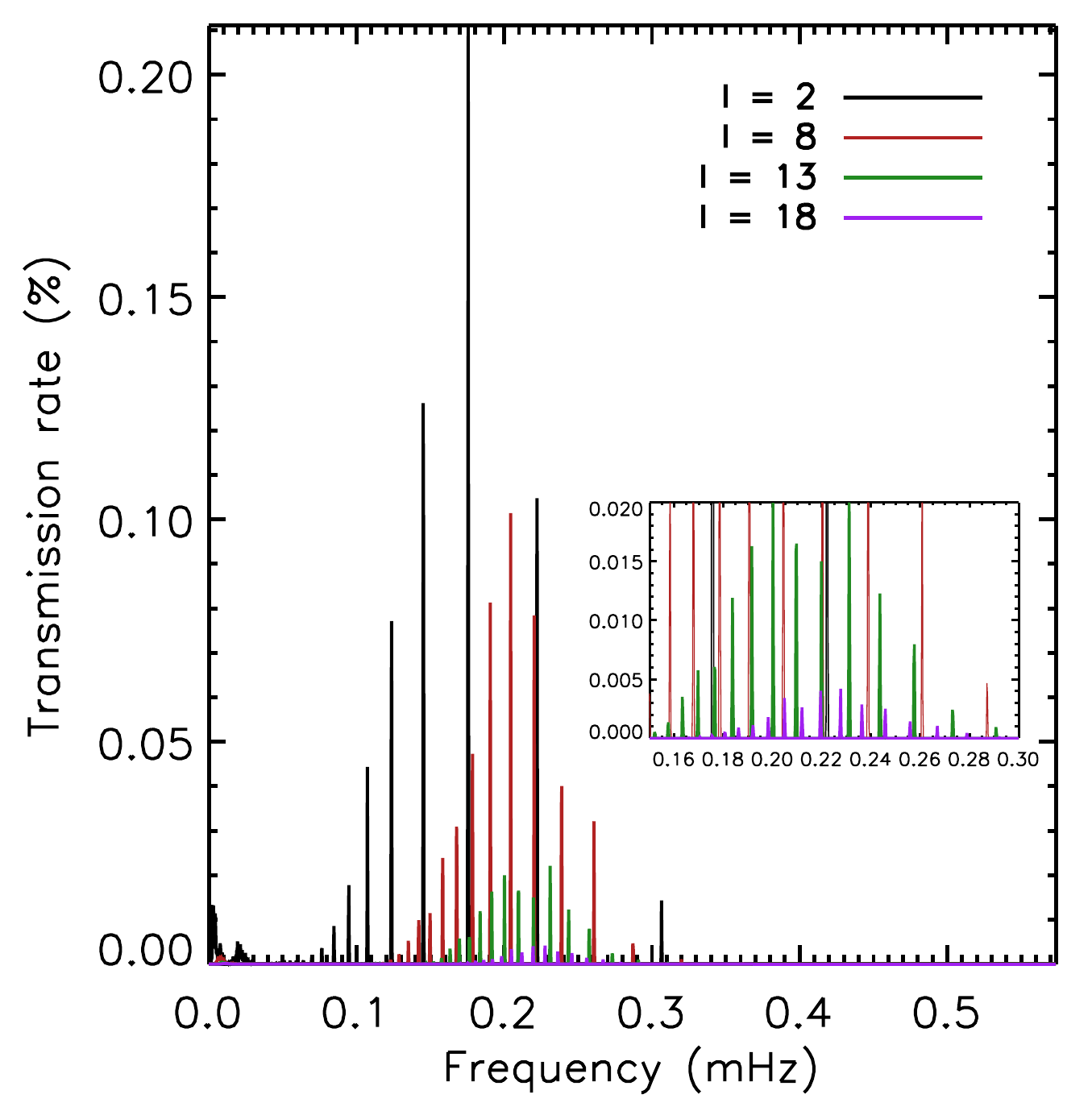}
   \caption{Transmission of energy from convective to radiative zone for
     different values of $\ell$. We retrieve the estimation of 0.1\% given in previous works for the most
    visible peak, {but the transmission rate decreases rapidly when
      increasing $\ell$.}}
  \label{fig:transmission}
\end{figure}

First of all, we have seen in Sec. \ref{sec:penetr-conv} that convective
plumes are slowed down by buoyancy when they enter the radiative region and that
a part of their kinetic energy is converted into gravity waves.
A long series of papers attempted to quantify the excitation of
gravity waves by convective penetration
\citep{Press1981,1991A&A...252..179Z,1992A&A...257..763A,1994SoPh..152..241A,1996SoPh..169..245S,2003A&A...405.1025T}. In
particular, \citet{1994SoPh..152..241A} evaluates the energy density in the
waves at about 0.1\% of the typical kinetic energy density in the
convective zone. We used the same method by comparing the kinetic energy density in
the convection zone at 0.73$R_\odot$ with the energy of gravity waves at
0.6$R_\odot$. For a given value of $\ell$, we thus define a transmission rate $T_\ell$ as
\begin{equation}
  \label{eq:30}
  T_\ell(\omega) = \frac{E(r_0=0.73R_\odot,\ell,\omega)}{E(r_0=0.6R_\odot,\ell,\omega)}\hbox{.}
\end{equation}
We plot $T_\ell$ for $\ell=$ 2, 8, 13 and 18 in the top panel of Fig.~\ref{fig:transmission}. For
$\ell$=2 and $\ell$=8, we find a transmission rate that is close to the one
predicted by \citet{1994SoPh..152..241A}, with a main peak at
0.21\%. The transmission rate is lower for higher values of $\ell$ and becomes
very small for $\ell$=18. These results are unchanged by choosing other
depths than 0.73$R_\odot$ and 0.6$R_\odot$, and by staying above and
below the tachocline. We thus
deduce that the convective kinetic energy is mainly distributed to low
orders $\ell$ and orders $n$ less than ten corresponding to the range
of frequencies $[0.10,0.25]$ mHz. \\\newline

\begin{figure}[h]
  \centering
  \includegraphics[width=0.5\textwidth]{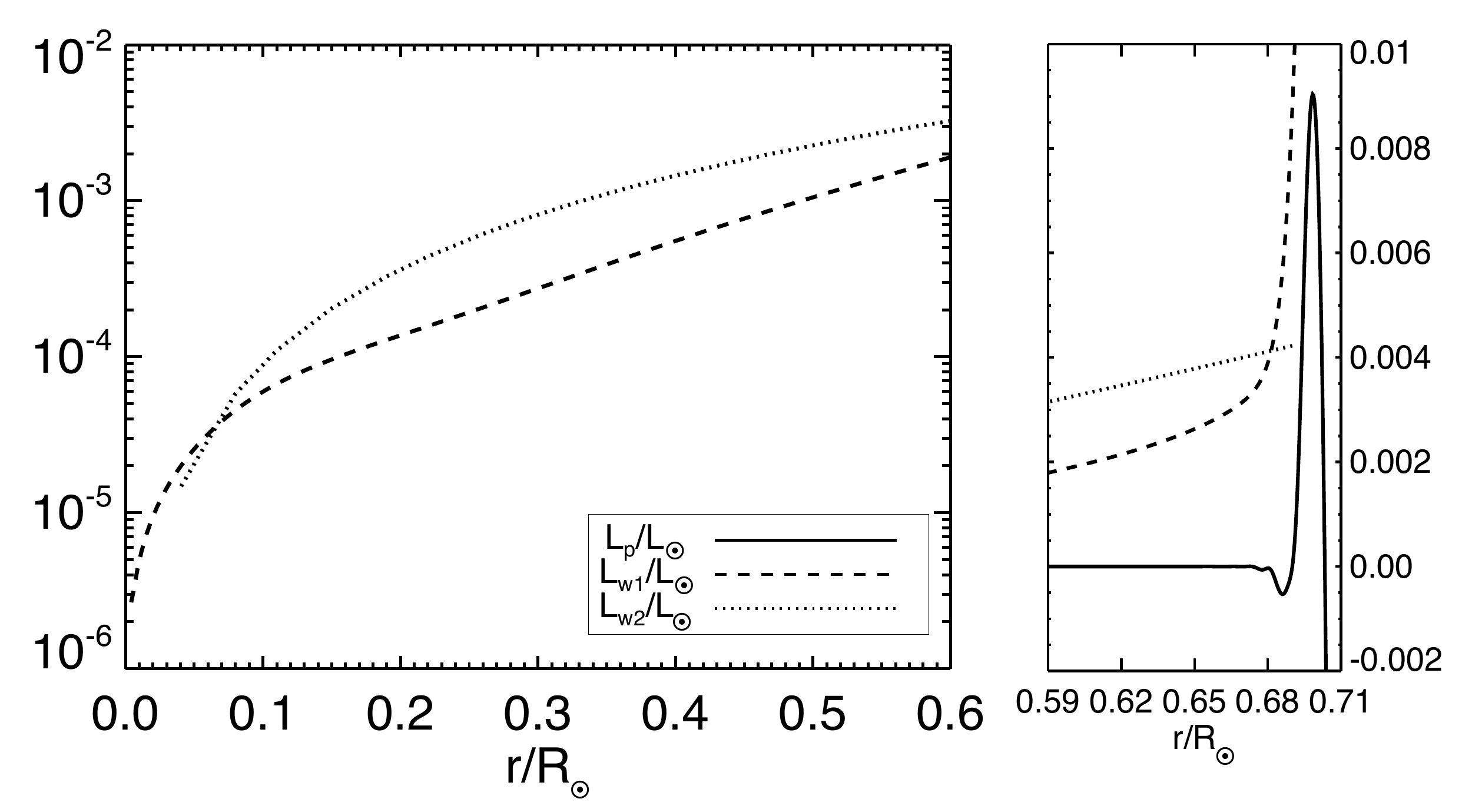}
  \caption{Fraction of the total luminosity converted into
    waves. {At the base of the convecting zone, the luminosity
      carried by waves is about 0.4\% $L_\odot$, which is coherent with the
      amount of energy transmitted from convection to waves (Fig. \ref{fig:transmission}).}}
  \label{fig:fluxwaves}
\end{figure}

But it is also interesting to estimate the energy flux carried by IGWs with
respect to the total luminosity. Several formula have been
used to calculate this flux. The first one is the flux associated with the
pressure fluctuations (acoustic flux) \citep{1978cup..book.....L,Mathis2009},
\begin{equation}
  \label{eq:32}
  \mathcal{F}_p = <\overline{V_r P}> \hbox{.}
\end{equation}
{It is directly linked to the angular momentum flux that
  characterizes the deposit (prograde) or extraction (retrograde) of
  angular momenum by waves \citep{ZahnTalonMatias1997}.}
Then, when $\omega_c \ll N$, one defines the total energy flux
$\mathcal{F}_{W1}$ carried by IGWs by 
\begin{equation}
  \label{eq:31}
 \mathcal{F}_{W1} \propto \frac{\omega_c}{N} \mathcal{F}_c \hbox{,}
\end{equation}
where $\omega_c$ is the convection frequency and $\mathcal{F}_c$ the
convective energy flux
\citep{1994ApJ...424..466G,1991ApJ...377..268G,2003AcA....53..321K}, and $\mathcal{F}_{W1}$
and $\mathcal{F}_{p}$ are theoretically equal. In our case, because the
enthalpy flux is nearly zero in the radiative zone (see flux balance in Fig. \ref{fig:profiles}), we take the maximum value for $\mathcal{F}_c$ in the
convective zone. Moreover, for the model \textit{ref}, we measure $\omega_c \approx $ 15 days, which verifies $\omega_c \ll N$ a posteriori.
Finally, we consider the flux $\mathcal{F}_{W2}$ given by
\citet{ZahnTalonMatias1997,2003AcA....53..321K} as
\begin{equation}
  \label{eq:33}
  \mathcal{F}_{W2} = \int_{\omega_c}^N \int_{k_h} \rho
  \frac{E(k_h,\omega)}{k_h \omega} V_{\mathrm{gr}}(k_h,\omega) dk_h d\omega \hbox{,}
\end{equation}
where$V_{\mathrm{gr}}(k_h,\omega)$ is the group velocity
\begin{equation}
  \label{eq:34}
  V_{\mathrm{gr}}(k_h,\omega) = \frac{\sqrt{N^2-\omega^2}}{N^2}\frac{\omega^2}{\sqrt{\ell(\ell+1})}r \hbox{,}
\end{equation}
and $E(k_h,\omega)$ the quantity defined by Eq.~\eqref{eq:5}.
We plot in Fig.~\ref{fig:fluxwaves} the comparison between those three
fluxes - $\mathcal{F}_p$, $\mathcal{F}_{W1}$ and $\mathcal{F}_{W2}$ -
converted into luminosity - $L_p$, $L_{W1}$, and $L_{W2}$ - and
divided by the solar total luminosity $L_\odot$.
The righthand part of the figure is a zoom in the top region of the radiative
zone. The linear vertical scale shows that the three luminosities are
comparable around $4 \times 10^{-3} L_\odot$ in the region of excitation of
IGWs. However, the acoustic flux drops rapidly and becomes extremely small
for $r<0.68R_\odot$. The lefthand panel of Fig. \ref{fig:fluxwaves} shows the
two remaining fluxes in the whole radiative zone with a logarithmic
vertical scale. We see that $\mathcal{F}_{W1}$ and $\mathcal{F}_{W2}$
remain similar, decreasing from $10^{-3}$ to $10^{-5}$ $L_\odot$. Thus, it seems that we find a consistency between
the percentage of the solar luminosity carried by IGWs at the beginning of
their propagation and the energy transmitted from convection to waves.

\subsection{Spatial radiative damping}
\label{sec:radiative-damping}
Gravity waves are damped during their propagation. According to
\citet{ZahnTalonMatias1997}, the amplitude of a gravity wave propagating in a non adiabatic
medium is damped by a factor $e^{-\tau/2}$ where
\begin{equation}
  \label{eq:25}
  \tau\left(r,\ell,\omega\right) =
  \left[\ell(\ell+1)\right]^{\frac{3}{2}}\int_{r}^{r_{\rm
      CZ}}\kappa\frac{N^3}{\omega^4}\frac{\mathrm dr'}{r'^3} \hbox{.}
\end{equation}
Here we take $r_{CZ} = 0.69R_\odot$, the radius
where waves are excited (i.e., at the interface between convective and radiative
zones). That formula was obtained in the linear regime and under the
assumption $\omega \ll N$. To compare this prediction with our simulation,
we take $E_0(\omega)=E(r=0.69R_\odot,\ell =2,\omega)$ as a starting amplitude. We then
look at the evolution with depth of this initial amplitude by taking
\begin{equation}
  \label{eq:26}
   E_{\mathrm{damp}}(r,\omega) = E_0(\omega) \times e^{-\tau(r,2,\omega)}\hbox{.}
\end{equation}

\begin{figure}
  \begin{center}
  \includegraphics[width=0.45\textwidth]{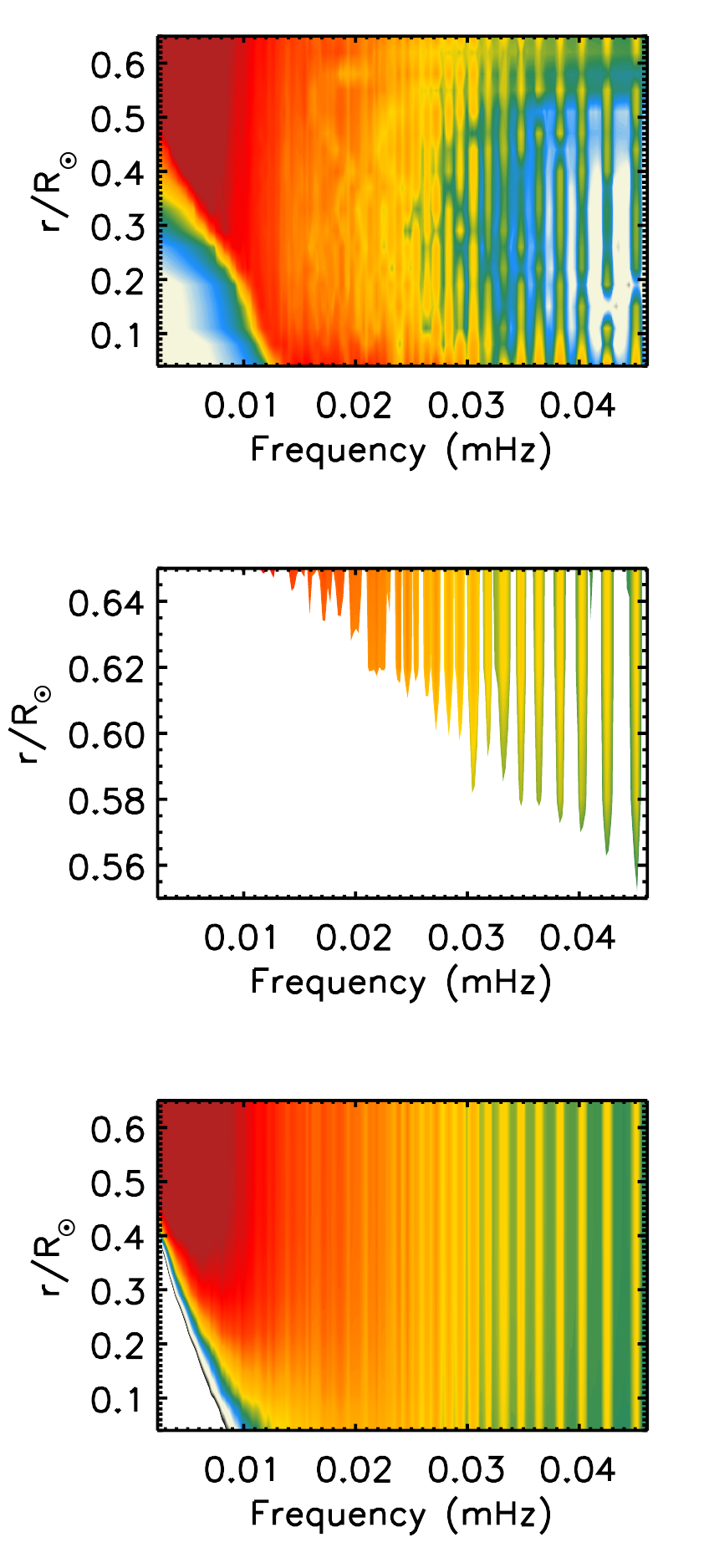}
  \caption{Comparison of simulated data with the theoretical radiative
    damping. \textbf{Top}: Spectrum obtained in the simulation (model
    \textit{ref}, $\ell$=2). \textbf{Middle}: Effect of a radiative
  damping with the theoretical coefficient $1/\omega^4$. The initial
  amplitude is the same as in the upper panel. \textbf{Bottom}: Effect of radiative
  damping with $1/\omega^3$.}
  \label{fig:damping}
\end{center}
\end{figure}

\begin{figure}[h]
  \centering
  \includegraphics[width=0.45\textwidth]{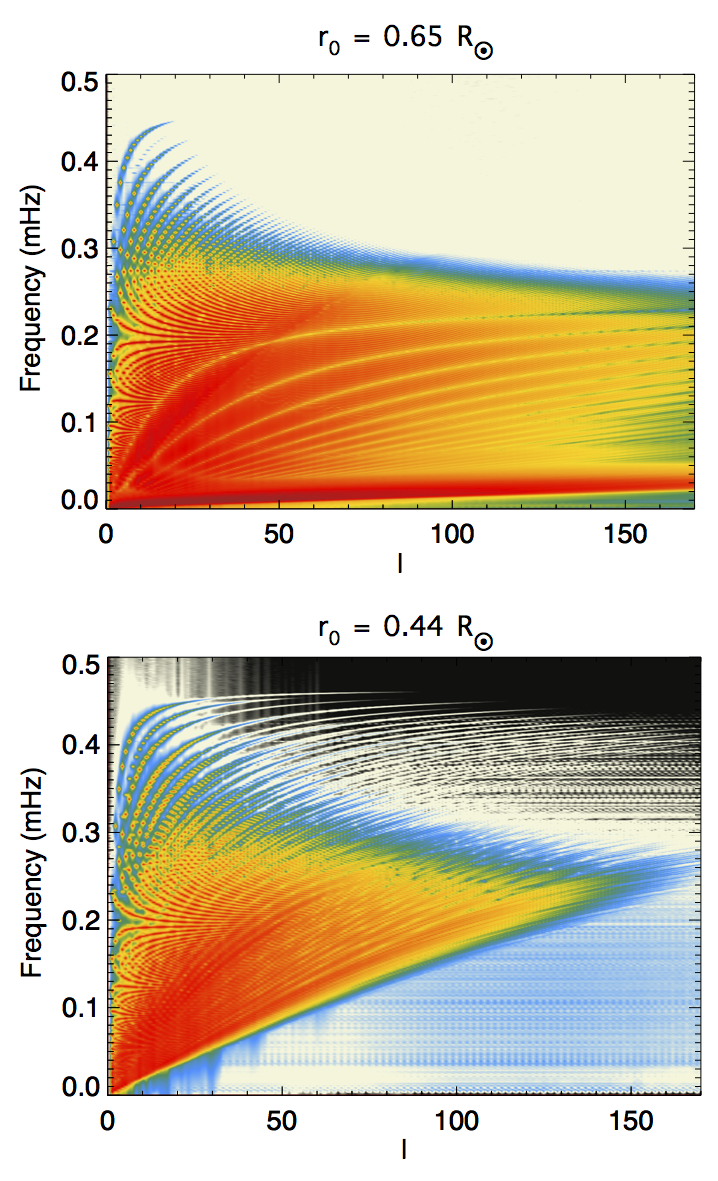} 
  \caption{Spectra obtained with a 40-day sequence of model
    \textit{sem-lin} at $r_0=0.65R_\odot$ (top panel) and $r_0=0.44R_\odot$
    (bottom panel). {Although the horizontal $N_\theta=512$ resolution of this
      model allows reaching $\ell_\mathrm{max}=340$ (see Eq. \eqref{eq:4}), we
      have cut the horizontal axis to $\ell=170$ to focus on the difference
      between both spectra. At low frequency, the waves visible at
      $r_0=0.65R_\odot$ have been totally damped at $r_0=0.44R_\odot$.}}
  \label{fig:nonlin_spectra}
\end{figure}

We represent $E_{\mathrm{damp}}$ as a function of $r/R_\odot$ and $\omega$ in Fig
\ref{fig:damping}. The top panel shows the spectrum obtained by ASH. We can
understand this figure as Fig. \ref{fig:wave_fun2} seen from 
above (top panel). The difference is the narrow horizontal range, because the maximum
frequency here is $0.045$ mHz $= N_{\mathrm{max}}/10$ to respect the hypothesis
$\omega \ll N$. The vertical lines correspond to eigenfunctions,
particularly visible on the right, and when frequency decreases, the modes become
closer and closer. This top panel is taken as a reference in this
discussion. The middle panel represents $ E_{damp}(r,\omega)$ with $\tau$ 
calculated with Eq. \eqref{eq:25}. The profiles of $\kappa$, $\nu$, and $N$
are those presented in Sect. \ref{sec:numerical-model}. We observe that the
amplitude drops much faster than in the top panel. In the bottom panel,
however, we have calculated the damping rate by replacing $\omega^4$ in
Eq.~\eqref{eq:25} by $\omega^3$. The attenuation obtained is much more in
accordance with the one predicted by the simulation. The same behavior has been observed by \citet{Rogers:2013ui} in their 2D
nonlinear simulations. We suspect that the nonlinear interactions between waves explain this difference between simulations and theory. Of course, we
do not attempt here to redefine the formulation of the damping rate, just giving a simple trend with $\omega$.\\\newline
To test this hypothesis, we use the model \textit{sem-lin} described in
Sect. \ref{sec:non-line-inter-1}. The wave spectra obtained are
represented in Fig. \ref{fig:nonlin_spectra} for two depths,
$r_0=0.65R_\odot$(top) and $r_0=0.44R_\odot$(bottom). The detailed study of these results
and the characterization of nonlinear interactions will be the object of a
forthcoming paper. Here, the point of interest lies in the fact that the
amount of energy visible in the top spectra at low frequency has totally disappeared in the bottom
one. By measuring the damping rate with the same method as in
Fig. \ref{fig:damping}, we obtain good agreement with a coefficient
${1}/{\omega^{3.6}}$ in Eq. \eqref{eq:25} instead of ${1}/{\omega^{4}}$. Although we do not retrieve the predicted behaviour exactly, we are clearly closer
than with the fully nonlinear model. The conclusion here is that wave-wave
and/or wave-fluid nonlinear interactions play a very important role, which
seems to be weighed against the linear radiative damping. Since thermal
damping is one of the processes (with for instance corotation resonances and wave breaking) responsible for the angular momentum
transport by IGWs in stars, this result has to be considered in, for example, stellar evolution codes that modeled this
transport \citep{Charbonnel:2013df,2013A&A...558A..11M}.

\subsection{Sensitivity to physical parameters}
\label{sec:sens-phys-param}

\begin{figure}
  \centering
  \includegraphics[width=0.5\textwidth]{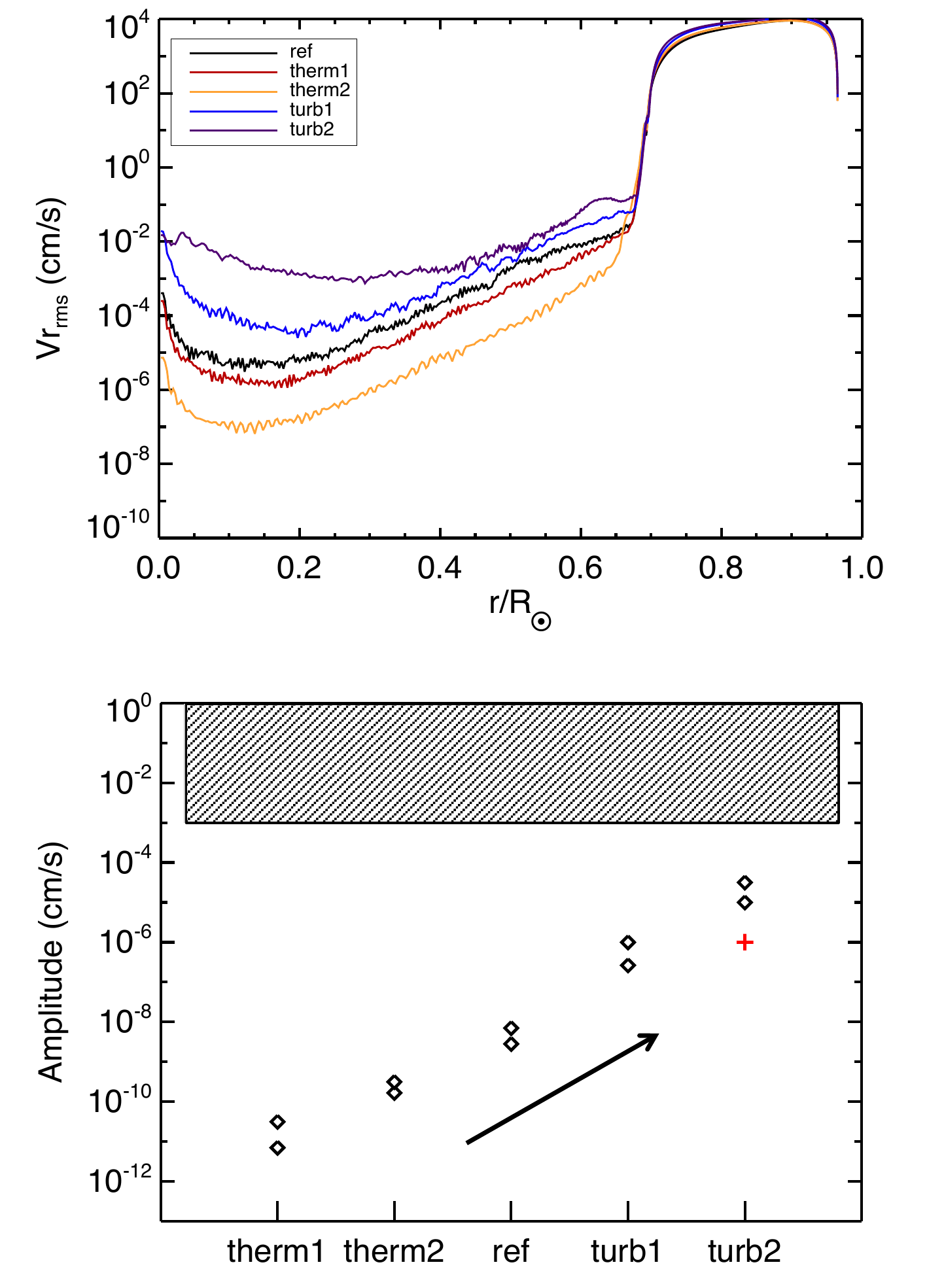}
  \caption{\textbf{Top}: Radial rms velocity as a function of normalized radius for five
    different models whose characteristics are summarized in
    Tab~\ref{tab:param_models}. \textbf{Bottom}: Amplitude of the most powerful peak in each
    model. We find that some waves in \textit{turb2} have sufficient
    amplitudes to go through the convective zone and emerge on the surface
    of the star (top boundary at $0.97R_\odot$). The hatched zone
    represents the values of surface solar g-modes predicted in the literature.}
  \label{fig:comp_models}
\end{figure}

The last remaining question concerns the amplitude of g-modes in the
radiative zone, but also at the surface of the Sun. Thus, we finish this
paper by comparing the different models introduced in Sect. \ref{sec:model}
and the corresponding IGWs' amplitudes. In the top panel of
Fig. \ref{fig:comp_models}, we show the rms radial velocities as a function
of the normalized radius for each model. We clearly
see the drop of velocity at the interface between radiative and
convective zones, around $0.7R_\odot$. Moreover, as expected by the choice
of the diffusivity coefficients and by the observation of the overshoot
region (see Sect. \ref{sec:penetr-conv}), \textit{turb2} (violet) is the model with
the highest rms velocity in the radiative zone. Then comes \textit{turb1}
(blue), \textit{ref} (black), \textit{therm2} (red), and \textit{therm1}
(orange). The order of these curves is directly related to the values of
the thermal diffusivities $\kappa$ (see Fig. \ref{fig:profiles}).\\\newline
For each model, we have calculated a spectrum at $r_0=0.66R_\odot$ and reported in the bottom
panel of Fig.~\ref{fig:comp_models} the amplitudes in
$cm/s$ of the highest and lowest peaks (black diamonds) of modes ($\ell$=1,2,3
and $n$=1-10). These modes are the ones identified in
Sect. \ref{sec:energy-transfer-from-1} as the most excited by the
convection. They are thus the best candidates for a possible detection at
the surface of the Sun. The amplitudes shown in black diamonds are the
excitation amplitudes. In fact, since the background noise level increases
in the convective zone (granulation), we are able to detect g-modes at $r_{\mathrm{top}}=0.97R_\odot$
in model \textit{turb2} only. The amplitude of the most visible peak at the
surface is indicated by the red cross in Fig. \ref{fig:comp_models}. At least, the
hatched zone in the top of the figure points to the values of surface solar
g-modes $\ell$=1 predicted in the literature \citep{2010A&ARv..18..197A}. The
optimistic and the pessimistic values differ by three orders of magnitude:
from 1 $cm/s$ 1 to $10^{-3}$ $cm/s$. We see that for all our models, the
amplitudes of the measured waves are much smaller, and we do not even
consider the atmosphere of the Sun. Nevertheless, the increasing tendency gives an
positive perspective since it indicates that the more we increase the
turbulence, the more g-modes are powerful. Thus, a possible way to reach
realistic amplitudes could be to model more and more turbulent
convective zones and to lower the thermal diffusivity $\kappa$ in the
radiative zone. 

\section{Conclusion}

In this paper, we have presented the first study of IGWs stochastic excitation and
propagation in a 3D spherical Sun using a realistic stratification in the
radiative zone and a nonlinear coupling between radiative and convective
zones. {This configuration allows a direct comparison with
  seismic studies}. These results are extremely rich, and we stand yet at the beginning
of their exploration and comprehension. \\\newline
Since \citet{BrunZahn2006}, the ASH code has entered a new area because it
is no longer dedicated to the study of convective envelopes alone
\citep{2000ApJ...533..546E,BrunToomre2002}. The nonlinear coupling with the
inner radiative zone opens up a large field of investigation. We
presented two recent improvements in the ASH code that have a strong impact
on our study of gravity waves. 
\begin{itemize}
\item On one hand, the implementation of the LBR equations \citep{Brown:2012bd}
ensures the right conservation of energy in the radiative zone and allows
IGWs' frequencies and amplitudes to be computec with a better accuracy. 
\item On the
other hand, the extension of the computational domain to $r=0$ by imposing
special boundary conditions \citep{2007PhRvE..75b6303B} largely improves
the treatment of g-modes since we now model the entire radiative cavity
without any central cutoff \citep{Brun:2011bl,2012sf2a.conf..289A}.
\end{itemize}
We then discussed the convective
overshoot observed in our models and related this process to the
excitation of a large spectrum of IGWs, in agreement with both fluid mechanics
and stellar oscillations theory predictions. This spectrum extends from zero to
the maximum of the BV frequency ($\sim$0.45mHz), which implies that both
propagative (low-frequency) and standing waves (high-frequency) must be
represented. Using our raytracing code
\citep[e.g.,][]{goughHouches,Christensen-Dalsgaard97lecturenotes} also
contributes to improving our understanding and illustrates the behavior of IGWs as propagative waves,
their group and phase velocity, and their location in the 3D sphere. This underlines the
complementarity between our simulations of the Sun and
linear and asymptotic theories and models. \\
\newline
The properties of the spectrum of IGWs presented in
this paper are multiple. To understand its structure, we
decomposed it into its spatial and temporal parts, and retrieved the results
of \citet{Belkacem:2009wl} predicting that the frequency spectrum was
better fitted with a Lorentzian-like function rather than with a Gaussian
function. We also showed the quick drop of energy with increasing
wavenumbers $k_h$. Then, we presented the changes in this spectrum
as a function of the depth and proposed a distinction between propagative
waves and g-modes. Indeed, this subject is rather hazy in the literature,
and it is sometimes difficult to place the limit between both
types. Although they correspond to the same physical process, only g-mode frequencies are
described by integers $n$. We then discussed some
important properties relative to g-modes. 
\begin{itemize}
\item We applied the same method as
\citet{Garcia:2007iq} to detect g-modes signatures at the surface of the
Sun and confirmed that the stratification chosen in the model plays an
important role in the calculation of g-modes frequencies.
\item We also had a
look at the impact of the rotation on g-modes, whose frequencies were splitted
with respect to their azimuthal number $m$. We showed that the precision of
the inversion process strongly depends on the radial order of the modes
that are considered and that one must take at least up to $n$=25 to get a precision of 
5\% in the estimation of the rotation rate.
\item Finally, we explained that the energy is not equally distributed into values of $m$
  but is instead distributed in high $m$. That shows that the assumption
  made in several codes, supposing an equal distribution of the energy must be
  treated with caution. Moreover, since high $m$ modes are located close to
  the equator, these results could orient the research of g-modes at
  the surface of the Sun. {This last result, in particular, could not have
  been obtained without taking the three dimensions of the problem into account.}
\end{itemize}
Finally, we dealt with the energy transferred from the convection to IGWS and
then carried by them. 
\begin{itemize}
\item We showed that the different formula supplied by the
literature to estimate this energy give a comparable estimation of the
percentage of the solar luminosity carried by waves. Indeed, we found that
about 0.4\% of the solar luminosity is converted into waves at the
interface between radiative and convective zones.
\item We pointed out that the radiative damping predicted by the linear theory is
  much stronger than the one observed and partly explained this difference by
  considering the impact of the nonlinear processes.
\item Concerning, finally, the amplitude of g-modes that could be detected at
  the surface of the Sun, we are not yet able to reach the required domain
  of parameters but we showed a promising trend toward a good
  estimation of these amplitudes.
\end{itemize}

Our results are of interest for several astrophysical
applications. The part concerning g-modes is directly related to
helioseimology. The asteroseismology community can {be concerned
  by a better understanding of the waves and} also seeing
that other types of stars can be simulated by the ASH code. Concerning low-frequency propagating IGWs, our work provides new
information about the radiative damping and the related effect of
nonlinearities to be considered. The spectra presented {and
  the radiative damping found} can be
implemented in stellar evolution codes to provide a more realistic
repartition of energy, especially concerning the distribution accross $m$
components.\\\newline

Finally, some perspectives of this work are identifiable and will be the
object of future works. We presented in
Sect. \ref{sec:rotational-splitting} our first results concerning the
effect of the rotation on IGWs. Following
\citet{DintransRieutord2000}, \citet{Ballot:2010jy}, and \citet{Rogers:2013ui}, it could be possible to
study the behavior of IGWs in rapidly rotating stars \citep{Mathis:2013wv} and the transport of
angular momentum by gravito-inertial waves
\citep{Mathis:2008ba,Mathis2009}. Also of great interest could be the
addition of a magnetic field in the simulations to characterize its impact
on IGWs
\citep{1992ApJ...395..307G,2010MNRAS.401..191R,MathisDeBrye2011,Mathis:2012tn}. { Indeed, the presence of a magnetic field will modify the dispersion relation. If
  its amplitude is high enough, we can anticipate that a large-scale magnetic field trapped in the radiative 
  zone will have a significant impact on the propagation of IGWs, such as wave reflexions, filtering, and frequency shift. Particularly, for waves frequencies close
  to the Alfven frequency, IGWs will be trapped vertically, while for frequencies below the inertial frequency ($2\Omega$) some equatorial trapping
  will occur. Moreover, we could expect that a time-dependent magnetic field generated by dynamo action would modulate the waves' signal.}\\\newline

{This work thus constitutes a first cornerstone where the completementary
use of 3D nonlinear simulations and of asymptotic theories allows
bringing the study of the excitation, propagation, and damping of gravity waves
in stellar interiors to a new level of understanding. Morever, the
potential application to other types of rotating and possibly magnetic
stars open a new window in theoretical asteroseismology in the whole HR
diagram.}

\section*{Acknowledgments}
{ {We thank the referee G. Glatzmaier for his remarks and suggestions that improved the original manuscript.} We thank
  J. Christensen-Dalsgaard for inviting L.A. in Aarhus, for letting us use the latest version of the ADIPLS 
  code and for useful discussions about the physics of waves. {We are especially grateful to N. Featherstone for his dedication to optimizing ASH
    and releasing the 2.0 version, and for his help in implementing the regularity conditions at $r=0$.} We also thank R. Garcia for much advice about data analysis,
  K. Augustson , B. Brown, and M. Miesch for discussion relative to the ASH code and its treatment of internal waves, and
  A. Strugarek for regular discussions regarding whole Sun models. We are grateful to
  B. Hindman, M. Lebars, T. Rogers, R. Samadi, J. Toomre and J.-P. Zahn for fruitful discussions during the
  preparation of this paper. We acknowledge funding by the ERC grant
  STARS2 207430 (www.stars2.eu), by the CNES for the Golf/SoHO
  instrument, IRSES, SPACEINN program, and CNRS Physique théorique et
  ses interfaces program. The simulations were performed using HPC resources of GENCI 1623 and PRACE
1069 projects.}

\section*{Appendix A: Inner boundary conditions}

{The authors are keen to thank N. Featherstone for the time spent with A.S. Brun to develop and test the full sphere version of ASH (as
  explained in this appendix), which
  allowed a more precise analysis of the gravity wave's properties in the whole radiative interior.}
We here follow the method proposed by \citet{BaylissPHD} to obtain the inner boundary conditions for ${Z}$ and ${W}$.
For the sake of clarity in the following equations, we introduce the notation $\vec{X}=\vec{\bar\rho\vec{\mathrm{v}}}$ and do not write the subscripts
$\ell,m$ if there is another subscript.\\\newline
Starting from the poloidal-toroidal decomposition of $\vec{X}=\vec{\bar\rho\vec{\mathrm{v}}}$ (Eq. \eqref{eq:29}), we project ${Z}$ and ${W}$ on the
spherical harmonics basis
\begin{equation}
  \label{eq:280}
  \left\{
      \begin{array}{ll}
   {Z}(r,\theta,\varphi)&=\displaystyle\sum_{\ell m}{Z}_{\ell m}(r)\mathrm{Y}_{\ell m}(\theta,\varphi) \hbox{,}\\
  {W}(r,\theta,\varphi)&=\displaystyle\sum_{\ell m}{W}_{\ell m}(r)\mathrm{Y}_{\ell m}(\theta,\varphi) \hbox{,}
    \end{array}
    \right.
\end{equation}
which leads to
\begin{eqnarray}
  \label{eq:290}
  \vec{X}&=&\sum_{\ell m}\left(\underbrace{\frac{l(l+1)}{r^2}{W}_{\ell m}\mathrm{Y}_{\ell m}}_{{{X_{{W}r}}}}\right)\vec{e}_r\\\nonumber
 &+&\sum_{\ell m}\left(\underbrace{\frac{1}{r}\partial_r({W}_{\ell m})\partial_\theta(\mathrm{Y}_{\ell m})}_{{{X_{{W}\theta}}}}+\underbrace{\frac{1}{r\sin\theta}{Z}_{\ell m}\partial_\varphi(\mathrm{Y}_{\ell m})}_{{{X_{{Z}\theta}}}}\right)\vec{e}_\theta\\
 &+& \sum_{\ell m}\left(\underbrace{\frac{1}{r\sin\theta}\partial_r({W}_{\ell m})\partial_\varphi(\mathrm{Y}_{\ell m})}_{{{X_{{W}\varphi}}}}-\underbrace{\frac{1}{r}{Z}_{\ell m}\partial_\theta(\mathrm{Y}_{\ell m})}_{{{-X_{{Z}\varphi}}}}\right)\vec{e}_\varphi \hbox{.}\nonumber 
\end{eqnarray}
We then expand ${Z}$ and ${W}$ in Taylor series in the vicinity of $r=0$

\begin{equation}
  \label{eq:300}
  \left\{
      \begin{array}{ll}
  {W}_{\ell m}(r)  &={W}_0+r{W}'_0+\displaystyle\frac{r^2}{2!}{W}''_0+\displaystyle\frac{r^3}{3!}{W}^{(3)}_0+\displaystyle\frac{r^4}{4!}{W}^{(4)}_0 \hbox{,}\\
{Z}_{\ell m}(r) &={Z}_0+r{Z}'_0+\displaystyle\frac{r^2}{2!}{Z}''_0+\displaystyle\frac{r^3}{3!}{Z}^{(3)}_0+\displaystyle\frac{r^4}{4!}{Z}^{(4)}_0 \hbox{,}
    \end{array}
    \right.
\end{equation}
where ${W}_0 \equiv {W}_{\ell m}\lvert_{r=0}$, ${W}'_0\equiv\displaystyle\frac{\partial{W}_{\ell m}}{\partial r}\biggr\lvert_{r=0}$, ${W}''_0
\equiv\displaystyle\frac{\partial^2{W}_{\ell m}}{\partial r^2}\biggr\lvert_{r=0}$... \\\newline
Replacing these developments in Eq.~\eqref{eq:290}, we obtain the expression of $\vec{X}$ as a function of the derivatives of
  ${W}_{\ell m}$ and ${Z}_{\ell m}$ at $r=0$. \\\newline
The central boundary conditions come from the fact that no cartesian component can depend on the angles $\theta$ and $\varphi$. In this case, the vector $\vec{X}$ would be
multi valued at the origin. Thus, we write the cartesian components of $\vec{X}$ as a function of its spherical components

\begin{equation}
  \label{eq:3800}
  \left\{
      \begin{array}{ll}
  {X}_x &=X_r\sin\theta\cos\varphi + X_\theta\cos\theta\cos\varphi -X_\varphi\sin\varphi \hbox{,}\\
  {X}_y &=X_r\sin\theta\sin\varphi + X_\theta\cos\theta\sin\varphi +X_\varphi\cos\varphi \hbox{,}\\
  {X}_z&=X_r\cos\theta -X_\theta\sin\theta \hbox{,}
    \end{array}
    \right.
\end{equation}
where $X_r=X_{{W}r}$, ${{X_\theta}}={{X_{{W}\theta}}}+{{X_{{Z}\theta}}}$ and
$X_\varphi={{X_{{W}\varphi}}}+{{X_{{Z}\varphi}}}$, and then
rewrite ${X}_x$, ${X}_y$, and ${X}_z$ by replacing ${W}_{\ell m}$ and ${Z}_{\ell m}$ with their Taylor developments. Terms of first
order and above in $r$ will be zero as $r \to 0$, but other terms that depend on $\theta$ and $\varphi$ must be set to zero by the choice of the values of ${W}_0$,
${W}'_0$, ${W}''_0$, ${Z}_0$, ${Z}'_0$, and ${Z}''_0$. \\\newline

For $\ell=1$ and $m=0$, we obtain 
\begin{eqnarray*}
  \label{eq:37}
  X_x(r,\theta,\varphi)\biggr\lvert_{1,0}&=&-\frac{1}{2}\sqrt{\frac{3}{\pi}}\sin\theta\sin\varphi{Z}'_{0}\\
                                 &&-\sqrt{\frac{3}{\pi}}\frac{\sin\theta}{2r}\left(-\cos\theta\cos\varphi{W}'_0+\sin\varphi{Z}_0\right)\\
                                 &&+\sqrt{\frac{3}{\pi}}\frac{1}{r^2}\cos\theta\sin\theta\cos\varphi{W}_0\\
                                 &&-\frac{r\sin\theta}{4\sqrt{3\pi}}\left(\cos\theta\cos\varphi{W}^{(3)}_0+3\sin\varphi{Z}''_0\right)\\
                                 &&-\frac{r^2\sin\theta}{8\sqrt{3\pi}}\left(\cos\theta\cos\varphi{W}^{(4)}_0+2\sin\varphi{Z}^{(3)}_0\right)\\
                                 &&-\frac{r^3}{16\sqrt{3\pi}}\sin\theta\sin\varphi{Z}^{(4)}_0 \hbox{.}
\end{eqnarray*}

\begin{itemize}
\item The first term on the righthand side is constant with respect to $r$, but since $X_x$ cannot depend on $\theta$ and $\varphi$, we must impose
  ${Z}'_{0}=0$.
\item The two following terms diverge when $r \to 0$, so we must impose ${W}'_0={Z}_0=0$ and ${W}_0=0$.
\item The remaining terms tend to 0 with $r$.
\end{itemize}

Then, 
\begin{eqnarray*}
  \label{eq:7}
    X_y(r,\theta,\varphi)\biggr\lvert_{1,0}&=&\frac{1}{2}\sqrt{\frac{3}{\pi}}\sin\theta\cos\varphi{Z}'_{0}\\
                                   &&+\sqrt{\frac{3}{\pi}}\frac{\sin\theta}{2r}\left(\cos\varphi{Z}_0+\cos\theta\sin\varphi{W}'_0\right)\\
                                   &&+\sqrt{\frac{3}{\pi}}\frac{1}{r^2}\cos\theta\sin\theta\sin\varphi{W}_0\\
                                   &&+\frac{r\sin\theta}{4\sqrt{3\pi}}\left(3\cos\varphi{Z}^{(2)}_0-\cos\theta\sin\varphi{W}^{(3)}_0\right)\\
                                   &&+\frac{r^2\sin\theta}{8\sqrt{3\pi}}\left(2\cos\varphi{Z}^{(3)}_0-\cos\theta\sin\varphi{W}^{(4)}_0\right)\\
                                   &&+\frac{r^3}{16\sqrt{3\pi}}\sin\theta\cos\varphi{Z}^{(4)}_0 \hbox{.}
\end{eqnarray*}
The conditions to impose are ${Z}'_{0}=0$ for the constant term, ${Z}_{0}={W}'_{0}=0$ for the term varying in $1/r$ and
${W}_{0}=0$ for the one varying in $1/r^2$. Finally, 
\begin{eqnarray*}
  \label{eq:8}
    X_z(r,\theta,\varphi)\biggr\lvert_{1,0}&=&\frac{1}{2}\sqrt{\frac{3}{\pi}}{W}''_{0}\\
                                   &&+\sqrt{\frac{3}{\pi}}\frac{1}{4r}\left(3+\cos 2\theta\right){W}'_0\\
                                   &&+\sqrt{\frac{3}{\pi}}\frac{1}{2r^2}\left(1+\cos 2\theta\right){W}_0\\
                                   &&+\frac{r}{8\sqrt{3\pi}}\left(5-\cos 2\theta\right){W}^{(3)}_0\\
                                   &&+\frac{r^2}{16\sqrt{3\pi}}\left(3-\cos 2\theta\right){W}^{(4)}_0 \hbox{.}\\
\end{eqnarray*}
This time, the constant term does not vary with $\theta$ or $\varphi$, so there is no need to nullify it. The divergent terms impose the conditions
${W}_{0}=0$ and ${W}'_{0}=0$.\\
We obtain the conditions for $\ell \ne 1$ by applying the same calculation to, for example, $\ell=2$ and $m=0$:

\begin{eqnarray*}
  \label{eq:2800}
    X_x(r,\theta,\varphi) \biggr\lvert_{2,0}&=&-\frac{3\sin\theta}{4}\sqrt{\frac{5}{\pi}}\left(\sin^2\theta\cos\varphi{W}''_0+2\cos\theta\sin\varphi{Z}'_0\right)\\
                                 &&-\frac{3\sin\theta}{2r}\sqrt{\frac{5}{\pi}}\left(-\cos2\theta\cos\varphi{W}'_0+\cos\theta\sin\varphi{Z}_0\right)\\
                                 &&-\frac{3}{2r^2}\sqrt{\frac{5}{\pi}}\sin\theta\cos\varphi{W}_0\left(1-3\cos^2\theta\right)\\
                                 &&-\frac{r\sin\theta}{4}\sqrt{\frac{5}{\pi}}\left(\cos\varphi{W}^{(3)}_0+3\cos\theta\sin\varphi{Z}''_0\right)\\
                                 &&-\frac{r^2\sin\theta}{16}\sqrt{\frac{5}{\pi}}\left((1+\cos^2\theta)\cos\varphi{W}^{(4)}_0\right.\\
                                 &&\hbox{\hspace{4cm}}+\left.4\cos\theta\sin\varphi{Z}^{(3)}_0\right)\\
                                 &&-\frac{r^3}{16}\sqrt{\frac{5}{\pi}}\cos\theta\sin\theta\sin\varphi{Z}^{(4)}_0\hbox{,}
\end{eqnarray*}

\begin{eqnarray*}
  \label{eq:2900}
    X_y(r,\theta,\varphi)\biggr\lvert_{2,0}&=&\frac{3\sin\theta}{4}\sqrt{\frac{5}{\pi}}\left(2\cos\theta\cos\varphi{Z}'_0+\sin^2\theta\sin\varphi{W}''_0\right)\\
                                 &&+\frac{3\sin\theta}{2r}\sqrt{\frac{5}{\pi}}\left(-\sin^2\theta\sin\varphi{W}'_0+\cos\theta\cos\varphi{Z}_0\right)\\
                                 &&-\frac{3\sin\theta}{2r^2}\sqrt{\frac{5}{\pi}}\left(1-3\cos^2\theta\right)\sin\varphi{W}_0\\
                                 &&+\frac{r\sin\theta}{4}\sqrt{\frac{5}{\pi}}\left(3\cos\theta\cos\varphi{Z}''_0-\sin\varphi{W}^{(3)}_0\right)\\
                                 &&-\frac{r^2\sin\theta}{16}\sqrt{\frac{5}{\pi}}\left((1+\cos^2\theta)\sin\varphi{W}^{(4)}_0\right.\\
                                 &&\hbox{\hspace{4cm}}-\left.4\cos\theta\cos\varphi{Z}^{(3)}_0\right)\\
                                 &&+\frac{r^3}{16}\sqrt{\frac{5}{\pi}}\cos\theta\sin\theta\cos\varphi{Z}^{(4)}_0\hbox{,}
\end{eqnarray*}

and

\begin{eqnarray*}
  \label{eq:3000}
    X_z(r,\theta,\varphi) \biggr\lvert_{2,0}&=&\frac{3\cos\theta}{8}\sqrt{\frac{5}{\pi}}\left(3+\cos 2\theta\right) {W}''_{0}\\
                                   &&+\frac{3\cos\theta}{2r}\sqrt{\frac{5}{\pi}}\left(1+\cos 2\theta\right) {W}'_{0}\\
                                   &&+\frac{3\cos\theta}{4r^2}\sqrt{\frac{5}{\pi}}\left(1+3\cos 2\theta\right) {W}_{0}\\
                                   &&+\frac{r\cos\theta}{2}\sqrt{\frac{5}{\pi}}{W}^{(3)}_{0}\\
                                   &&+\frac{r^2\cos\theta}{32}\sqrt{\frac{5}{\pi}}\left(5-\cos 2\theta\right){W}^{(4)}_{0}\hbox{.}
\end{eqnarray*}

Finally, the conditions to impose at
$r=0$ are
\begin{itemize}
\item ${Z}_0={Z}'_{0}=0$ and ${W}_0={W}'_0=0$ for $\ell=1$,
\item ${Z}_0={Z}'_{0}=0$ and ${W}_0={W}'_0={W}''_0=0$ for $\ell \ne 1$.
\end{itemize}
Considering the order of the equations verified by $\mathrm{Z_{\ell m}}$ and ${W}_{\ell m}$, we can impose only one condition for
$\mathrm{Z_{\ell m}}$ and two for $\mathrm{W_{\ell m}}$ at each limit of the domain. Thus, we have made the choice to distinguish between
$\ell = 1$ and $\ell \ne 1$ by imposing (set 1):
\begin{itemize}
\item ${Z}_0=0$ for all $\ell$,
\item ${W}_0={W}'_0=0$ for $\ell=1$,
\item ${W}_0={W}''_0=0$ for $\ell \ne 1$,
\end{itemize}
but another possible set is (set 2):
\begin{itemize}
\item ${Z}_0=0$ for all $\ell$,
\item ${W}_0={W}'_0=0$ for all $\ell$.
\end{itemize}

\begin{figure*}[]
  \centering
  \includegraphics[width=0.95\textwidth]{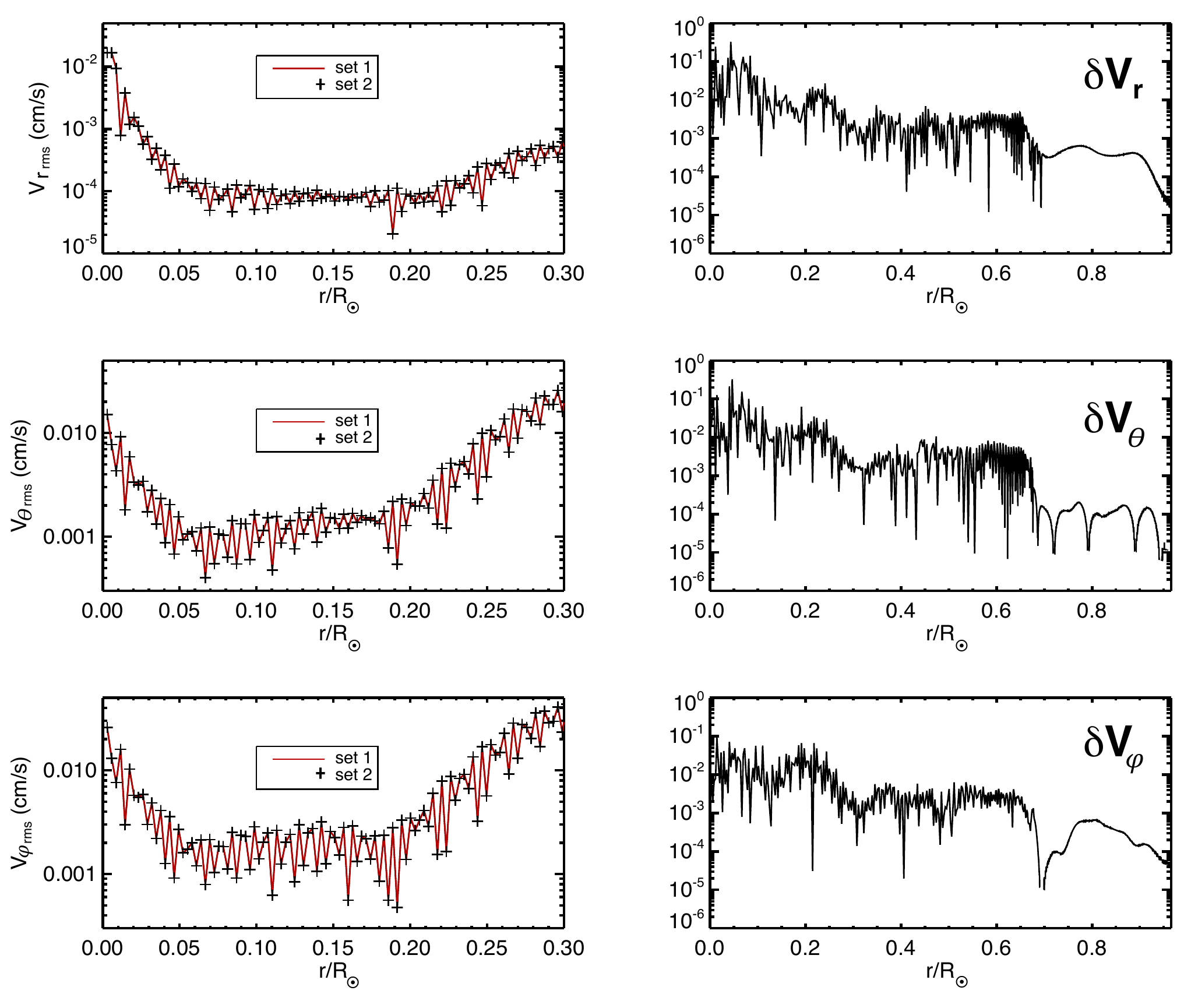}
  \caption{\textbf{Left}: Radial and horizontal rms velocities as a function of the normalized radius for model \textit{ref} (red line, calculated with set 1) and the
    same model (black crosses) calculated with set 2 at time t $\approx$ 90 days. Both curves are almost perfectly superimposed. \textbf{Right}: Difference between both models in percent:
    $\biggr\lvert\displaystyle\frac{\mathrm{v}_1-\mathrm{v}_2}{\mathrm{v}_1}\biggr\rvert \times 100$, where $\mathrm{v}_1$ corresponds to model \textit{ref}
    and $\mathrm{v}_2$ to the other one. The difference at $r=0$ is about 0.1\% and much less in the rest of the sphere.} 
  \label{fig:tests_BC0}
\end{figure*}

In Fig. \ref{fig:tests_BC0}, we compare the model \textit{ref} developed in the current article with another model calculated with set 2 (every other parameter
is identical). The radial and horizontal rms velocities in both models differ from less than 0.1\%. \\\newline
{For a better numerical stability, a more stringent condition could be to impose $Z, W \to r^{l+1}$ as $r \to 0$ \citep{2007PhRvE..75b6303B,2007JCoPh.227.1209L,Glatzmaiernook}.}

\bibliographystyle{aa}  
\bibliography{ABM2013_bib.bib} 

\begin{thebibliography}{119}
\expandafter\ifx\csname natexlab\endcsname\relax\def\natexlab#1{#1}\fi

\bibitem[{Aerts {et~al.}(2010)Aerts, Christensen-Dalsgaard, \&
  Kurtz}]{aerts2010asteroseismology}
Aerts, C., Christensen-Dalsgaard, J., \& Kurtz, D. D.~W. 2010,
  {Asteroseismology}, Astronomy and Astrophysics Library (Springer London,
  Limited)

\bibitem[{Alvan {et~al.}(2012)Alvan, Brun, \& Mathis}]{2012sf2a.conf..289A}
Alvan, L., Brun, A.~S., \& Mathis, S. 2012, in SF2A-2012: Proceedings of the
  Annual meeting of the French Society of Astronomy and Astrophysics. Eds.: S.
  Boissier, 289--293

\bibitem[{Alvan {et~al.}(2013)Alvan, Mathis, \&
  Decressin}]{2013A&A...553A..86A}
Alvan, L., Mathis, S., \& Decressin, T. 2013, Astronomy and Astrophysics, 553,
  86

\bibitem[{Andersen(1994)}]{1994SoPh..152..241A}
Andersen, B.~N. 1994, Solar Physics (ISSN 0038-0938), 152, 241

\bibitem[{Andersen(1996)}]{1996A&A...312..610A}
Andersen, B.~N. 1996, Astronomy and Astrophysics, 312, 610

\bibitem[{Andreassen {et~al.}(1992)Andreassen, Andersen, \&
  Wasberg}]{1992A&A...257..763A}
Andreassen, O., Andersen, B.~N., \& Wasberg, C.~E. 1992, Astronomy and
  Astrophysics (ISSN 0004-6361), 257, 763

\bibitem[{{Appourchaux} {et~al.}(2010){Appourchaux}, {Belkacem}, {Broomhall},
  {Chaplin}, {Gough}, {Houdek}, {Provost}, {Baudin}, {Boumier}, {Elsworth},
  {Garc{\'{\i}}a}, {Andersen}, {Finsterle}, {Fr{\"o}hlich}, {Gabriel}, {Grec},
  {Jim{\'e}nez}, {Kosovichev}, {Sekii}, {Toutain}, \&
  {Turck-Chi{\`e}ze}}]{2010A&ARv..18..197A}
{Appourchaux}, T., {Belkacem}, K., {Broomhall}, A.-M., {et~al.} 2010, \aapr,
  18, 197

\bibitem[{Baldwin {et~al.}(2001)Baldwin, Gray, Dunkerton, Hamilton, Haynes,
  Randel, Holton, Alexander, Hirota, Horinouchi, Jones, Kinnersley, Marquardt,
  Sato, \& Takahashi}]{2001RvGeo..39..179B}
Baldwin, M.~P., Gray, L.~J., Dunkerton, T.~J., {et~al.} 2001, Reviews of
  Geophysics, 39, 179

\bibitem[{{Ballot} {et~al.}(2010){Ballot}, {Ligni{\`e}res}, {Reese}, \&
  {Rieutord}}]{Ballot:2010jy}
{Ballot}, J., {Ligni{\`e}res}, F., {Reese}, D.~R., \& {Rieutord}, M. 2010,
  \aap, 518, A30

\bibitem[{Bayliss(2006)}]{BaylissPHD}
Bayliss, R.~A. 2006, PhD thesis, University of Wisconsin-Madison

\bibitem[{Bayliss {et~al.}(2007)Bayliss, Forest, Nornberg, Spence, \&
  Terry}]{2007PhRvE..75b6303B}
Bayliss, R.~A., Forest, C.~B., Nornberg, M.~D., Spence, E.~J., \& Terry, P.~W.
  2007, Physical Review E, 75, 26303

\bibitem[{Beck {et~al.}(2012)Beck, Montalban, Kallinger, De~Ridder, Aerts,
  Garcia, Hekker, Dupret, Mosser, Eggenberger, Stello, Elsworth, Frandsen,
  Carrier, Hillen, Gruberbauer, Christensen-Dalsgaard, Miglio, Valentini,
  Bedding, Kjeldsen, Girouard, Hall, \& Ibrahim}]{2012Natur.481...55B}
Beck, P.~G., Montalban, J., Kallinger, T., {et~al.} 2012, Nature, 481, 55

\bibitem[{{Bedding} {et~al.}(2011){Bedding}, {Mosser}, {Huber},
  {Montalb{\'a}n}, {Beck}, {Christensen-Dalsgaard}, {Elsworth},
  {Garc{\'{\i}}a}, {Miglio}, {Stello}, {White}, {De Ridder}, {Hekker}, {Aerts},
  {Barban}, {Belkacem}, {Broomhall}, {Brown}, {Buzasi}, {Carrier}, {Chaplin},
  {di Mauro}, {Dupret}, {Frandsen}, {Gilliland}, {Goupil}, {Jenkins},
  {Kallinger}, {Kawaler}, {Kjeldsen}, {Mathur}, {Noels}, {Aguirre}, \&
  {Ventura}}]{Bedding:2011te}
{Bedding}, T.~R., {Mosser}, B., {Huber}, D., {et~al.} 2011, \nat, 471, 608

\bibitem[{Belkacem {et~al.}(2009)Belkacem, Mathis, Goupil, \&
  Samadi}]{Belkacem:2009cc}
Belkacem, K., Mathis, S., Goupil, M.~J., \& Samadi, R. 2009, Astronomy and
  Astrophysics, 508, 345

\bibitem[{{Belkacem} {et~al.}(2009){Belkacem}, {Mathis}, {Goupil}, \&
  {Samadi}}]{Belkacem:2009wl}
{Belkacem}, K., {Mathis}, S., {Goupil}, M.~J., \& {Samadi}, R. 2009, \aap, 508,
  345

\bibitem[{Belkacem {et~al.}(2009)Belkacem, Samadi, Goupil, Dupret, Brun, \&
  Baudin}]{2009A&A...494..191B}
Belkacem, K., Samadi, R., Goupil, M.~J., {et~al.} 2009, Astronomy and
  Astrophysics, 494, 191

\bibitem[{Berthomieu \& Provost(1990)}]{1990A&A...227..563B}
Berthomieu, G. \& Provost, J. 1990, Astronomy and Astrophysics (ISSN
  0004-6361), 227, 563

\bibitem[{Berthomieu \& Provost(1991)}]{1991SoPh..133..127B}
Berthomieu, G. \& Provost, J. 1991, IRIS /International Research on the
  Interior of the Sun/ Workshop, 133, 127

\bibitem[{Booker \& Bretherton(1967)}]{Booker:1967wd}
Booker, J. \& Bretherton, F. 1967, Journal of Fluid Mechanics, 27, 513

\bibitem[{{Braginsky} \& {Roberts}(1995)}]{1995GApFD..79....1B}
{Braginsky}, S.~I. \& {Roberts}, P.~H. 1995, Geophysical and Astrophysical
  Fluid Dynamics, 79, 1

\bibitem[{{Brown} {et~al.}(2012){Brown}, {Vasil}, \& {Zweibel}}]{Brown:2012bd}
{Brown}, B.~P., {Vasil}, G.~M., \& {Zweibel}, E.~G. 2012, \apj, 756, 109

\bibitem[{Browning {et~al.}(2004)Browning, Brun, \& Toomre}]{BBT2004}
Browning, M., Brun, A.~S., \& Toomre, J. 2004, ApJ, 601, 512

\bibitem[{Brun {et~al.}(2002)Brun, Antia, Chitre, \& Zahn}]{BrunAl2002}
Brun, A.~S., Antia, H.~M., Chitre, S.~M., \& Zahn, J.-P. 2002, Astronomy and
  Astrophysics, 391, 725

\bibitem[{Brun {et~al.}(2004)Brun, Miesch, \& Toomre}]{BMT2004}
Brun, A.~S., Miesch, M.~S., \& Toomre, J. 2004, ApJ, 614, 1073

\bibitem[{Brun {et~al.}(2011)Brun, Miesch, \& Toomre}]{Brun:2011bl}
Brun, A.~S., Miesch, M.~S., \& Toomre, J. 2011, ApJ, 742, 79

\bibitem[{Brun \& Toomre(2002)}]{BrunToomre2002}
Brun, A.~S. \& Toomre, J. 2002, ApJ, 570, 865

\bibitem[{Brun {et~al.}(1998)Brun, Turck-Chi{\`e}ze, \& Morel}]{Brun1998}
Brun, A.~S., Turck-Chi{\`e}ze, S., \& Morel, P. 1998, ApJ, 506, 913

\bibitem[{Brun \& Zahn(2006)}]{BrunZahn2006}
Brun, A.~S. \& Zahn, J.-P. 2006, Astronomy and Astrophysics, 457, 665

\bibitem[{Chanmugam(1972)}]{1972NPhS..236...83C}
Chanmugam, G. 1972, Nature Physical Science, 236, 83

\bibitem[{Charbonnel {et~al.}(2013)Charbonnel, Decressin, Amard, Palacios, \&
  Talon}]{Charbonnel:2013df}
Charbonnel, C., Decressin, T., Amard, L., Palacios, A., \& Talon, S. 2013,
  Astronomy and Astrophysics, 554, A40

\bibitem[{Charbonnel \& Talon(2005)}]{CharbonnelTalon2005}
Charbonnel, C. \& Talon, S. 2005, Science, 309, 2189

\bibitem[{Christensen-Dalsgaard(1997)}]{Christensen-Dalsgaard97lecturenotes}
Christensen-Dalsgaard, J. 1997, Lecture Notes on Stellar Oscillations

\bibitem[{Christensen-Dalsgaard {et~al.}(1995)Christensen-Dalsgaard, Bedding,
  \& Kjeldsen}]{1995ApJ...443L..29C}
Christensen-Dalsgaard, J., Bedding, T.~R., \& Kjeldsen, H. 1995, Astrophysical
  Journal, 443, L29

\bibitem[{Clune {et~al.}(1999)Clune, Elliott, Miesch, Toomre, \&
  Glatzmaier}]{CluneAl1999}
Clune, T., Elliott, J., Miesch, M.~S., Toomre, J., \& Glatzmaier, G.~A. 1999,
  Parallel Computing, 25, 361

\bibitem[{De~Cat {et~al.}(2011)De~Cat, Uytterhoeven, Guti{\'e}rrez-Soto,
  Degroote, \& Sim{\'o}n-D{\'\i}az}]{DeCat:2011gq}
De~Cat, P., Uytterhoeven, K., Guti{\'e}rrez-Soto, J., Degroote, P., \&
  Sim{\'o}n-D{\'\i}az, S. 2011, Stellar Instability and Evolution, 6, 433

\bibitem[{Deheuvels {et~al.}(2012)Deheuvels, Garcia, Chaplin, Basu, Antia,
  Appourchaux, Benomar, Davies, Elsworth, Gizon, Goupil, Reese, Regulo, Schou,
  Stahn, Casagrande, Christensen-Dalsgaard, Fischer, Hekker, Kjeldsen, Mathur,
  Mosser, Pinsonneault, Valenti, Christiansen, Kinemuchi, \&
  Mullally}]{2012ApJ...756...19D}
Deheuvels, S., Garcia, R., Chaplin, W.~J., {et~al.} 2012, The astrophysical
  journal, 756, 19

\bibitem[{Dintrans {et~al.}(2005)Dintrans, Brandenburg, Nordlund, \&
  Stein}]{2005A&A...438..365D}
Dintrans, B., Brandenburg, A., Nordlund, {\AA}., \& Stein, R.~F. 2005,
  Astronomy and Astrophysics, 438, 365

\bibitem[{Dintrans \& Rieutord(2000)}]{DintransRieutord2000}
Dintrans, B. \& Rieutord, M. 2000, Astronomy and Astrophysics, 354, 86

\bibitem[{Domingo {et~al.}(1995)Domingo, Fleck, \&
  Poland}]{1995SSRv...72...81D}
Domingo, V., Fleck, B., \& Poland, A.~I. 1995, Space Science Reviews, 72, 81

\bibitem[{Duez \& Mathis(2010)}]{Duez:2010hg}
Duez, V. \& Mathis, S. 2010, Astronomy and Astrophysics, 517, A58

\bibitem[{Dunkerton(1997)}]{1997JGR...10226053D}
Dunkerton, T.~J. 1997, Journal of Geophysical Research, 102, 26053

\bibitem[{Elliott {et~al.}(2000)Elliott, Miesch, \&
  Toomre}]{2000ApJ...533..546E}
Elliott, J., Miesch, M.~S., \& Toomre, J. 2000, The astrophysical journal, 533,
  546

\bibitem[{Fritts {et~al.}(1998)Fritts, Vadas, \&
  Andreassen}]{1998A&A...333..343F}
Fritts, D.~C., Vadas, S.~L., \& Andreassen, O. 1998, Astronomy and
  Astrophysics, 333, 343

\bibitem[{Garaud \& Garaud(2008)}]{2008MNRAS.391.1239G}
Garaud, P. \& Garaud, J.~D. 2008, Monthly Notices of the Royal Astronomical
  Society, 391, 1239

\bibitem[{Garcia {et~al.}(2007)Garcia, Turck-Chi{\`e}ze, Jim{\'e}nez-Reyes,
  Ballot, Pall{\'e}, Eff-Darwich, Mathur, \& Provost}]{Garcia:2007iq}
Garcia, R.~A., Turck-Chi{\`e}ze, S., Jim{\'e}nez-Reyes, S.~J., {et~al.} 2007,
  Science, 316, 1591

\bibitem[{Garcia~Lopez \& Spruit(1991)}]{1991ApJ...377..268G}
Garcia~Lopez, R.~J. \& Spruit, H.~C. 1991, Astrophysical Journal, 377, 268

\bibitem[{Gastine \& Dintrans(2010)}]{Gastine:2010ki}
Gastine, T. \& Dintrans, B. 2010, Astrophysics and Space Science, 328, 245

\bibitem[{Giorgetta {et~al.}(2002)Giorgetta, Manzini, \&
  Roeckner}]{2002GeoRL..29.1245G}
Giorgetta, M.~A., Manzini, E., \& Roeckner, E. 2002, Geophysical Research
  Letters, 29, 1245

\bibitem[{Glatzmaier(1984)}]{1984JCoPh..55..461G}
Glatzmaier, G.~A. 1984, Journal of Computational Physics (ISSN 0021-9991), 55,
  461

\bibitem[{{Glatzmaier}(2013)}]{Glatzmaiernook}
{Glatzmaier}, G.~A. 2013, {Introduction to Modeling Convection in Planets and
  Stars : Magnetic Filed, Density stratification, Rotation} (Princeton
  University Press)

\bibitem[{Goldreich \& Kumar(1990)}]{1990ApJ...363..694G}
Goldreich, P. \& Kumar, P. 1990, Astrophysical Journal, 363, 694

\bibitem[{Goldreich {et~al.}(1994)Goldreich, Murray, \&
  Kumar}]{1994ApJ...424..466G}
Goldreich, P., Murray, N., \& Kumar, P. 1994, Astrophysical Journal, 424, 466

\bibitem[{{Goode} \& {Thompson}(1992)}]{1992ApJ...395..307G}
{Goode}, P.~R. \& {Thompson}, M.~J. 1992, \apj, 395, 307

\bibitem[{Gough(1986)}]{gough1986seismology}
Gough, D. 1986, {Seismology of the Sun and the Distant Stars}, C: Nato advanced
  science institutes series (D. Reidel Publishing Company)

\bibitem[{Gough(1993)}]{goughHouches}
Gough, D. 1993, {In: Astrophysical Fluid Dynamics - Les Houches 1987}, 399--560

\bibitem[{Gough \& McIntyre(1998)}]{1998Natur.394..755G}
Gough, D. \& McIntyre, M. 1998, Nature, 394, 755

\bibitem[{{Green} {et~al.}(2003){Green}, {Fontaine}, {Reed}, {Callerame},
  {Seitenzahl}, {White}, {Hyde}, {{\O}stensen}, {Cordes}, {Brassard}, {Falter},
  {Jeffery}, {Dreizler}, {Schuh}, {Giovanni}, {Edelmann}, {Rigby}, \&
  {Bronowska}}]{Green:2002va}
{Green}, E.~M., {Fontaine}, G., {Reed}, M.~D., {et~al.} 2003, \apjl, 583, L31

\bibitem[{Hurlburt {et~al.}(1986)Hurlburt, Toomre, \&
  Massaguer}]{1986ApJ...311..563H}
Hurlburt, N.~E., Toomre, J., \& Massaguer, J.~M. 1986, Astrophysical Journal,
  311, 563

\bibitem[{Hurlburt {et~al.}(1994)Hurlburt, Toomre, Massaguer, \&
  Zahn}]{1994ApJ...421..245H}
Hurlburt, N.~E., Toomre, J., Massaguer, J.~M., \& Zahn, J.-P. 1994,
  Astrophysical Journal, 421, 245

\bibitem[{Jones {et~al.}(2011)Jones, Boronski, Brun, Glatzmaier, Gastine,
  Miesch, \& Wicht}]{Jones:2011in}
Jones, C.~A., Boronski, P., Brun, A.~S., {et~al.} 2011, Icarus, 216, 120

\bibitem[{Kiraga {et~al.}(2003)Kiraga, Jahn, Stepien, \&
  Zahn}]{2003AcA....53..321K}
Kiraga, M., Jahn, K., Stepien, K., \& Zahn, J.-P. 2003, Acta Astronomica, 53,
  321

\bibitem[{Kiraga {et~al.}(2005)Kiraga, Stepien, \& Jahn}]{Kiraga:2005wg}
Kiraga, M., Stepien, K., \& Jahn, K. 2005, Acta Astronomica, 55, 205

\bibitem[{Kumar \& Quataert(1995)}]{Anonymous:0GcaI-aC}
Kumar, P. \& Quataert, E. 1995, 1

\bibitem[{{Kumar} {et~al.}(1999){Kumar}, {Talon}, \&
  {Zahn}}]{1999ApJ...520..859K}
{Kumar}, P., {Talon}, S., \& {Zahn}, J.-P. 1999, \apj, 520, 859

\bibitem[{Landolt(1968)}]{1968ApJ...153..151L}
Landolt, A.~U. 1968, Astrophysical Journal, 153, 151

\bibitem[{Lantz(1992)}]{LantzPHD}
Lantz, S.~R. 1992, PhD thesis, Cornell University

\bibitem[{{Lecoanet} \& {Quataert}(2013)}]{Lecoanet:ws}
{Lecoanet}, D. \& {Quataert}, E. 2013, \mnras, 430, 2363

\bibitem[{Lighthill(1978)}]{1978cup..book.....L}
Lighthill, J. 1978, Cambridge

\bibitem[{Lighthill(1986)}]{LighthillBook}
Lighthill, J. 1986, Provided by the SAO/NASA Astrophysics Data System

\bibitem[{Livermore {et~al.}(2007)Livermore, Jones, \&
  Worland}]{2007JCoPh.227.1209L}
Livermore, P.~W., Jones, C.~A., \& Worland, S.~J. 2007, Journal of
  Computational Physics, 227, 1209

\bibitem[{Massaguer {et~al.}(1984)Massaguer, Latour, Toomre, \&
  Zahn}]{1984A&A...140....1M}
Massaguer, J.~M., Latour, J., Toomre, J., \& Zahn, J.-P. 1984, Astronomy and
  Astrophysics (ISSN 0004-6361), 140, 1

\bibitem[{Mathis(2009)}]{Mathis2009}
Mathis, S. 2009, Astronomy and Astrophysics, 506, 811

\bibitem[{Mathis \& de~Brye(2011)}]{MathisDeBrye2011}
Mathis, S. \& de~Brye, N. 2011, Astronomy and Astrophysics, 526, A65

\bibitem[{Mathis \& de~Brye(2012)}]{Mathis:2012tn}
Mathis, S. \& de~Brye, N. 2012, Astronomy and Astrophysics, 1

\bibitem[{Mathis {et~al.}(2013)Mathis, Decressin, Eggenberger, \&
  Charbonnel}]{2013A&A...558A..11M}
Mathis, S., Decressin, T., Eggenberger, P., \& Charbonnel, C. 2013, Astronomy
  and Astrophysics, 558, 11

\bibitem[{Mathis \& Neiner(2013)}]{Mathis:2013wv}
Mathis, S. \& Neiner, C. 2013, SF2A-2013: Proceedings of the Annual meeting of
  the French Society of Astronomy and Astrophysics, -1, 241

\bibitem[{Mathis {et~al.}(2008)Mathis, Talon, Pantillon, \&
  Zahn}]{Mathis:2008ba}
Mathis, S., Talon, S., Pantillon, F.~P., \& Zahn, J.-P. 2008, Solar Physics,
  251, 101

\bibitem[{{Mathis} \& {Zahn}(2004)}]{2004A&A...425..229M}
{Mathis}, S. \& {Zahn}, J.-P. 2004, \aap, 425, 229

\bibitem[{{Mathur} {et~al.}(2008){Mathur}, {Eff-Darwich}, {Garc{\'{\i}}a}, \&
  {Turck-Chi{\`e}ze}}]{Mathur:2008hs}
{Mathur}, S., {Eff-Darwich}, A., {Garc{\'{\i}}a}, R.~A., \& {Turck-Chi{\`e}ze},
  S. 2008, \aap, 484, 517

\bibitem[{Miesch {et~al.}(2000)Miesch, Elliott, Toomre, Clune, Glatzmaier, \&
  Gilman}]{2000ApJ...532..593M}
Miesch, M.~S., Elliott, J.~R., Toomre, J., {et~al.} 2000, The astrophysical
  journal, 532, 593

\bibitem[{{Montalban, J.}(1994)}]{Montalban:1994wx}
{Montalban, J.} 1994, Astronomy and Astrophysics, 281, 421

\bibitem[{{Moravveji} {et~al.}(2012){Moravveji}, {Moya}, \&
  {Guinan}}]{2012ApJ...749...74M}
{Moravveji}, E., {Moya}, A., \& {Guinan}, E.~F. 2012, \apj, 749, 74

\bibitem[{Mosser {et~al.}(2013)Mosser, Belkacem, \& Vrard}]{Mosser:2013uk}
Mosser, B., Belkacem, K., \& Vrard, M. 2013, astro-ph

\bibitem[{{Mosser} {et~al.}(2013){Mosser}, {Samadi}, \&
  {Belkacem}}]{Mosser:2013vu}
{Mosser}, B., {Samadi}, R., \& {Belkacem}, K. 2013, 25

\bibitem[{Müller {et~al.}(1986)Müller, Holloway, Henyey, \&
  Pomphrey}]{ROG:ROG1096}
Müller, P., Holloway, G., Henyey, F., \& Pomphrey, N. 1986, Reviews of
  Geophysics, 24, 493

\bibitem[{Neiner {et~al.}(2012)Neiner, Floquet, Samadi, Espinosa~Lara,
  Fr{\'e}mat, Mathis, Leroy, de~Batz, Rainer, Poretti, Mathias, Guarro~Fl{\'o},
  Buil, Ribeiro, Alecian, Andrade, Briquet, Diago, Emilio, Fabregat,
  Guti{\'e}rrez-Soto, Hubert, Janot-Pacheco, Martayan, Semaan, Suso, \&
  Zorec}]{2012A&A...546A..47N}
Neiner, C., Floquet, M., Samadi, R., {et~al.} 2012, Astronomy and Astrophysics,
  546, 47

\bibitem[{Plumb \& McEwan(1978)}]{1978JAtS...35.1827P}
Plumb, R.~A. \& McEwan, A.~D. 1978, Journal of Atmospheric Sciences, 35, 1827

\bibitem[{Press(1981)}]{Press1981}
Press, W.~H. 1981, ApJ, 245, 286

\bibitem[{Provost \& Berthomieu(1986)}]{1986A&A...165..218P}
Provost, J. \& Berthomieu, G. 1986, Astronomy and Astrophysics (ISSN
  0004-6361), 165, 218

\bibitem[{Rogers \& Glatzmaier(2005{\natexlab{a}})}]{RogersGlatzmaier2005}
Rogers, T.~M. \& Glatzmaier, G.~A. 2005{\natexlab{a}}, Monthly Notices of the
  Royal Astronomical Society, 364, 1135

\bibitem[{Rogers \& Glatzmaier(2005{\natexlab{b}})}]{2005ApJ...620..432R}
Rogers, T.~M. \& Glatzmaier, G.~A. 2005{\natexlab{b}}, The astrophysical
  journal, 620, 432

\bibitem[{Rogers \& Glatzmaier(2006)}]{2006ApJ...653..756R}
Rogers, T.~M. \& Glatzmaier, G.~A. 2006, The astrophysical journal, 653, 756

\bibitem[{{Rogers} {et~al.}(2006){Rogers}, {Glatzmaier}, \&
  {Jones}}]{Rogers:2006ks}
{Rogers}, T.~M., {Glatzmaier}, G.~A., \& {Jones}, C.~A. 2006, \apj, 653, 765

\bibitem[{{Rogers} {et~al.}(2013){Rogers}, {Lin}, {McElwaine}, \&
  {Lau}}]{Rogers:2013ui}
{Rogers}, T.~M., {Lin}, D.~N.~C., {McElwaine}, J.~N., \& {Lau}, H.~H.~B. 2013,
  \apj, 772, 21

\bibitem[{Rogers \& MacGregor(2010)}]{2010MNRAS.401..191R}
Rogers, T.~M. \& MacGregor, K.~B. 2010, Monthly Notices of the Royal
  Astronomical Society, 401, 191

\bibitem[{{Rogers} {et~al.}(2008){Rogers}, {MacGregor}, \&
  {Glatzmaier}}]{Rogers:2008bl}
{Rogers}, T.~M., {MacGregor}, K.~B., \& {Glatzmaier}, G.~A. 2008, \mnras, 387,
  616

\bibitem[{Saikia {et~al.}(2000)Saikia, Singh, Chan, Roxburgh, \&
  Srivastava}]{2000ApJ...529..402S}
Saikia, E., Singh, H.~P., Chan, K.~L., Roxburgh, I.~W., \& Srivastava, M.~P.
  2000, The astrophysical journal, 529, 402

\bibitem[{{Samadi} {et~al.}(2003){Samadi}, {Nordlund}, {Stein}, {Goupil}, \&
  {Roxburgh}}]{Samadi:2003ir}
{Samadi}, R., {Nordlund}, {\AA}., {Stein}, R.~F., {Goupil}, M.~J., \&
  {Roxburgh}, I. 2003, \aap, 403, 303

\bibitem[{Schatzman(1993)}]{1993A&A...279..431S}
Schatzman, E. 1993, Astronomy and Astrophysics (ISSN 0004-6361), 279, 431

\bibitem[{Schatzman(1996)}]{1996SoPh..169..245S}
Schatzman, E. 1996, Solar Physics, 169, 245

\bibitem[{Shiode {et~al.}(2013)Shiode, Quataert, Cantiello, \&
  Bildsten}]{Shiode:2013kp}
Shiode, J.~H., Quataert, E., Cantiello, M., \& Bildsten, L. 2013, Monthly
  Notices of the Royal Astronomical Society, 430, 1736

\bibitem[{Spiegel \& Zahn(1992)}]{Spiegel:1992tr}
Spiegel, E. \& Zahn, J.-P. 1992, Astronomy and Astrophysics, 265, 106

\bibitem[{Strugarek {et~al.}(2011{\natexlab{a}})Strugarek, Brun, \&
  Zahn}]{2011A&A...532A..34S}
Strugarek, A., Brun, A.~S., \& Zahn, J.-P. 2011{\natexlab{a}}, astro-ph, 34

\bibitem[{Strugarek {et~al.}(2011{\natexlab{b}})Strugarek, Brun, \&
  Zahn}]{2011AN....332..891S}
Strugarek, A., Brun, A.~S., \& Zahn, J.-P. 2011{\natexlab{b}}, Astronomische
  Nachrichten, 332, 891

\bibitem[{{Talon} \& {Charbonnel}(2003)}]{2003A&A...405.1025T}
{Talon}, S. \& {Charbonnel}, C. 2003, \aap, 405, 1025

\bibitem[{Talon \& Charbonnel(2005)}]{Talon:2005iu}
Talon, S. \& Charbonnel, C. 2005, Astronomy and Astrophysics, 440, 981

\bibitem[{{Talon} \& {Charbonnel}(2008)}]{2008A&A...482..597T}
{Talon}, S. \& {Charbonnel}, C. 2008, \aap, 482, 597

\bibitem[{Thompson {et~al.}(2003)Thompson, Christensen-Dalsgaard, Miesch, \&
  Toomre}]{2003ARA&A..41..599T}
Thompson, M.~J., Christensen-Dalsgaard, J., Miesch, M.~S., \& Toomre, J. 2003,
  Annual Review of Astronomy {\&}Astrophysics, 41, 599

\bibitem[{Turck-Chi{\`e}ze {et~al.}(2004)Turck-Chi{\`e}ze, Garcia, Couvidat,
  Ulrich, Bertello, Varadi, Berthomieu, Brun, Lopes, Pall{\'e}, Provost,
  Robillot, Kosovichev, Gabriel, \& Roca~Cortes}]{TurckChieze:2004vw}
Turck-Chi{\`e}ze, S., Garcia, R., Couvidat, S., {et~al.} 2004, The
  astrophysical journal, 604, 455

\bibitem[{Unno {et~al.}(1989)Unno, Osaki, Ando, Saio, \&
  Shibahashi}]{1989nos..book.....U}
Unno, W., Osaki, Y., Ando, H., Saio, H., \& Shibahashi, H. 1989, Nonradial
  oscillations of stars

\bibitem[{Vauclair(2005)}]{2005EAS....17..133V}
Vauclair, G. 2005, EAS Publications Series, 17, 133

\bibitem[{Voisin(1991)}]{1991JFM...231..439V}
Voisin, B. 1991, Journal of Fluid Mechanics, 231, 439

\bibitem[{Waelkens(1991)}]{1991A&A...246..453W}
Waelkens, C. 1991, Astronomy and Astrophysics (ISSN 0004-6361), 246, 453

\bibitem[{Warner \& Robinson(1972)}]{1972NPhS..239....2W}
Warner, B. \& Robinson, E.~L. 1972, Nature Physical Science, 239, 2

\bibitem[{{Winget} \& {Kepler}(2008)}]{Winget:2008dz}
{Winget}, D.~E. \& {Kepler}, S.~O. 2008, \araa, 46, 157

\bibitem[{Zahn(1983)}]{1983apum.conf..253Z}
Zahn, J.-P. 1983, Saas-Fee Advanced Course 13, 253

\bibitem[{Zahn(1991)}]{1991A&A...252..179Z}
Zahn, J.-P. 1991, Astronomy and Astrophysics (ISSN 0004-6361), 252, 179

\bibitem[{Zahn(1992)}]{1992A&A...265..115Z}
Zahn, J.-P. 1992, Astronomy and Astrophysics (ISSN 0004-6361), 265, 115

\bibitem[{Zahn {et~al.}(1997)Zahn, Talon, \& Matias}]{ZahnTalonMatias1997}
Zahn, J.-P., Talon, S., \& Matias, J. 1997, Astronomy and Astrophysics, 322,
  320

\end{thebibliography}
 
\end{document}